\documentclass[aps,prd,secnumarabic,twocolumn,nofootinbib,superscriptaddress,floatfix]{revtex4}

\usepackage{blindtext}

\usepackage{graphics}
\usepackage{graphicx}

\usepackage{cancel}

\usepackage{color}

\usepackage{graphicx}

\usepackage{amssymb}

\usepackage{amsmath}

\usepackage{epstopdf}

\usepackage{xspace}

\usepackage{mathbbol}

\usepackage{wasysym} 

\usepackage{bbm}     

\usepackage{rotating}

\usepackage{dcolumn}% Align table columns on decimal point

\usepackage{bm}% bold math

\usepackage{longtable}% Include figure files

\usepackage{multirow}

\usepackage{enumitem}

\begin{document}

%%%%%%%%%%%%%%%%%%%%%%
%\input{Title}

%\setpagewiselinenumbers
%\modulolinenumbers[5]
%\switchlinenumbers
%\linenumbers

\title{
{\rm
%{\hbox to\textwidth{\hfill \today}}
{\hbox to\textwidth{\hfill ~~}}
{\hbox to\textwidth{\hfill arXiv:15zz.xxyy}}
{\hbox to\textwidth{\hfill ~~}}
{\hbox to\textwidth{\hspace{0.3\textwidth}{ SI-HEP-2015-02, QFET-2015-02, DF/4/2015,  MITP/15-052}
    }}
%{\hbox to\textwidth{\hspace{0.5\textwidth}{preprint number 3}
%      {\hfill preprint number n}}}
%{\hbox to\textwidth{\hfill OTHER-preprint}}
~\\
~\\
}
\Large Flavour, Electroweak Symmetry Breaking and Dark Matter: state of  the art and future prospects}
%%%%%%%%%%%%%%

%\input{Authors}
\author{Giulia Ricciardi}
\affiliation{Dipartimento di Fisica, Universit\`a  degli  Studi di Napoli Federico II, and   Istituto  Nazionale  di  Fisica  Nucleare, Sezione di Napoli,
Complesso Universitario di Monte Sant'Angelo, Via Cintia,
I-80126 Napoli, Italy}

\author{Alexandre Arbey}
\affiliation{
Universit\'e de Lyon, Universit\'e Lyon 1, F-69622 Villeurbanne Cedex, 
France; Centre de Recherche Astrophysique de Lyon, CNRS, UMR 5574, 
Saint-Genis Laval Cedex, F-69561, France; Ecole Normale Sup\'erieure de 
Lyon, France
and CERN Theory Division, Physics Department, CH-1211 Geneva 23, Switzerland}

\author{Enrico Bertuzzo}
\affiliation{IFAE, Universitat Aut\`onoma de Barcelona, 08193 Bellaterra, Barcelona, Spain}

\author{Adri\'an Carmona}
\affiliation{Institute for Theoretical Physics, 
ETH Zurich, 8093 Zurich, Switzerland}

\author{Radovan Derm\' \i\v sek}
\affiliation{Physics Department, Indiana University, Bloomington, IN 47405, USA}

\author{Tobias Huber}
\affiliation{Theoretische Physik 1, Naturwissenschaftlich-Technische Fakult\"at, Universit\"at Siegen,  D-57068 Siegen, Germany}

\author{Tobias Hurth}
\affiliation{PRISMA Cluster of Excellence and Institute for Physics (THEP),
Johannes Gutenberg University, D-55099 Mainz, Germany}

\author{Yuval Grossman}
\affiliation{Laboratory for Elementary-Particle Physics, Cornell University, Ithaca, N.Y., USA}

\author{J\"orn Kersten}
\affiliation{University of Bergen, Department of Physics and Technology, PO Box
7803, 5020 Bergen, Norway}

\author{Enrico Lunghi}
\affiliation{Physics Department, Indiana University, Bloomington, IN 47405, USA}

\author{Farvah Mahmoudi}
\affiliation{
Universit\'e de Lyon, Universit\'e Lyon 1, F-69622 Villeurbanne Cedex, 
France; Centre de Recherche Astrophysique de Lyon, CNRS, UMR 5574, 
Saint-Genis Laval Cedex, F-69561, France; Ecole Normale Sup\'erieure de 
Lyon, France
and CERN, CH-1211 Geneva, Switzerland}

\author{Antonio Masiero}
\affiliation{Dipartimento  di  Fisica  ed  Astronomia  G.Galilei,
Universit\`a   degli  Studi  di  Padova,  and  Istituto  Nazionale  di  Fisica  Nucleare,
Sezione  di  Padova,  Via  Marzolo  8,  35131  Padova,  Italy}

\author{Matthias Neubert}
\affiliation{PRISMA Cluster of Excellence and Institute for Physics (THEP),
Johannes Gutenberg University, D-55099 Mainz, Germany}

\author{William Shepherd}
\affiliation{Department of Physics and Santa Cruz Institute for Particle Physics, University of California Santa Cruz, California USA}

\author{Liliana Velasco-Sevilla}
\affiliation{University of Bergen, Department of Physics and Technology, PO Box
7803, 5020 Bergen, Norway}.

\begin{abstract}
%\newpage
%\begin{center}
%{\large Abstract}
%\end{center}
%~\\\noindent 
%document about
 %interesting topics we have discussed
With the discovery of the Higgs boson the Standard Model has become a complete and comprehensive theory, which has been verified with unparalleled precision and in principle might be valid at all scales. However, several reasons remain why we firmly believe that there should be physics beyond the Standard Model.
Experiments such as the LHC, new $B$ factories, and earth- and space-based astro-particle experiments provide us with unique opportunities to discover a coherent framework for many of the long-standing puzzles of our field. 
Here we  explore several significant  interconnections between the physics of the Higgs boson, the physics of flavour, and the experimental clues we have about dark matter.

\end{abstract}
%\PACS{
%      {PACS-key}{discribing text of that key}   \and
%      {PACS-key}{discribing text of that key}
%     } % end of PACS codes
%} %end of abstract

%\date{\today}

\maketitle

%\footnotetext[1]{Editors}
%\footnotetext[2]{Section Coordinators}
%\footnotetext[3]{Now at ...}

\renewcommand{\thefootnote}{\arabic{footnote}}

{\small \tableofcontents}

%%%%%%%%%%%%%
%%%%%%%%%%%%%%%
%\input{Introduction}
\section{Preface$^1$}
\label{sec:intro}
 
\addtocounter{footnote}{1}
\footnotetext{Contributing authors: Matthias Neubert and Giulia Ricciardi}

%The intention of this program was to explore in depth
It is interesting to explore the various interconnections between the physics of the Higgs boson, the physics of flavour, and the experimental clues we have about dark matter. All of these fields are at the boundary of the Standard Model, and many connections between them exist. With the discovery of the Higgs boson the Standard Model has become a complete and comprehensive theory, which has been verified with unparalleled precision and in principle might be valid at all scales. However, several reasons remain why we firmly believe that there should be physics beyond the Standard Model. Observational facts, such as the strong evidence for existence of dark matter and dark energy, neutrino masses and the cosmic matter-antimatter asymmetry are not explained by the Standard Model. Also, we are lacking a compelling theory of flavour, which can explain the striking patterns and hierarchies seen in the spectrum of fermion masses and mixings. The possibility of a unification of the fundamental forces (including gravity) is still to be proved or disproved. In low-energy Supersymmetry, new particles near the TeV scale are required for a successful unification of the gauge coupling constants. A possible signature of unification at a scale around $10^{16}$\,GeV would be proton decay mediated by new heavy particles. 
%(scalars or gauge bosons),  related to the unified physics at that scale, which do not respect the baryon and lepton number symmetries. For a mediator of mass approximately $10^{16}$\,GeV we expect a proton lifetime in the ballpark of $10^{34}$ years, which is in principle experimentally accessible. Another signature of unification would be the observation of neutron--antineutron oscillations, given a unified symmetry breaking down to an intermediate symmetry subsequently spontaneously broken with the breaking of the baryon number of two units (for example, $SO(10)\to SU(4)_{PS}\times SU(2)_L\times SU(2)_R\to SU(3)_c\times SU(2)_L\times U(1)_Y$). 
When the Standard Model comes in touch with much more massive particles related to a new energy scale, the gauge hierarchy problem becomes a twofold puzzle. We are not only concerned with the origin of scales $M_{\rm GUT}$ or $M_{\rm Planck}$ much larger than the electroweak scale, but also with the stabilization of the Higgs mass near the weak scale at any order in perturbation theory. 
%New scales have also been assumed by the analysis of recent (and controversial) BICEP2 data, suggesting inflation at about $10^{16}$\,GeV.
A crucial question in this context in that about the fundamental mechanism behind electroweak symmetry breaking. Precision measurements of the properties of the discovered Higgs boson, including its couplings to the gauge bosons and fermions of the Standard Model, may open a portal to discover some physics beyond the Standard Model, and it is not unlikely that this new physics might be connected to the dark sector of the Universe.

Indeed, many research themes exploring the physics beyond the Standard Model are related to several aspects within a global approach, involving electroweak symmetry breaking, flavour phenomena and dark matter. Trivial examples are the connections between dark matter and electroweak symmetry breaking given by the possibility of the existence of weekly interacting massive particles (WIMPs) or by searches for dark matter at colliders. Experiments such as the LHC, new $B$ factories, and earth- and space-based astro-particle experiments provide us with unique opportunities to discover a coherent framework for many of the long-standing puzzles of our field. Hopefully, some signatures of new physics will be identified in the coming years, and it will then be important to delineate the ensuing implications. 
%Higher intensity is the path to follow for the exploration of the flavour structure with the study of rare or forbidden decays, both in the quark and in the lepton sector, of tiny deviations from the SM expectations, unification, undiscovered symmetries, the search for other sources of CP violation, the possibility of a weakly coupled hidden sector that is related to dark matter, and so on.
Here,  some of these interdisciplinary aspects will be examined.

%
%{\bf [Should we include some nice photos of the workshop here?]}
%

%%%%%%%%%%%%%%
%%%%%%%%%%%%%
%\input{Masiero}
\section{Susy prospects for the next LHC run and  Dark  Mattern}
\label{sec:EFT}
\subsection{Introduction$^2$}

\addtocounter{footnote}{1}

\footnotetext{Contributing author: Antonio Masiero}

After the discovery of a light scalar of $\sim 125$  GeV with all the right features to represent the Higgs boson of the Standard Model (SM), the central open issue related to the electroweak symmetry breaking remains its naturalness. Namely, is such 125 GeV scalar mass resulting from a purely accidental and extremely precise tuning of parameters at the far ultraviolet scale (maybe, the grand unification scale or the Planck scale) or is it the fruit of some yet unknown dynamics active at the electroweak (ELW) scale? 

If we take the latter road, namely we invoke some dynamics to stabilise the ELW symmetry breaking scale, then quite a few options have been widely scrutinised in the literature: supersymmetry (SUSY), compositeness for the Higgs boson, extra-dimensions, quantum gravity at the ELW scale are some of the natural solutions that attracted most attention. Undoubtedly, SUSY, or, more precisely, low-energy SUSY sticks out among them as that which succeeded to produce complete particle model extensions of the SM to be tested in high-energy and high-intensity facilities. 

Such plus of low-energy SUSY has proven to become along the years also its major drawback. The failure of many experiments conducted at several facilities worldwide, in particular those at LEP and LHC at CERN, to, directly or indirectly, reveal the presence of SUSY partner particles has severely undermined the initial enthusiasm surrounding SUSY at “the” solution of the gauge hierarchy problem at  its start in the ‘80s. 

Indeed, already at the end of the LEP activity, and even more with the advent of LHC, it has become clear that some level of tuning of the parameters (i.e.. of unnaturalness) has to be inherently present in any kind of “natural” dynamics stabilising the Higgs mass at the O(100 GeV) scale. The reason is quite simple: barring very specific (and contrived) constructions one can generally expect that in order to stabilise the Higgs mass at $O$ (100 GeV) the new dynamics should be produced by particles and, in general, new physics beyond the SM (BSM) present at that energy scale. However, all the searches for such $O$(100 GeV) new physics have been so far unsuccessful  both in flavour experiments (in particular those concerning GIM-suppressed flavour changing neutral current and CP violating process) and in high-energy searches. 

As we said above, low-energy SUSY extensions of the SM were particularly suitable to be tested by all such experiments looking for deviations from the SM physics given the fact that they could give rise to concrete, complete physics models. Roughly speaking, after the first run of the LHC (at 7 and 8 TeV), one should conclude that coloured SUSY particles should be heavier than 1 TeV, whilst for SUSY particles with only electroweak interactions the bound remains in the hundreds of GeV range. Though, important caveats exist: the most remarkable one, concerns the scalar partner of the top quark, the stop, which could still be much lighter that 1 TeV in very peculiar realizations of the SUSY particle spectrum. 

The value of the Higgs mass which was experimentally found contributed to add to the above mentioned tension between a dynamical explanation of the ELW scale and the request of having a natural way of achieving its stabilisation. Indeed, in the minimal  SUSY extension of the SM, the MSSM (Minimal Supersymmetric Standard Model), where the minimal number of superfields  strictly needed to supersymmetrise the SM  is introduced, the lightest scalar boson (corresponding to the Higgs SM) is predicted to have a mass of the order of that of the Z boson at the tree level, i.e. before taking into account the radiative correction. The value of 125 GeV can be obtained for the lightest scalar in the MSSM only in a very restricted area of the SUSY parameters (indeed, about 135 GeV is the maximum possible value which can be obtained exploring the entire huge parameter space of the MSSM).

Still remaining inside the "minimality" of the SUSY version of the SM,
% (i.e., using only those superfields which are strictly needed to supersymmetrize the SM), 
there is ample freedom in constructing the specific SUSY model. Indeed, the class of SUSY models respecting such "minimality" criterion, the MSSM
% the so-called Minimal SUSY SM (MSSM), 
have O(100) free parameters. A reasonable restriction on such enormous parameter space is i) to go from O(100) to O(10) parameters regulating the relevant masses and mixings of the SUSY particle spectrum and ii) impose that the values of such free parameters lead to the construction of phenomenologically viable models. These phenomenologically allowed MSSM (denoted by pMSSM) couple minimality with more freedom than in the constrained MSSM. The pMSSM will be discussed in what follows. 

%The MSSM, in spite of being the minimal SUSY extension of the SM, has still $O$(100) free parameters; 
Adding rather drastic assumptions  (for instance, universality of the gaugino and sfermion masses) to the minimality of the MSSM, one can construct new versions of the MSSM with a much smaller number of free parameters . For instance, the so-called Constrained MSSM (CMSSM) or the minimal SUGRA model,  have only 4 or 5 free parameters. The value of the Higgs mass of 125 GeV, combined with  all the other existing constraints, makes it even more difficult to 
construct phenomenologically viable very constrained versions of the MSSM. Even the MSSM can still survive, but at the price of going to a very restrictive corner of the SUSY parameter space. The critical point in minimal SUSY extensions of the SM where the Higgs scalar sector is represented by two iso-doublet Higgs superfields is the absence of quartic scalar terms with a free parameter, i.e. the analogue  of the        $\lambda\, H^4$ term of the SM Higgs potential. 

That’s why great interest has arisen about non-minimal SUSY extensions of the SM where it is possible to have such quartic terms.  
The most studied class of such non-minimal MSSM adds to the above two Higgs superfield doublets also a singlet scalar N. Coupling N to the mentioned two Higgs doublets can yield the quartic Higgs term allowing for a larger SUSY Higgs mass at tree level. Section \ref{Supersymmetryandnaturalness} will discuss the phenomenology of such a non-minimal SUSY model. Interestingly enough, such a model can more easily pass all the usual high-nergy and high-intensity tests. The severe constraints that have to be applied on the SUSY parameter space of the minimal case can now be relaxed; the price to pay is some degree of further complication of the model with the introduction of a new singlet superfield whose presence is not dictated by the mere request of supersymmetrizing the SM. 

The main reason motivating the introduction of low-energy SUSY in the phenomenological arena at the beginning of the ‘80s was the search for a dynamics ensuring an ultraviolet cutoff at the ELW scale for the SM. Soon after the construction of the first low-energy SUSY realizations, it was realized that the presence of the SM SUSY partners at the ELW scale was entailing two consequences of utmost relevance:  grand unification, i.e. a common value of the electroweak and strong gauge coupling constants, was successfully achieved and an interesting candidate for cold dark matter (CDM) was emerging from the SUSY particle spectrum. 
Indeed, such two relevant implications of low-energy SUSY turn out to be somewhat linked one to the other. Asking for the supersymmetrization of the SM encounters a major block: the presence of ELW scale SUSY particles naturally yields a unbearably fast proton decay. To ensure baryon number conservation an additional discrete symmetry has to be imposed, the so-called R parity, distinguishing ordinary particles from their superpartners. 
Then, proton decay proceeds through the exchange of super-heavy SUSY particles present at the above mentioned grand unification scale and a sufficiently long proton lifetime can be enforced. 
But the presence of R yields another major consequence: the lightest SUSY particle cannot decay and if it is neutral and interacts only weakly, it can represent a Weakly Interacting Massive Particle (WIMP), an interesting candidate of CDM. 

In  Section \ref{Supersymmetricdarkmatter} we consider the particularly interesting case where such SUSY WIMP is represented by the lightest neutral SUSY fermion, the so-called lightest neutralino. 

Such a DM candidate can be searched for in three different ways: direct searches through the recoil of target nuclei hit by the cosmic neutralino, indirect searches through the study of products (mainly gamma rays, antiparticles and neutrinos) of neutralino annihilation and its production in collisions at the LHC. The combination of the important new bounds coming from these search roads with the constraints on the SUSY parameter space from SUSY particle searches in the first 7 and 8 TeV LHC run is providing a new interesting picture of the DM SUSY issue.  

The next subsections  offer an insight on where we stand on the issue of low-energy SUSY, in particular after the 8 TeV LHC run. They show that the possibility of coping with the gauge hierarchy problem through the dynamics of a low-energy SUSY extension of the SM is still well alive, although we are likely to be forced to give up the simplest SUSY model constructions of the last three decades. The newly started run of LHC at 13-14 TeV will be able to shed precious light on the existence of SUSY at the ELW scale. As shown in this Section, if not finding the much sought for SUSY particles, at least such new LHC run, together with the relentless (direct and indirect) searches for dark matter, are going to very strongly define the space of phenomenologically viable low-energy SUSY models.

%%%%%%%%%%%%%%

\subsection{Supersymmetry and naturalness$^3$}
\label{Supersymmetryandnaturalness}

\addtocounter{footnote}{-1}

\footnotetext{Contributing author: Enrico Bertuzzo}

If one firm conclusion can be drawn examining the outcome of LHC-I, it is that our concept of naturalness is becoming more and more at odds with experiments. Although no firm theorem about naturalness and fine tuning can be stated, the physical implications of the problem can be clearly stated: whenever an elementary scalar is present in a theory, radiative corrections tend to push its mass to values only a loop factor below the theory's cut off (provided the scalar interacts with particles living at the cutoff). It is then clear that {\it light} scalars imply either a low cut off, or the need of precise cancellations between the different contributions entering in the determination of the physical scalar mass. 
The more precise these cancellations, the more the theory is tuned. With the Higgs boson discovery the SM is now complete and self consistent. The problem however persists, since we can expect gravity to play a role at very short distances of order the Plank mass: what is then keeping the Higgs boson mass at $126$ GeV, {\it i.e.} $16$ orders of magnitude smaller than the Plank scale?\\

A possible solution is to make the SM supersymmetric: the unification of scalars and fermions in a unique symmetry multiplet allows to reduce the mass sensitivity to the cut off. In this way, the scalar mass can be much lighter than the cut off itself. The immediate drawback is that no unbroken SUSY multiplet has been observed in nature, signaling that SUSY must be broken. SUSY breaking reintroduces the problem back: the scalar mass is now quadratically sensitive to the scale of SUSY breaking. The hope now is that such a scale can be low enough to avoid the fine tuning problem.

Since no SUSY partners have been observed at the LHC, however, the conclusion that can be drawn is rather firm: in the minimal SUSY extension of the SM (MSSM), a fine tuning parameter of \emph{at least} $\Sigma_v \simeq 100$ is needed~\cite{Hall:2011aa} (see Eq.~\ref{eq:FT_measure1} below for the definition of $\Sigma_v$) in order to accommodate for $m_h = 126$ GeV. The MSSM is thus tuned at best at the percent level. 
Of course it may well be that this is the level at which nature is tuned; on the other hand, such a large level of tuning can be seen as a motivation to seek for more natural SUSY extensions of the SM. As we are going to see, one of the general prices to pay is minimality (in the sense of particle content and/or symmetry structure). 

In what follows, we will discuss two conceptually different frameworks (without any attempt at completeness): the well known case of the NMSSM (based mainly on~\cite{Gherghetta:2012gb}), in which the fine tuning is improved raising the Higgs boson mass at tree level, and SUSY models with Dirac Gauginos (based on \cite{Bertuzzo:2014bwa}), in which instead the fine tuning is ameliorated via additional loop contributions to the Higgs boson mass.

\subsubsection{Framework 1: NMSSM}

By definition, the NMSSM  is obtained from the MSSM by adding a chiral singlet $S$ to the particle content. Among the many supersymmetric interactions that can be written between $S$ and the Higgs doublet, we will focus on the case in which all the superfields are charged under a $\mathbb{Z}_3$ symmetry. In this so called ``scale invariant'' NMSSM the relevant superpotential is
\begin{equation}
 W = \lambda S H_u H_d + \frac{k}{3} S^3\; ,
\end{equation}
while the soft SUSY breaking potential is given by
\begin{eqnarray}
 V_{SSB} &=& m_{H_u}^2 |H_u|^2 + m_{H_d}^2 |H_d|^2 + m_S^2 |S|^2 \nonumber \\
 && + \left( a_\lambda S H_u H_d + \frac{a_k}{3} S^3 + h.c. \right) \; .
\end{eqnarray}
The couplings $\lambda$ and $k$ will be required to be perturbative up to the cutoff scale $\Lambda_{mess}$, which will be taken to be relatively 
low, in the range $\Lambda_{mess} = 20 - 1000$ TeV. The scale $\Lambda_{mess}$ could be associated with the mass of messengers fields that 
communicate SUSY breaking to the visible sector, or may be interpreted as the scale at which the NMSSM fields emerge from an underlying strong sector.\\

The improved naturalness is due to the additional tree level quartic coupling for the Higgs doublets produced by the singlet $F$-term. Indeed, once $H_u$ and $H_d$ acquire respectively vevs $v_u$ and $v_d$, the mass of the Higgs-like scalar ({\it i.e.} the scalar acquiring the electroweak vev $v$, defined as usual in terms of $\tan\beta = v_u/v_d$ as $h = \cos\beta H_d^0 + \sin\beta H_u^0$) is given by
\begin{equation}
 m_h^2 = m_Z^2 \cos^2 2\beta + \lambda^2 v^2 \sin^2 2\beta \, ,
\end{equation}
so that for moderate $\tan\beta$ it can be raised above the MSSM limit, reducing the sensitivity of the physical mass to loop corrections.

Let us notice that $m_h$ is not the mass of the lightest physical scalar: in general important mixing terms with the singlet and the orthogonal 
doublet are present. 
As a consequence, two physical quantities may require a relevant tuning to be kept stable under variation 
of the underlying parameters: the EW scale $v$ (connected to $m_h$) and the physical lightest scalar mass $m_{s1}$, which takes into account 
the mixing between the Higgs-like scalar and the other scalars.

Following~\cite{Gherghetta:2012gb}, we will use the usual logarithmic measure for the fine tuning in both cases:
\begin{equation}\label{eq:FT_measure1}
 \Sigma_v \equiv \max_i \left| \frac{d \log v^2}{d \log \xi_i} \right|, 
 ~~~\Sigma_h \equiv \max_i \left| \frac{d \log m_{s1}^2}{d \log \xi_i} \right|\, .
\end{equation}
For $\Sigma_v$ the relevant parameters (to be evaluated at the scale $\Lambda_{mess}$) are 
$\xi_i = \left\{ m_{H_u}^2 \right.$, $m_{H_d}^2$, $m_S^2$, $\lambda$, $k$, $a_\lambda$, $a_k$, $m_{Q_3}^2$, $m_{u_3}^2$, $m_{d_3}^2$, $A_t$, $A_b$, 
$M_1$, $M_2$, $\left. M_3 \right\}$, 
while for $\Sigma_h$ the list reads $\xi_i = \left\{\lambda \right.$, $k$, $a_\lambda$, $a_k$, $m_{Q_3}^2$, $m_{u_3}^2$, $m_{d_3}^2$, $A_t$, $A_b$, 
$M_1$, $M_2$, $\left. M_3 \right\}$. 
In the latter case, the soft masses $m_{H_u}^2$, $m_{H_d}^2$ and $m_S^2$ have been traded for the vevs $v_u$, $v_d$ and $v_S$ using the minimum equations, 
and the vevs are kept fixed since the associated tuning on $v$ is already taken into account in $\Sigma_v$.

In the computation of the fine tuning, the 1-loop Coleman-Weinberg (CW) potential is used, 
with renormalization scale fixed at $m_{soft} = \sqrt{m_{Q_3} m_{u_3}}$ 
in order to make the approximation more scale independent. 
The RGEs taking into account the running from $\Lambda_{mess}$ down to $m_{soft}$ are 
solved in the leading log approximation; however, since the $\lambda$ and $k$ couplings run quickly, in their case the full numerical solution is used.
To make evident that the parameters entering in the CW potential (and hence the minimum conditions) are computed at 
$m_{soft}$, with $\Lambda_{mess}$ feeding through the RG equations, we can rewrite Eq.~\ref{eq:FT_measure1} using explicitly the chain rule:
\begin{equation}\label{eq:FT_measure_chain}
 \Sigma_v = \max_i \left| \sum_j \frac{\xi_i(\Lambda_{mess})}{v^2} \frac{d v^2}{ d\xi_j(m_{soft})} \frac{d \xi_j(m_{soft})}{d \xi_i(\Lambda_{mess})} \right|
\end{equation}
Considering that by construction the two measures of fine tuning are independent, we choose to quantify the total tuning with the product 
\begin{equation}
\Sigma_{tot}=\Sigma_v \Sigma_h\, .
\end{equation}
\begin{figure}[tb]
\begin{center}
 \includegraphics[scale=0.35]{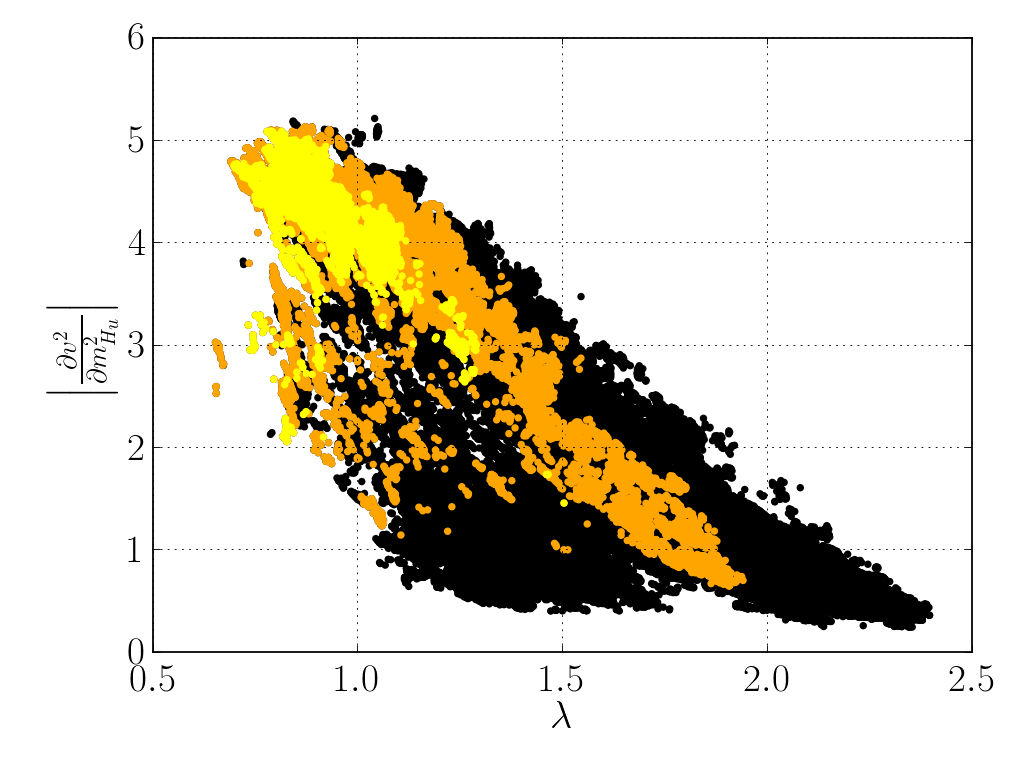}\\
 \includegraphics[scale=0.35]{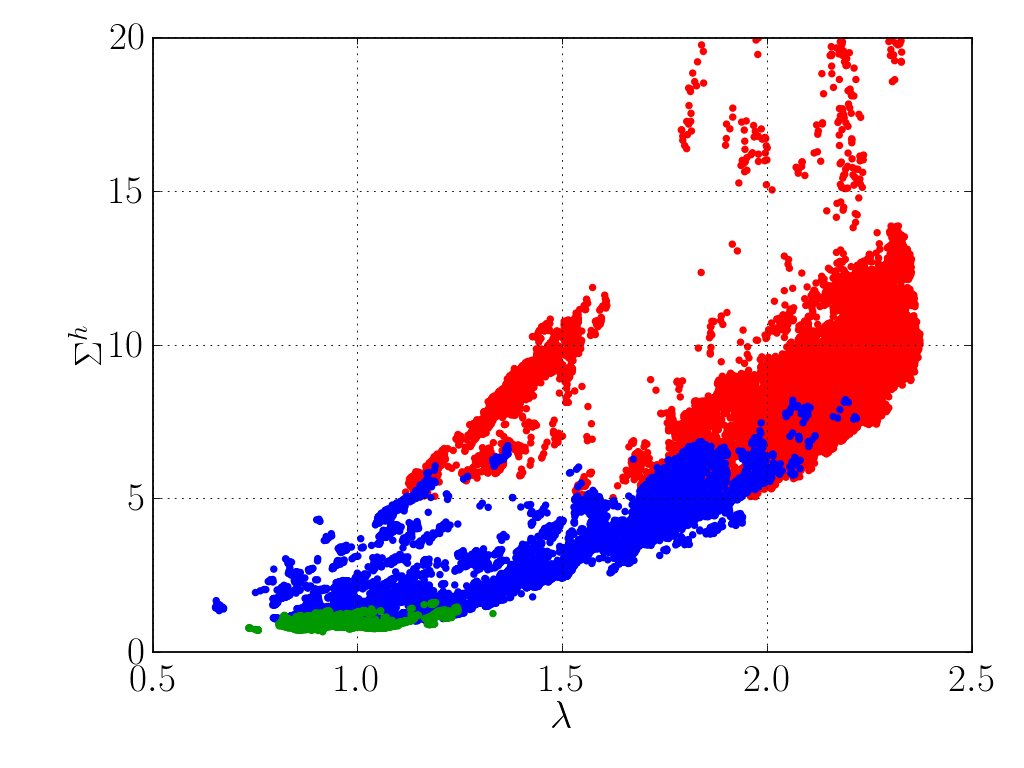}\\
 \includegraphics[scale=0.35]{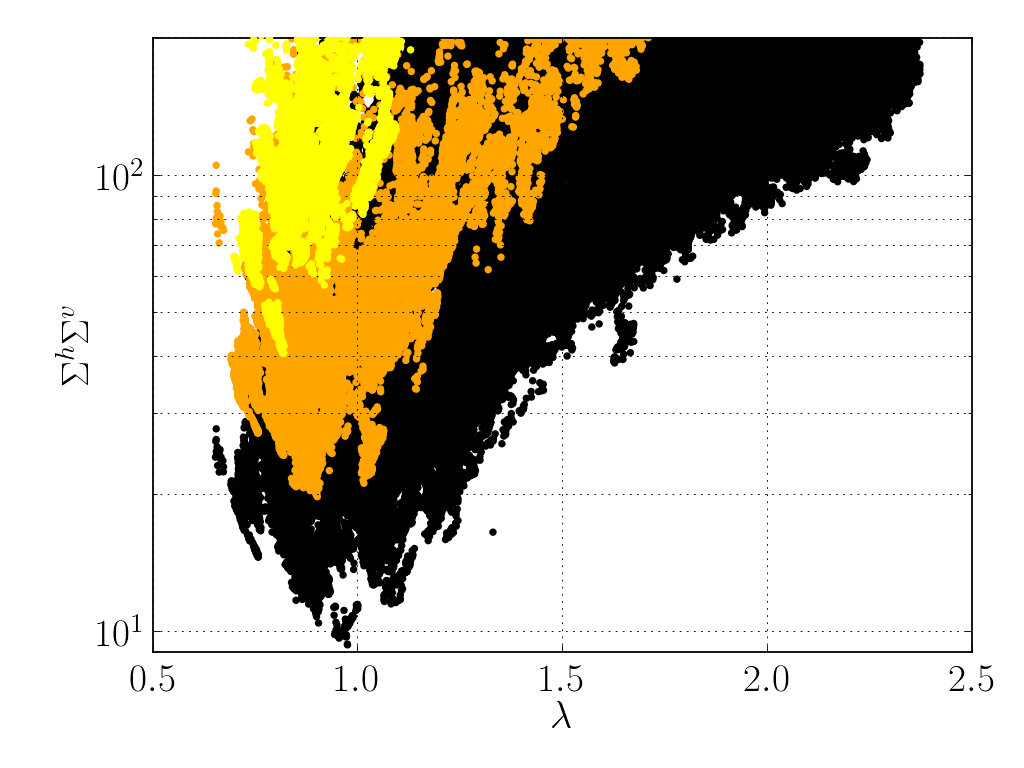}
 \caption{\label{fig:tuning_lambda} Dependence of $dv^2/d m_{H_u}^2$ (the quantity feeding into Eq.~\ref{eq:FT_measure_chain}, upper panel), 
 $\Sigma_h$ (middle panel) and of the combined tuning $\Sigma_v \Sigma_h$ (lower panel) on $\lambda$. 
 The black, orange and yellow points correspond to $\Lambda_{mess} = 20, 100, 1000$ TeV, respectively. 
 The green, blue and red points correspond, for fixed $\Lambda_{mess} = 20$ TeV, to a combined tuning $\Sigma_{tot}$ better than $5\%$, 
 between $1\%$ and $5\%$ and worse than $1\%$, respectively.~\cite{Gherghetta:2012gb}}
 \end{center}
 \end{figure}
 \begin{figure}[tb]
  \begin{center}
   \includegraphics[scale=0.35]{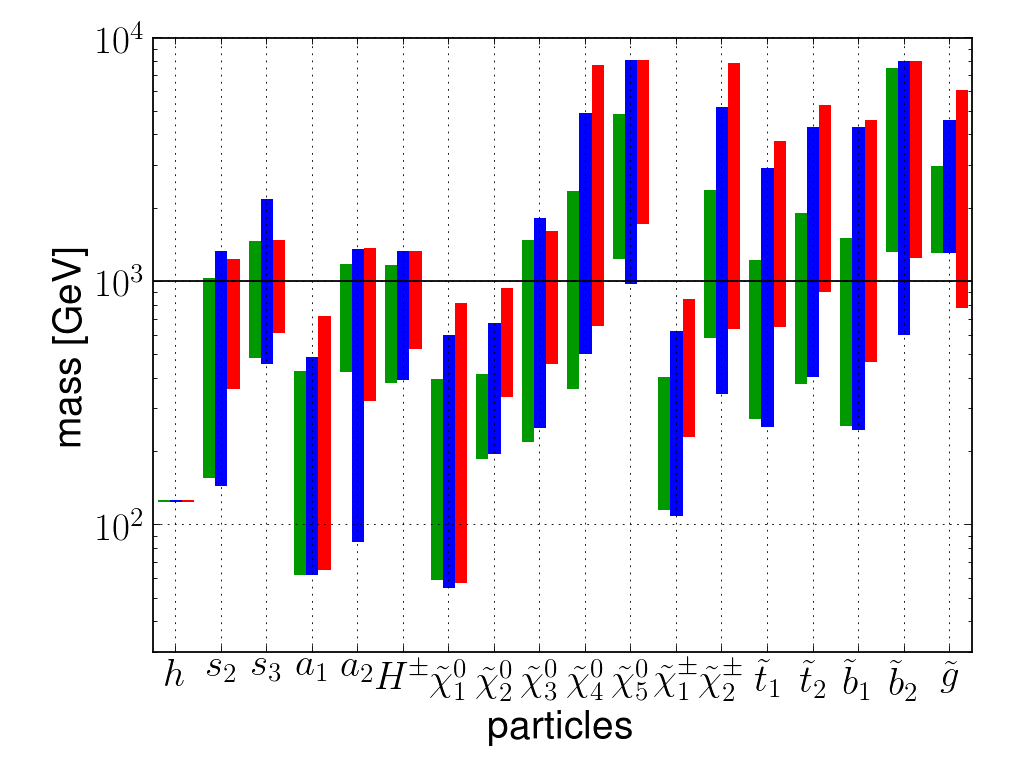}
  \end{center}
  \caption{\label{fig:spectrumNMSSM} Particle spectrum with $\Lambda_{mess}=20$ TeV. The green, blue and red points correspond to a combined tuning better than $5\%$, between $1\%$ and $5\%$ and worse 
 than $1\%$, respectively.~\cite{Gherghetta:2012gb}}
 \end{figure}
We present in Fig.~\ref{fig:tuning_lambda} the dependence of the separate measures $\Sigma_v$ and $\Sigma_h$ on $\lambda$. 
All the points shown satisfy the following phenomenological requirements (see~\cite{Gherghetta:2012gb} for details): 
\begin{itemize}
 \item All the LHC measurements on $R_X = \frac{\sigma(h) \times BR(h \rightarrow X)}{\sigma(h_{SM}) \times BR(h_{SM} \rightarrow X)}$ 
 with the exception of the $R_{\gamma\gamma}$ data;
 \item limits on the decays of heavier CP even scalars;
 \item latest SUSY searches at LHC8;
 \item electroweak precision measurements;
 \item flavour constraints (mass differences in the B system, charged and neutral B decays);
 \item (considering an LSP neutralino) not overclosure of DM density (WMAP7) and direct detection limits.
\end{itemize}
As can be seen, large values of $\lambda$ are preferred by $\Sigma_v$ but are disfavored by $\Sigma_h$. 
This can be understood as follows: large values of $\lambda$ help 
to reduce the derivative $dv^2/dm_{H_u}^2(m_{soft})$ appearing in Eq.~\ref{eq:FT_measure_chain}, 
reducing in this way the sensitivity to radiative corrections. On the other hand, large values of 
$\lambda$ increase too much the tree level mass of the Higgs-like scalar, so that a tuning among 
parameters is needed to obtain the correct mass mixing to bring it down to 125 GeV.~\footnote{See~\cite{Bertuzzo:2014sma} for an example of models in which the Higgs boson mass is basically untuned.}
Putting all together, the neat result is that the minimum amount of tuning is obtained for $\lambda \simeq 1$ and a low cutoff 
scale, $\Lambda_{mess} = 20$ TeV (see Fig.~\ref{fig:tuning_lambda}, lower panel). 
In particular, we can see that $\Sigma_{tot} \gtrsim 10, 20, 40$ for $\Lambda_{mess} = 20, 100, 1000$ TeV, respectively.

The consequences of the previous results on the ``most natural'' sparticle spectrum that we can expect at the 
LHC is summarized in Fig.~\ref{fig:spectrumNMSSM} for different values of the combined tuning. 

As expected, when we insist on naturalness many particles are expected to be below the TeV (usually independently on tuning), 
even in the colored sector. 
Notice however that, unlike what happens in the MSSM, the lightest stop can have a mass slightly above 1 TeV without a significant 
detriment in the tuning, and the same is essentially true also for the gluino.

As expected, for small total tuning the colored sector tends to be generically lighter, although there are also light colored particles in the more 
tuned region.

\subsubsection{Framework 2: Dirac Gauginos and $R$-symmetry}

We turn now to a different extension of the MSSM, in which additional matter 
is added to obtain Dirac instead of Majorana Gaugino masses.

The mechanism behind the improved naturalness is twofold: 
Dirac gaugino masses are generated through supersoft operators, which give only
finite contributions to scalar masses~\cite{Fox:2002bu}. In particular, the gluino 
contribution to the Higgs mass is less important than in the MSSM case, relaxing the naturalness bound. 
Moreover, the additional particle content needed to build Dirac gaugino masses can give sizable 
loop contribution to the Higgs boson mass, diminishing in this way the sensitivity to the individual contribution. 

It must be stressed, however, that scalar masses for the adjoint scalars are not supersoft. Indeed, they contribute at the two loop 
level to the RGEs of the sfermion masses~\cite{Arvanitaki:2013yja, Csaki:2013fla}. In particular, there are regions in 
the parameter space in which the squark masses become tachyonic, breaking charge and color. In addition, the masses 
of the CP-odd scalars may become tachyonic already at tree level, triggering again charge and/or color breaking. 
In what follows, we will always restrict to regions in parameter space in which this is not the case.

Models with Dirac gauginos are also interesting from a purely phenomenological
point of view: first of all, squark pair production is suppressed at
the LHC due to the absence of Majorana mass insertions, softening the experimental limits~\cite{Kribs:2012gx}.
Moreover, Dirac gaugino masses are compatible with a
global $U(1)_R$ symmetry, which would be otherwise broken by the
Majorana mass.  The $R$-symmetry can be used as an alternative to 
$R$-parity to forbid operators leading to proton
decay~\cite{Hall:1990dga,Hall:1990hq}, but has far richer
consequences.  Indeed, the absence of $A$-terms, $\mu$ term and
Majorana gaugino masses has a beneficial effect on the SUSY
flavour problem~\cite{Kribs:2007ac}.

A peculiar aspect of $R$-symmetric models is the Higgs sector particle
content. Various possibilities are summarized in Table~\ref{tab:Higgs_content}. In particular, to avoid spontaneous $R$-symmetry breaking, the scalars charged under 
$U(1)_R$ are all assumed to be inert, while the active doublets (singlets under $U(1)_R$) take part in electroweak symmetry breaking. 
Let us notice that in this framework it is possible to have a combination of sneutrinos playing the role of $H_d^0$~\cite{Frugiuele:2011mh}~\footnote{In this 
framework also 
neutrino masses can be accommodated.~\cite{Bertuzzo:2012su}}, 
and it is even possible to eliminate completely any Higgs boson from the spectrum, with only the sneutrinos taking part in EWSB~\cite{Riva:2012hz}.
\\
\begin{table}[tb]
\begin{center}
\begin{tabular}{c|c|c}
 Active ($R=0$) & Inert ($R=2)$ & Reference \\
 \hline
 $H_u$, $H_d$ & $R_u$, $R_d$ & \cite{Kribs:2007ac}\\
 $H_u$, $\tilde L$ & $R_d$ & \cite{Frugiuele:2011mh}\\
 $H_u$ & $R_d$ & \cite{Davies:2011mp}\\
 $\tilde L$ & $\times$ & \cite{Riva:2012hz}\\
\end{tabular}
\caption{\label{tab:Higgs_content} Examples of $R$-symmetric Higgs sectors.}
\end{center}
\end{table}

Let us now consider in detail the case of the Supersymmetric One Higgs Doublet Model (SOHDM)~\cite{Davies:2011mp}. As noted in~
\cite{Bertuzzo:2014bwa}, this is also representative of the large $\tan\beta$ limit of the models presented in~\cite{Kribs:2007ac,Frugiuele:2011mh}.
The superpotential is given by 
\begin{equation}
 W \supset \sqrt{2} \lambda_T H_u T R_d + \lambda_S S H_u R_d + \mu H_u R_d
\end{equation}
where $S$ and $T$ are the adjoint superfields associated with $U(1)_Y$ and $SU(2)_L$, respectively. 
Let us recall that we will assume $\langle R_d^0 \rangle =0$ in order to avoid spontaneous $U(1)_R$ breaking. 
\begin{figure}[tb]
\begin{center}
 \includegraphics[width=.4\textwidth]{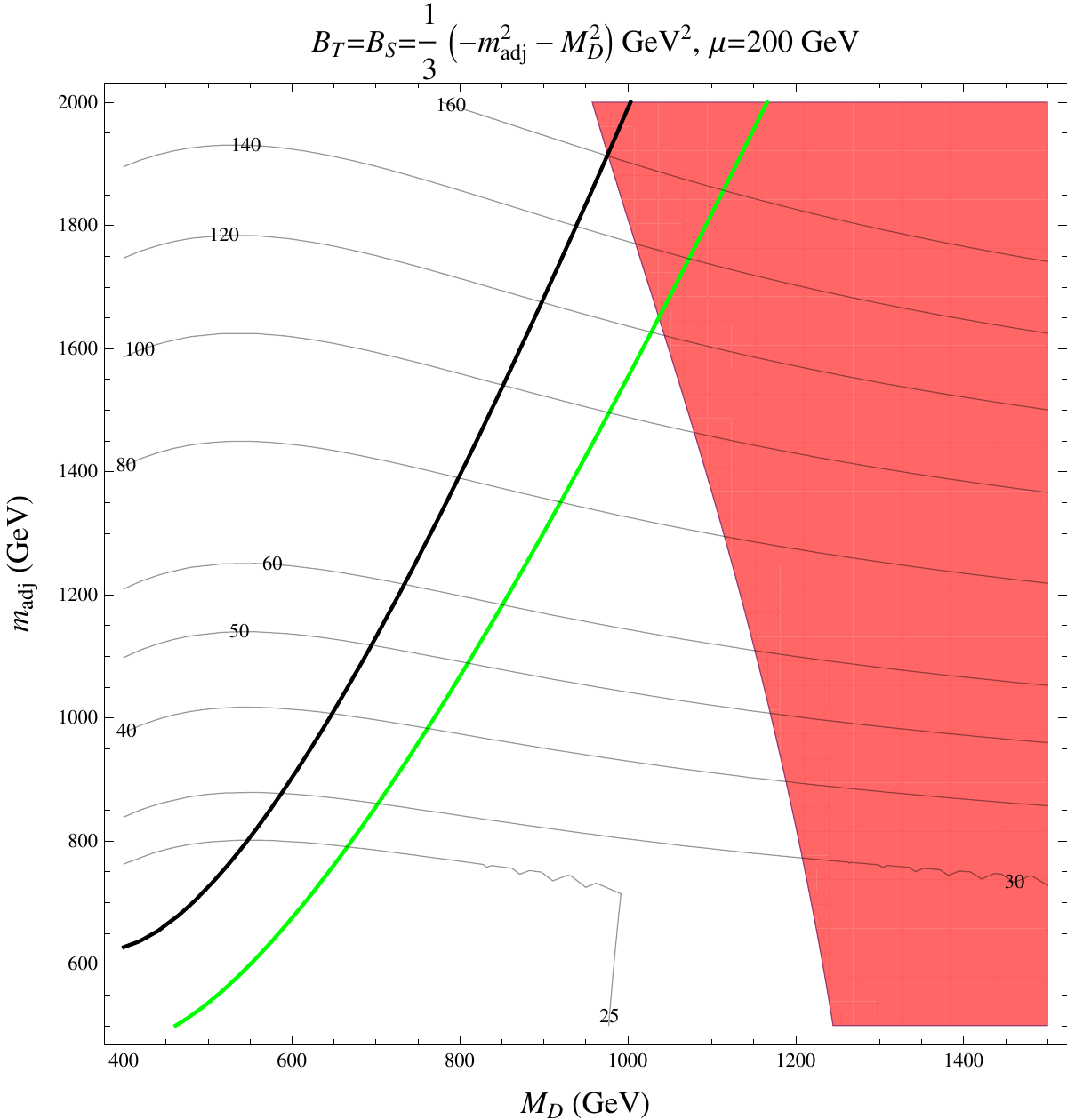}\\
 \includegraphics[width=.4\textwidth]{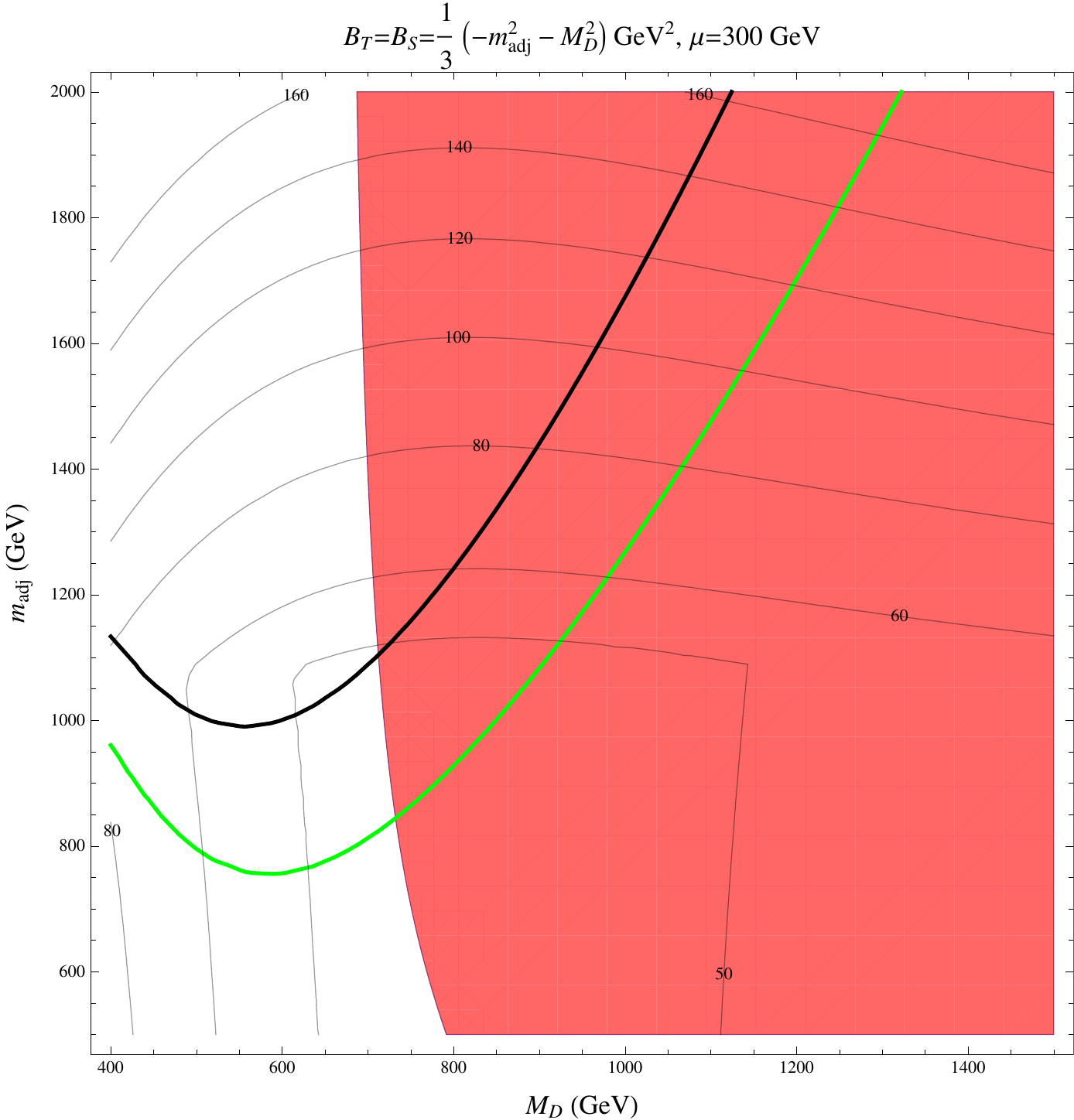}
 \caption{\label{fig:Dirac_gauginos} Higgs boson mass $m_h=125$ GeV (black and green thick
  lines) and fine-tuning parameter $\Delta$ (thin lines), as a function of
  $M_D=M_{\tilde W} = M_{\tilde B}$ and $m_{adj} = m_T = m_S=m_{R_d}$, 
  for $B_T=B_S= - \frac{1}{3} (m^2_{Adj}+M_D^2) $.  We fix
  $\lambda_T = 1 = -\lambda_S$. The upper (black) curve refers to a common stop
  mass of $m_{stop} = 300$ GeV, the lower (green) curve to $m_{stop} = m_{adj}$. Upper panel: $\mu=200$ GeV; Lower panel: $\mu=300$ GeV. 
  The red region is allowed at $95\%$ C.L. by EWPM ($T < 0.2.$)}
 \end{center}
 \end{figure}
The main features of the scalar spectrum are summarized in the following:
\begin{itemize}
 \item any NMSSM-like tree-level enhancement is forbidden by the $R$-symmetry, so that 
 as in the MSSM the mass of the Higgs-like scalar is given by
 \begin{equation}
  m_h^2 = m_Z^2 \cos^2 2 \beta
 \end{equation}
\item once all the neutral fields get a vev (including the singlet and the neutral triplet), there is a non 
vanishing mixing between the Higgs-like boson and the additional states, so that 
the lightest eigenvalue has a mass which is lowered by eigenvalue repulsion. An 
approximate expression, valid in the limit in which all the Dirac masses are smaller than the scalar adjoint masses,  is given by
\begin{equation}
 \begin{array}{ccl}
  (m_{\rm lightest}^2)_{tree} &\simeq & m_Z^2 -v^2 \frac{(-\sqrt{2} g M_{\tilde W} +2 \lambda_T \mu)^2}{m^2_{T_R}} \\
 && \mbox{} -v^2 \frac{(\sqrt{2} g' M_{\tilde B} +2 \lambda_S \mu)^2}{m^2_{S_R}}\; ,
 \end{array}
\end{equation}
where $m^2_{T_R} = 4 M_{\tilde W} + m^2_T + 2 B_T$ and $m^2_{S_R} = 4 M_{\tilde B} + m^2_S + 2 B_S$ are the masses 
of the real adjoint scalars before EWSB.
\end{itemize}
At the loop level, the relevant contributions are given by:
\begin{itemize}
 \item Stop contributions, as in the MSSM~\cite{Carena:1995wu}:
 \begin{equation}
  \begin{array}{ccl}
   V_{Higgs}^{CW} &\supset& \frac{1}{4} \left[ \frac{3}{16 \pi^2} y_t^2 \left(y_t^2-  \frac{m_Z^2}{2v^2}\right) \log\frac{M^2}{m_t^2}  \right. \\
&& \left. \mbox{} + \frac{3 y_t^4}{(16\pi^2)^2} \left(\frac{3}{2} y_t^2 - 32 \pi \alpha_3(m_t) \right) \log^2\frac{M^2}{m_t^2}   \right] h_u^4 \, , 
  \end{array}
 \end{equation}
 with $M$ a common stop mass scale;
 \item Adjoint scalars and fermions contributions. The complete expression can be simplified in two limiting cases, according to the 
 hierarchy between the Dirac mass scale $M_D$ and the scalar adjoint mass scale $m_{adj}$: 
 \begin{enumerate}
 \item $\mu \ll M_D \ll m_{adj}$:
 \begin{equation}
  \begin{array}{ccl}
   V_{Higgs}^{CW} &\supset & \frac{1}{4} \bigg[ \frac{5 \lambda_T^4 + 2 \lambda_T^2 \lambda_S^2 + \lambda_S^4}{16\pi^2} \log\frac{m_{adj}^2}{M_D^2}  \\
                  && \phantom{\frac{1}{4} \bigg[ }  + \frac{\lambda_S^2 \lambda_T^2 }{16\pi^2}\bigg] h_u^4 \, ,
  \end{array}
 \end{equation}
 \item $ m_{adj} \ll M_D $
 \begin{equation}
  \begin{array}{ccl}
   V^{CW}_{Higgs} &\supset & \frac{1}{4} \bigg[ 
			      \frac{-5 \lambda_T^4 -\lambda_S^4+ 2 \lambda_T^2 \lambda_S^2}{32 \pi^2} \log\frac{M_D^2}{Q^2} \\
		  && \phantom{\frac{1}{4} \bigg[} +\frac{\lambda_T^2 \lambda_S^2}{8\pi^2} \bigg] h_u^4   \\
  \end{array}
 \end{equation}
 \end{enumerate}
\end{itemize}
Some comments are now in order. First of all, in the region in which the additional contributions are comparable to the stop one, 
$|\lambda_T| \simeq |\lambda_S| \simeq 1$, we expect the last contribution ($ m_{adj} \ll M_D $ region) to be negative.
The hierarchy $M_D \ll m_{adj}$ is thus preferred to increase the Higgs boson quartic. 
Moreover, in the stop case there is an important negative two loop contribution proportional to $\alpha_3$, 
which reduces the effectiveness of the stop contribution. On the contrary, we do not expect the two loop contribution in the singlet and 
triplet case to be so important (since they are not proportional to $\alpha_3$), making more efficient the boost to the Higgs quartic.\\

There is however a potential drawback: $\lambda_T$ and $\lambda_S$ break custodial symmetry, so that we need to worry about potentially large contribution 
to Electroweak Precision Measurements (EWPM) for $|\lambda_T| \simeq |\lambda_S| \simeq 1$. We included this constraint in our analysis. \\

The results of the full contributions to the Coleman-Weinberg potential are presented in Fig.~\ref{fig:Dirac_gauginos}, as a 
function of a common Dirac gaugino mass $M_D$ and of a common adjoint scalar mass $m_{adj}$. We use two different 
values of $\mu$: $\mu=200$ GeV (upper panel) and $\mu=300$ GeV (lower panel). The green (black) lines represent 
the contour of a 125 GeV Higgs for $m_{stop} = m_{adj}$ and $m_{stop}=300$ GeV, respectively. The scalar mass of the 
inert doublet is fixed to $m_{R_d} = m_{adj}$. The thin black contours show 
the values of the fine tuning parameter $\Sigma_v$ for $\Lambda_{mess} = 20$ TeV, 
while the red region is the one allowed at $95\%$ C.L. by electroweak precision data.

Let us comment on two counterintuitive features of the results: to achieve the correct Higgs mass with 
less tuning,  {\it heavier} stops and {\it heavier} Higgsinos are needed. This can be understood as 
follows: for the upper (black) curves, the lightness of the stops is such that the main boost to 
the Higgs quartic comes from the adjoint and inert fields. On the contrary, for the lower (green) curves 
the stop boost to the Higgs quartic gives a contribution comparable to those of the adjoint scalars. 
However, as shown in~\cite{Bertuzzo:2014bwa}, there is no 
worsening in the tuning for $m^2_{\tilde stop} = m^2_T = m^2_{R_d}$. In addition, 
the ``collective'' quartic enhancement in the lower curves allows for smaller soft SUSY breaking masses, implying less tuning.
Turning to the $\mu$ parameter, we stress that compatibility with EWPM for lighter 
Higgsinos require heavier gauginos ({\it i.e.} larger $M_D$). Considering that from the shape of the Higgs mass curves 
in Fig.~\ref{fig:Dirac_gauginos} it is clear that this requires heavier scalars to get $m_h = 125$ GeV, a worsening in the tuning 
is expected. This is indeed the case: for $\mu=300$ GeV compatibility between $m_h=125$ GeV and EWPM 
is achieved for $m_{adj} \gtrsim 800 - 1100$ GeV (for $m_{\tilde t} = m_{adj}$ or $300$ GeV, respectively), {\it i.e.} when the sensitivity 
is still dominated by $\mu$. On the contrary, for $\mu=200$ GeV the scalar masses are pushed up to $m_{adj} \gtrsim 1500-1900$ GeV 
(again for $m_{\tilde t} = m_{adj}$ or $300$ GeV, respectively), in a region in which the soft SUSY breaking masses dominate the tuning.

\subsection{Supersymmetric dark matter$^4$}
\label{Supersymmetricdarkmatter}

\addtocounter{footnote}{1}

\footnotetext{Contributing authors: Alexandre Arbey and Farvah Mahmoudi}

The lightest neutralino in the MSSM constitutes a prototype candidate for cold dark matter, provided R-parity is conserved. Here we review the constraints on neutralino dark matter from different sectors, namely flavour physics, Higgs and SUSY LHC searches and dark matter detection experiments.

\subsubsection{Dark Matter observables}

Dark matter searches can be divided into four different categories.

First, cosmological observations lead to the determination of the average dark matter density. This density can then be compared to the neutralino relic density, which is computed assuming the supersymmetric particles have been initially in thermal equilibrium, then annihilated and coannihilated with other supersymmetric particles until the freeze-out period, leaving only the stable neutralino to constitute dark matter. Hence, the relic density observable is sensitive to the annihilation of the lightest neutralinos as well as the coannihilation of the other light supersymmetric particles to SM particles. The relic density calculation is also sensitive to the properties of the Universe close to the time of freeze-out, which is generally considered to be radiation-dominated. The comparison with the dark matter density relies on the fact that the cosmological dark matter is composed of one single component. Alternative cosmological scenarios could however strongly alter the computed relic density \cite{Kamionkowski:1990ni,Moroi:1999zb,Giudice:2000ex,Profumo:2003hq,Gelmini:2006pq,Arbey:2008kv,Arbey:2009gt}.

Second, dark matter is clustered in halos around galaxies, and the solar system is travelling across the Milky Way halo. Since dark matter particles interact very weakly with matter, they generally cross through matter without interaction, but it is still possible that dark matter particles scatter with nuclear partons inside atoms. This is the principle of direct detection experiments, which aim to measure the recoil energy deposited by the interaction of neutralinos with nuclei of a gas or crystal, in order to reconstruct the scattering cross section of dark matter with protons and neutrons. The main uncertainty for this observable comes from the local density and velocity of dark matter close to the Earth.

Third, dark matter particles can annihilate into SM particles, which can modify the flux of photons, positrons/electrons, proton/anti-protons, etc., measured around Earth. The dark matter indirect detection experiments probe the cosmic ray fluxes, and detect deviation generated by dark matter annihilation. The clearest dark matter signal would be a definite line in the gamma ray spectra. Here the two main sources of uncertainty in addition to the astrophysical backgrounds are the density of dark matter in the annihilation region, and the propagation of the charged particles.

Finally, LHC can also probe the dark matter sector, through direct pair production of neutralinos. However, such processes would be completely invisible at the detectors. A hard single jet emitted by initial state particles can be used as a marker of the production of a pair of neutralinos, resulting in monojet signatures.

\subsubsection{Nature of the neutralino}

The neutralino can be a pure state of bino, wino or higgsino, or a mixed state, leading to diverse properties.

A pure bino neutralino has its couplings to the $Z$ and Higgs bosons suppressed. For this reason, it would be very difficult to detect it in direct and indirect detection experiments, as well as at the LHC. Moreover, because of the low annihilation rate, the relic density is expected to be too large. Therefore, to retrieve the observed dark matter density, another slightly heavier supersymmetric particle, such as a stau or a squark is required, that can coannihilate with the neutralino, in order to increase the effective (co-)annihilation rate.

\begin{figure}[t!]
 \begin{center}
  \includegraphics[width=8.cm]{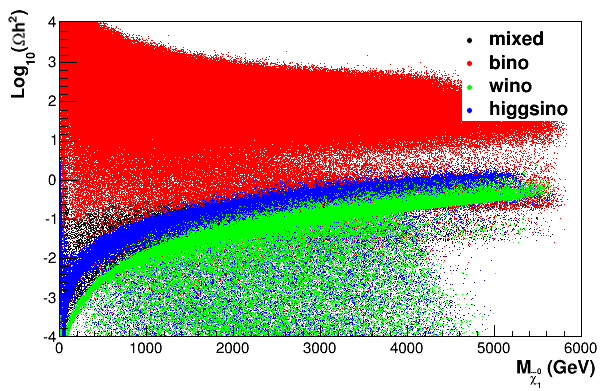}
 \end{center}
\caption{Relic density as a function of the neutralino mass, for different compositions of the neutralino.\label{fig:Oh2}}
\end{figure}

A pure wino or higgsino has also couplings to the $Z$ and Higgs bosons suppressed, to a lesser extent, leading to difficult direct and indirect detections. Concerning the relic density however, a pure wino is accompanied with a chargino, and a pure higgsino with a second neutralino and a chargino. For this reason, even if the other supersymmetric particles are much heavier, the correct amount of relic density can be achieved naturally for a wino of $\sim$2.3 TeV, or a higgsino of $\sim$1.2 TeV (see Fig.~\ref{fig:Oh2}).

For mixed state neutralinos, the couplings to the $Z$ and Higgs bosons can be large, leading to large scattering or annihilation cross sections, making a direct or indirect detection more likely. In addition, the correct dark matter relic density can be achieved even in absence of coannihilations. These scenarios however are becoming severely constrained by the direct detection experiments.

Fig.~\ref{fig:Oh2} shows the distribution of the relic density for the different types of neutralinos. The relic density is expected to be close to the observed cold DM density, $\Omega h^2 \sim 0.11$ \cite{Hinshaw:2012aka,Ade:2013zuv}. In general, the relic density increases with the neutralino mass. For a pure bino, the relic density is often too large, and only coannihilations can help reaching the right DM density. For higgsino and wino states on the contrary, the relic density is too small for light neutralinos, because of the coannihilation with the associated chargino or neutralino. For mixed states, the correct relic density can be obtained for any neutralino mass.

%%%%%%%%%%%%%%%%%%%%%%%%%%%%%%%%%%%%%%%%%%%%%%%%

\subsubsection{Constraints on the MSSM parameters}

\begin{table}[t]
\begin{center}
\begin{tabular}{|c|c|}
\hline
~~~~Parameter~~~~ & ~~~~~~~~~~Range~~~~~~~~~~\\
\hline\hline
$\tan\beta$ & [1, 60]\\
\hline
$M_A$ & [0, 3500]\\
\hline
$M_1$ & [-3500, 3500]\\
\hline
$M_2$ & [-3500, 3500]\\
\hline
$M_3$ & [0, 3500]\\
\hline
$A_d=A_s=A_b$ & [-10000, 10000]\\
\hline
$A_u=A_c=A_t$ & [-10000, 10000]\\
\hline
$A_e=A_\mu=A_\tau$ & [-10000, 10000]\\
\hline
$\mu$ & [-3500, 3500]\\
\hline
$M_{\tilde{e}_L}=M_{\tilde{\mu}_L}$ & [0, 3500]\\
\hline
$M_{\tilde{e}_R}=M_{\tilde{\mu}_R}$ & [0, 3500]\\
\hline
$M_{\tilde{\tau}_L}$ & [0, 3500]\\
\hline
$M_{\tilde{\tau}_R}$ & [0, 3500]\\
\hline
$M_{\tilde{q}_{1L}}=M_{\tilde{q}_{2L}}$ & [0, 3500]\\
\hline
$M_{\tilde{q}_{3L}}$ & [0, 3500]\\
\hline
$M_{\tilde{u}_R}=M_{\tilde{c}_R}$ & [0, 3500]\\
\hline
$M_{\tilde{t}_R}$ & [0, 3500]\\
\hline
$M_{\tilde{d}_R}=M_{\tilde{s}_R}$ & [0, 3500]\\
\hline
$M_{\tilde{b}_R}$ & [0, 3500]\\
\hline
\end{tabular}
 \end{center}
\caption{pMSSM parameter ranges (in GeV when applicable).\label{tab:pmssm}}
\end{table}

In the following, we consider the phenomenological MSSM with 19 parameters, which is the most general MSSM model with R-parity and CP conservation, and Minimal Flavour Violation at the weak scale \cite{Djouadi:1998di}. This model is flexible enough to allow for general studies of most of the MSSM neutralino dark matter scenarios, in particular because $M_1$, $M_2$ and $\mu$, the bino, wino and higgsino mass terms respectively, are independent, contrary to the usual constrained scenarios. The effect of CP-violation in the pMSSM has been recently studied in \cite{Arbey:2014msa}. Because of the large number of parameters only a combination of experimental analyses from different sectors can lead to strong constraints. In particular, we consider limits from flavour physics, Higgs physics, dark matter searches and LHC supersymmetric particle searches. In the following, the pMSSM parameters are varied in the ranges given in Table~\ref{tab:pmssm}, following the methodology of \cite{Arbey:2011un,Arbey:2011aa} to impose constraints from the above-mentioned sectors.

\begin{figure}[t!]
 \begin{center}
  \includegraphics[width=8.cm]{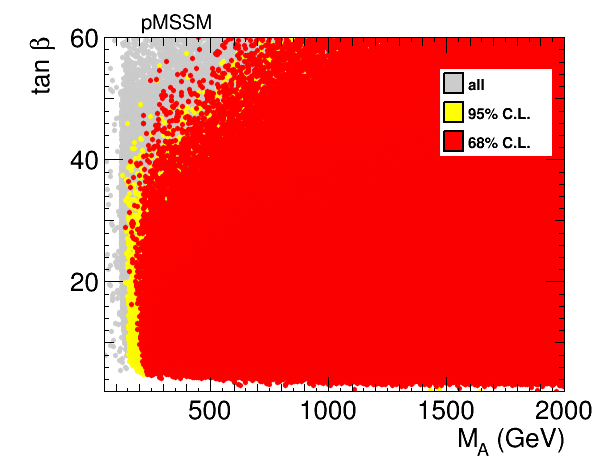}\\
  \includegraphics[width=8.cm]{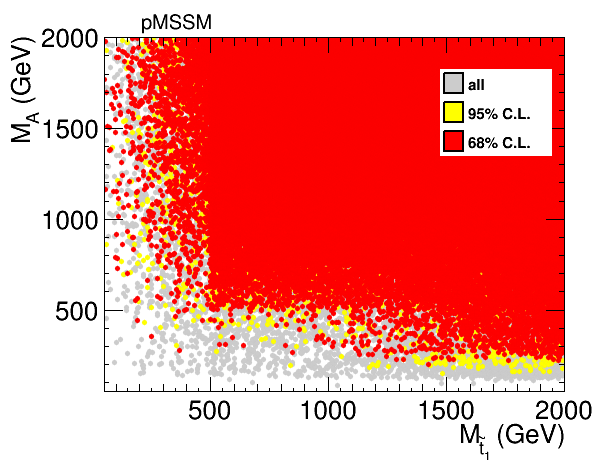}
 \end{center}
\caption{Constraints from a combination of flavour observables at 68\% and 95\% C.L. in the $(M_A,\tan\beta)$ (upper panel) and $(M_{\tilde{t}_1},M_A)$ (lower panel) pMSSM parameter planes \cite{Mahmoudi:2014mja}.\label{fig:flavour}}
\end{figure}

Flavour physics observables, such as rare decays, can impose stringent constraints. In particular, LHCb and CMS have observed for the first time the decay $B_s \to \mu^+ \mu^-$ \cite{Aaij:2013aka,Chatrchyan:2013bka}. This observable is particularly sensitive to $\tan\beta$ and the mass of the CP-odd Higgs, $M_A$ \cite{Arbey:2012ax}. Complementary information can also be obtained from the branching fraction of $b \to s \gamma$ and the angular observables of $B \to K^* \mu^+ \mu^-$, which are very sensitive to $\tan\beta$ and the chargino and stop masses \cite{Mahmoudi:2014mja}. Imposing flavour constraints restricts $\tan\beta$ to smaller values, and the CP-odd Higgs and stop masses to larger values. This is illustrated in Fig.~\ref{fig:flavour}, where the pMSSM points are projected on the $(M_A,\tan\beta)$ (upper panel) and $(M_{\tilde{t}_1},M_A)$ parameter planes. We see that the region with $\tan\beta > 40$ and $M_A <600$ GeV is strongly constrained, and the CP-odd Higgs and lightest stop masses cannot be simultaneously large, irrespectively of $\tan\beta$. In addition, since charginos are involved in these flavour decays at loop-level, constraints on the $M_2$, $\mu$ and $\tan\beta$ parameters can be deduced, leading to indirect constraints on the DM sector.

\begin{figure}[t!]
 \begin{center}
  \includegraphics[width=8.cm]{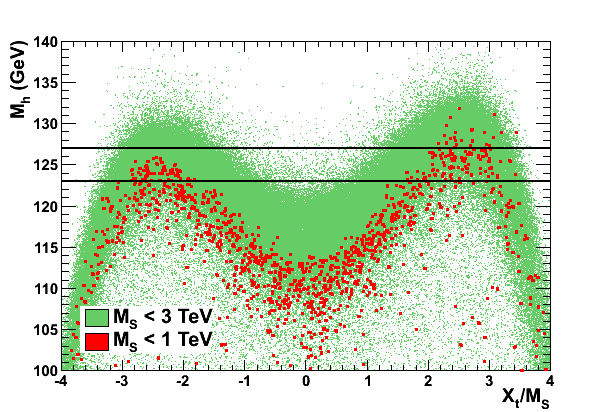}
 \end{center}
\caption{Lightest Higgs mass as a function of $X_t/M_S$ \cite{Arbey:2011ab}.\label{fig:mh}}
\end{figure}

The discovery of a Higgs boson at the LHC provides also very strong constraints on the MSSM. The lightest CP-even Higgs boson is generally considered as the discovered state. Its mass is given at the one loop level by
\begin{eqnarray}
M_h^2 &\approx & M_Z^2 \cos^2(2\beta)\label{mh}\\
&& + \frac{3\, \bar{m}_t^4}{2\pi^2 v^2\sin^ 2\beta} \left[ \log
\frac{M_S^2}{\bar{m}_t^2} + \frac{X_t^2}{M_S^2} \left( 1 -
\frac{X_t^2}{12\,M_S^2} \right) \right],  \nonumber
\end{eqnarray}
where $M_S$ is the SUSY breaking scale, defined as the geometric average 
of the two stop masses
\begin{equation}
M_S = \sqrt{ m_{\tilde t_1} m_{\tilde t_2}} 
\end{equation}
and $X_t = A_t -\mu \cot\beta$ is the mixing parameter in the stop sector. The requirement of $M_h\sim125$~GeV imposes very strong constraints on the pMSSM parameter space. In Fig.~\ref{fig:mh}, the distribution of the predicted Higgs mass is presented as a function of $X_t/M_S$. To reach large $M_h$ values, a large $M_S$, i.e. large stop mass, can be necessary. This condition can be slightly relaxed in the case of maximal mixing where $|X_t| \approx \sqrt6 M_S$.

\begin{figure}[t!]
 \begin{center}
  \includegraphics[width=8.cm]{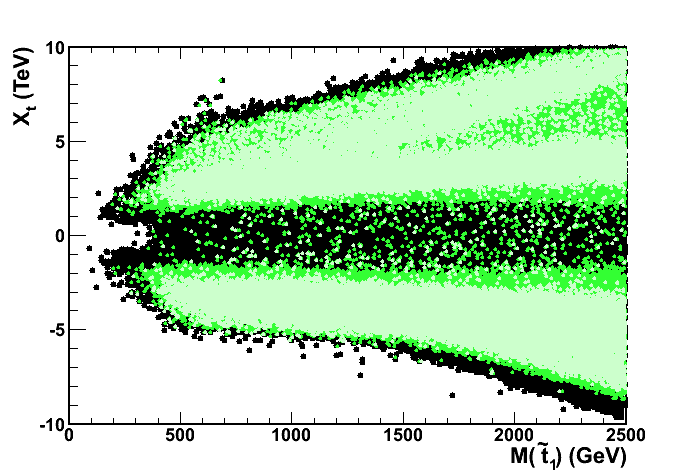}\\
  \includegraphics[width=8.cm]{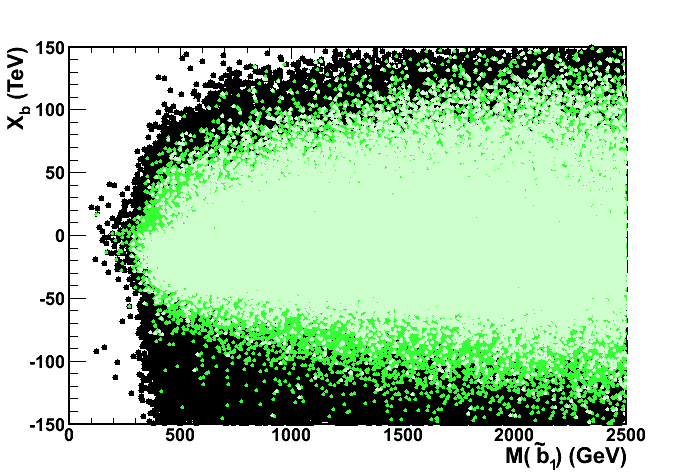}
 \end{center}
\caption{Distributions of the pMSSM points in the $(M_{\tilde{t}_1},X_t)$ (upper panel) and $(M_{\tilde{b}_1},X_b)$ (lower panel) parameter planes. The black dots show the accepted pMSSM points, those in dark (light) green the points compatible with the observed mass and rate constraints at 90\% (68\%) C.L.\cite{Arbey:2012bp}.\label{fig:hcouplings}}
\end{figure}

In addition to the mass, the measured Higgs couplings provide further constraints. In Fig.~\ref{fig:hcouplings}, the constraints from the Higgs couplings on the stop and sbottom sectors are presented. As can be seen, a large $X_t$ is favoured, and stop masses as light as 350 GeV can still be allowed. For the sbottoms, no specific mixing is favoured.

\begin{figure}[t!]
 \begin{center}
  \includegraphics[width=8.cm]{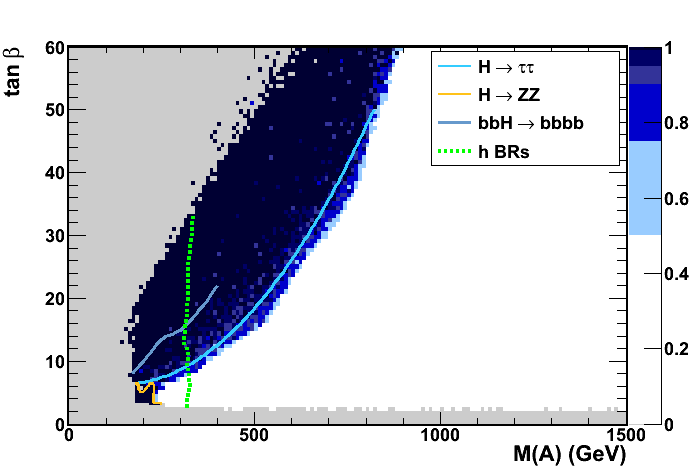}
 \end{center}
\caption{Combination of the expected constraints on the $(M_A,\tan\beta)$ parameter plane from the $\tau \tau$, $ZZ$ and $bb$ channels after the LHC 8 TeV run. The colour scale gives the fraction of pMSSM points excluded at each $M_A$ and $\tan \beta$ value. The contours show the 
limits corresponding to 95\% or more of the points excluded. The 90\% C.L.\ constraint from the Higgs signal strengths is also shown in the dotted green line. The grey region has no accepted pMSSM points after the BR($B_s \rightarrow \mu^+ \mu^-$), DM direct searches and Higgs mass constraints \cite{Arbey:2013jla}.\label{fig:heavyH}}
\end{figure}

Searches for heavier Higgs states also impose strong constraints on the SUSY parameter space. In Fig.~\ref{fig:heavyH}, we analyse the constraints from heavy Higgs searches on the pMSSM parameter points, in absence of flavour and dark matter constraints. We show that the constraints from the decay channels $H \to \tau\tau$, $ZZ$, $b\bar b$ disfavour the region at large $\tan\beta$ and small $M_A$. This region is also probed by the flavour observables and in particular BR($B_s\to \mu^+\mu^-$), and as we will see later by DM direct detection results.

\begin{figure}[t!]
 \begin{center}
  \includegraphics[width=8.cm]{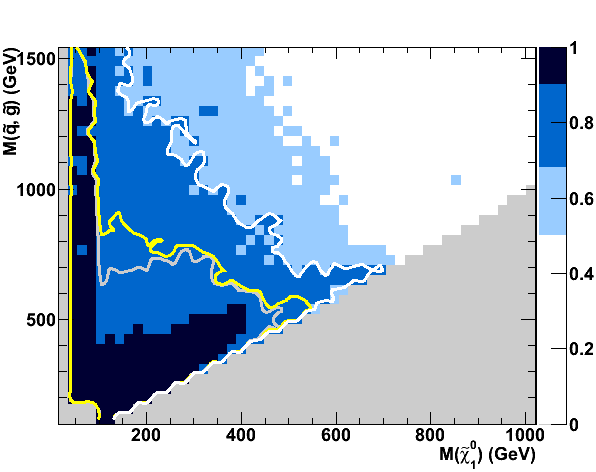} 
 \end{center}
\caption{Fraction of pMSSM points excluded by the combination of the LHC jets/leptons+MET, monojet analyses and direct DM searches in the $(M_{\tilde{q},\tilde{g}},M_{\tilde{\chi}})$ parameter plane. The lines give the parameter region where 68\% of the pMSSM points are excluded by the jets/leptons+MET searches alone (grey line), the combination with monojet searches (yellow line) and also with the direct DM LUX experiment (white line) \cite{Arbey:2013iza}.
\label{fig:monoj}}
\end{figure}

Direct searches for supersymmetry at the LHC set the strongest bounds on the mass of the supersymmetric particles. However, since the LHC is a hadron collider, the strong sector of the MSSM, i.e. the squark and gluino sector, is more deeply probed, and the electroweak sector, which is more correlated to DM, is less constrained. At the LHC, the main SUSY channels are searches with jets or leptons plus missing energy in the final states. However, if the mass splitting of the lightest neutralino with the searched supersymmetric particle is small, these searches lose their power, since most of the jets would become soft. From the point of view of the detectors, a DM particle is an invisible object which will leave no energy, leading to missing transverse energy or momentum. When neutralinos are produced, it would be possible to know that the process effectively occurs if an additional single hard jet is produced, leading to a so-called monojet signature. Nevertheless the production cross section of two neutralinos and one hard jet is very small. Yet, in case of production of a gluino or squark pair plus a jet, if the squarks and gluinos decay into soft jets, such a process would appear as a monojet \cite{Arbey:2013iza,Arbey:2015hca}. This generally happens when the mass splitting between the squark or gluino and the lightest neutralino is small. For this reason, monojet searches are complementary to the direct searches. In Fig.~\ref{fig:monoj}, the constraining power of the direct and monojet searches is demonstrated. Depending on the lightest neutralino mass, the masses of the squarks and gluinos can be probed up to 1500 GeV by the combination of the SUSY and monojet searches for light neutralinos of $\lesssim 100$~GeV, while this value can be reduced to less than 800 GeV for heavier neutralinos.

\begin{figure}[t!]
 \begin{center}
  \includegraphics[width=8.cm]{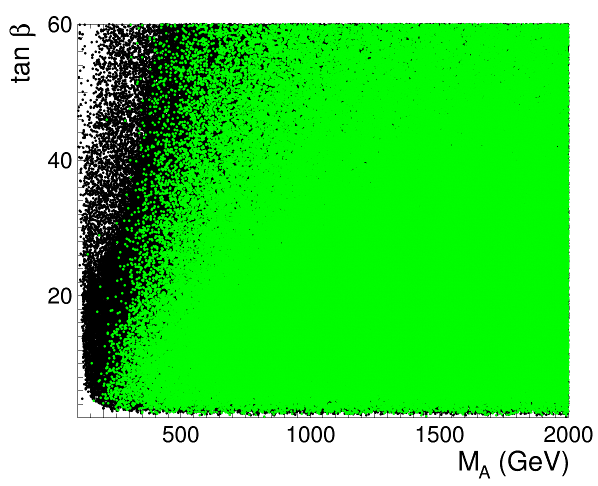} 
 \end{center}
\caption{Distribution of the pMSSM points in the $(M_A,\tan\beta)$ parameter plane, the black points are excluded at 90\% C.L. by the LUX exclusion limit.
\label{fig:ddmatb}}
\end{figure}

\begin{figure}[t!]
 \begin{center}
  \includegraphics[width=8.cm]{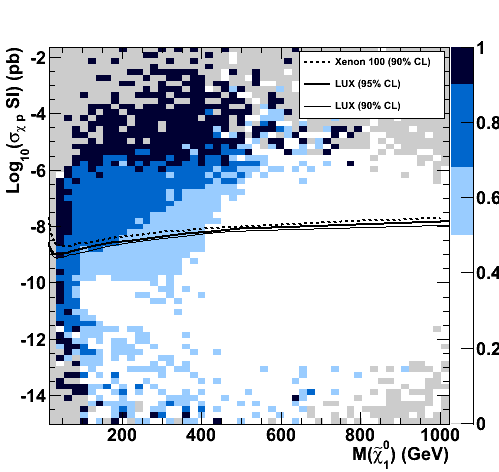} 
 \end{center}
\caption{Fraction of pMSSM points excluded by the combination of the SUSY and monojet searches
in the $(M_{\tilde{\chi}^0_1},\sigma_{\tilde{\chi}p})$ parameter plane. The solid and dashed lines correspond to the upper limits from direct detection experiments \cite{Arbey:2013iza}.
\label{fig:dd}}
\end{figure}

\begin{figure}[t!]
 \begin{center}
  \includegraphics[width=8.cm]{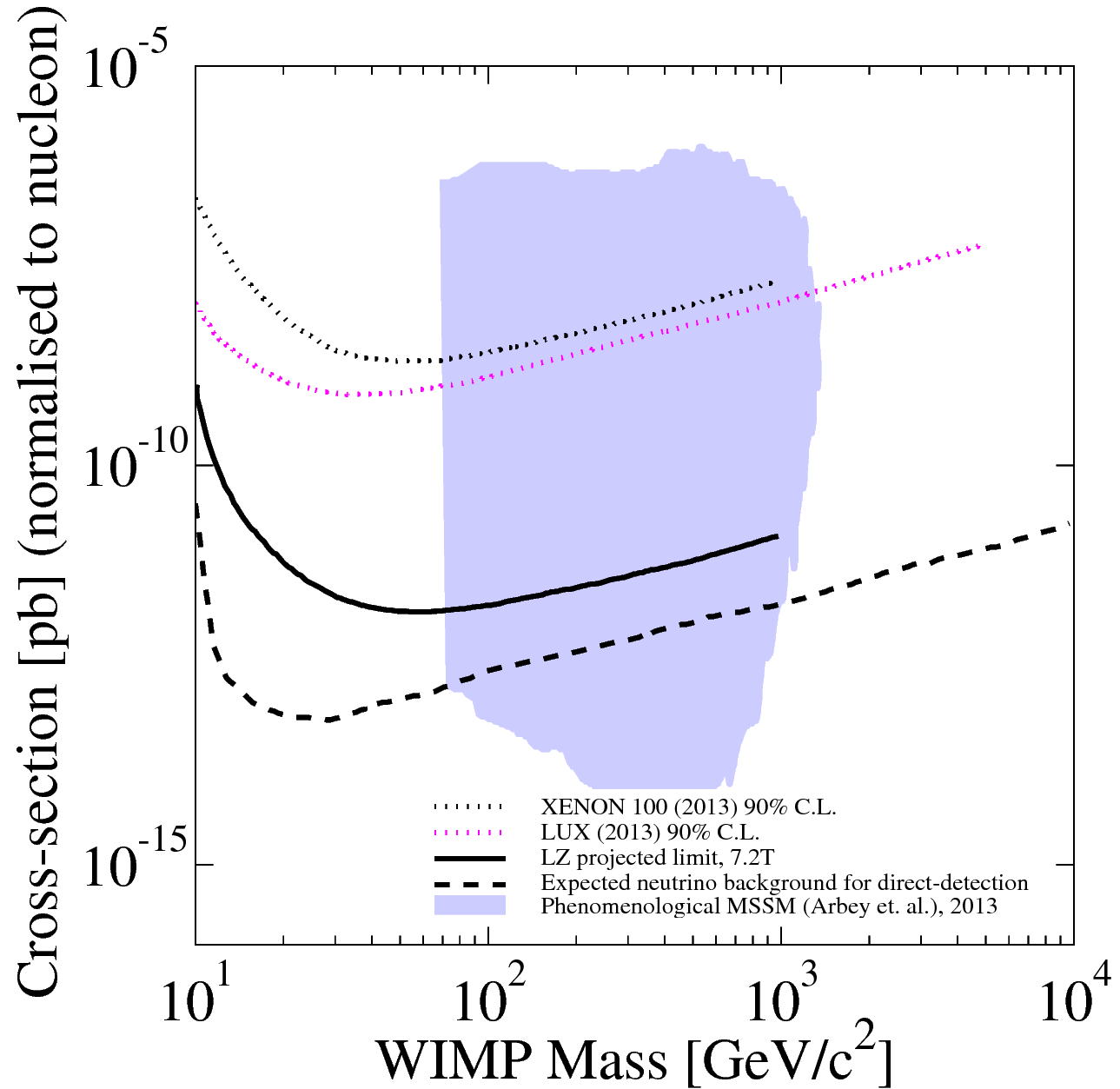} 
 \end{center}
\caption{Distribution of the pMSSM points in the $(M_{\tilde{\chi}^0_1},\sigma_{\tilde{\chi}p})$ parameter plane. The XENON-100 and LUX are providing the current best limits, the black line is a projection for the future LUX-ZEPLIN experiment, and the dashed line corresponds to the background due to the direct detection of neutrinos \cite{plotter}.
\label{fig:ddfuture}}
\end{figure}

Finally, dark matter experiments can also impose very strong constraints on the MSSM. Due to the large uncertainties in the dark matter densities and in the propagation of cosmic rays, we do not consider here DM indirect detection results. Imposing the dark matter density constraints leads to different conclusions on the nature of the neutralino, as seen in Fig.~\ref{fig:Oh2}. The dark matter direct detection experiments further constrain the MSSM. The XENON-100 \cite{Aprile:2012nq} and LUX  \cite{Akerib:2013tjd} collaborations currently provide the most constraining limits on the neutralino-nucleon spin-independent scattering cross sections. This scattering cross section is sensitive in particular to $M_A$ and $\tan\beta$, providing constraints complementary to the BR$(B_s\to\mu^+\mu^-)$ and $H\to\tau^+\tau^-$. This is illustrated in Fig.~\ref{fig:ddmatb}.
The additional constraint from LUX data on the pMSSM is also superimposed in Fig.~\ref{fig:monoj}, with the white line. In Fig.~\ref{fig:dd}, the XENON-100 and LUX limits are shown in the neutralino-nucleon scattering cross section vs. neutralino mass plane. We see that a substantial fraction of the pMSSM points is excluded by the LUX limits. Yet, many points with much smaller scattering cross sections remain, corresponding to the pure bino neutralinos, which have suppressed couplings to the $Z$ and Higgs bosons, leading to reduced cross sections. In the future, the sensitivity of direct detection will improve by large factors, as shown in Fig.~\ref{fig:ddfuture}. Even at the neutrino background limit, there remains still a pMSSM region compatible with the direct detection constraints.

%%%%%%%%%%%%%%%%%%%%%%%%%%%%%%%%%%%%%%%%%%%%%%%%

\subsection{Light neutralino and sbottom scenario}

We now investigate the possibility that light sparticles could have evaded the current experimental searches. It is known in particular that scenarios with compressed spectra, i.e. with small mass splittings, are particularly difficult to identify through SUSY direct searches, because of the associated soft jets and leptons. In addition, several dark matter direct detection experiments claimed for signal of light dark matter particles \cite{Bernabei:2010mq,Aalseth:2011wp,Angloher:2011uu,Ahmed:2010wy,Agnese:2013rvf}, today severely challenged by the LUX data.
In view of these data, in \cite{Arbey:2012na,Arbey:2013aba} the possibility of finding in the pMSSM a scenario with a light neutralino of about 10 GeV still consistent with all the current data was investigated.

For such a light neutralino, constraints from the previous electron-positron colliders have to be considered. A pure higgsino or wino state is always accompanied by a chargino of similar mass. Such a light chargino would have been discovered at LEP, even for small mass splittings between the chargino and the neutralino. Therefore, a light neutralino has to be a pure bino. As a consequence, this neutralino would have suppressed couplings to the Higgs and $Z$ bosons, lowering the decay fraction of the Higgs and $Z$ bosons to a pair of neutralinos so that it is consistent with the available measurements of the invisible decays.

\begin{figure}[t!]
 \begin{center}
  \includegraphics[width=8.cm]{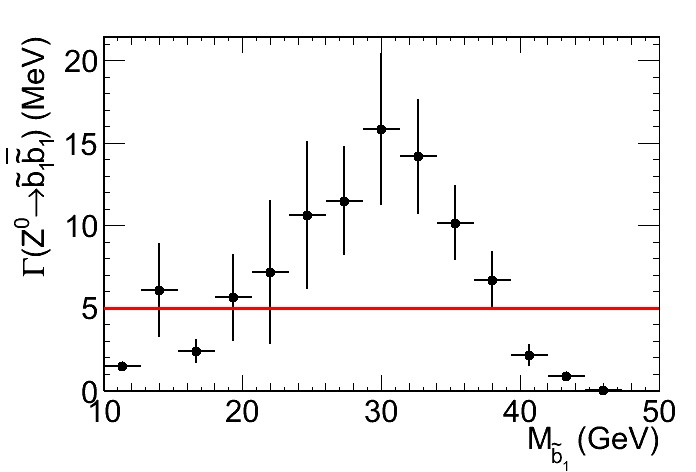} 
 \end{center}
\caption{Average $Z$ decay width to $\tilde b_1 \bar{\tilde b}_1$ as a function of the sbottom mass. The horizontal line corresponds to the LEP limit \cite{Arbey:2012na}.
\label{fig:width}}
\end{figure}

Concerning the relic density constraint, the bino neutralino alone would lead to a too large relic density, because the annihilation cross section of binos is small. As a consequence, it is necessary to have another light supersymmetric particle which could coannihilate with the neutralino in order to lead to the correct dark matter abundance. However, any charged or strongly interacting light particles should have been discovered at lepton colliders because of their small masses. This is clearly the case for the charged sleptons, charginos, wino, higgsino or wino-higgsino mixed state neutralinos, gluino or squarks of the first or second generations. A light stop or sbottom however could have escaped the constraints from LEP if their couplings to the $Z$ boson were suppressed, which is achievable if the stop or sbottom mixing makes them mostly right-handed. The mass of the light Higgs boson requires heavy stops, but a light sbottom is still possible. The discovered Higgs could still decay to right-handed sbottoms, it is therefore necessary to find a compromise in the sbottom mixing such that the coupling to the $Z$ boson is still suppressed while the coupling to the Higgs boson is reduced. This possibility can be achieved if the sbottom is of about 15 GeV. The decay width of the $Z$ boson to sbottoms is presented in Fig.~\ref{fig:width}.

\begin{figure}[t!]
 \begin{center}
  \includegraphics[width=8.cm]{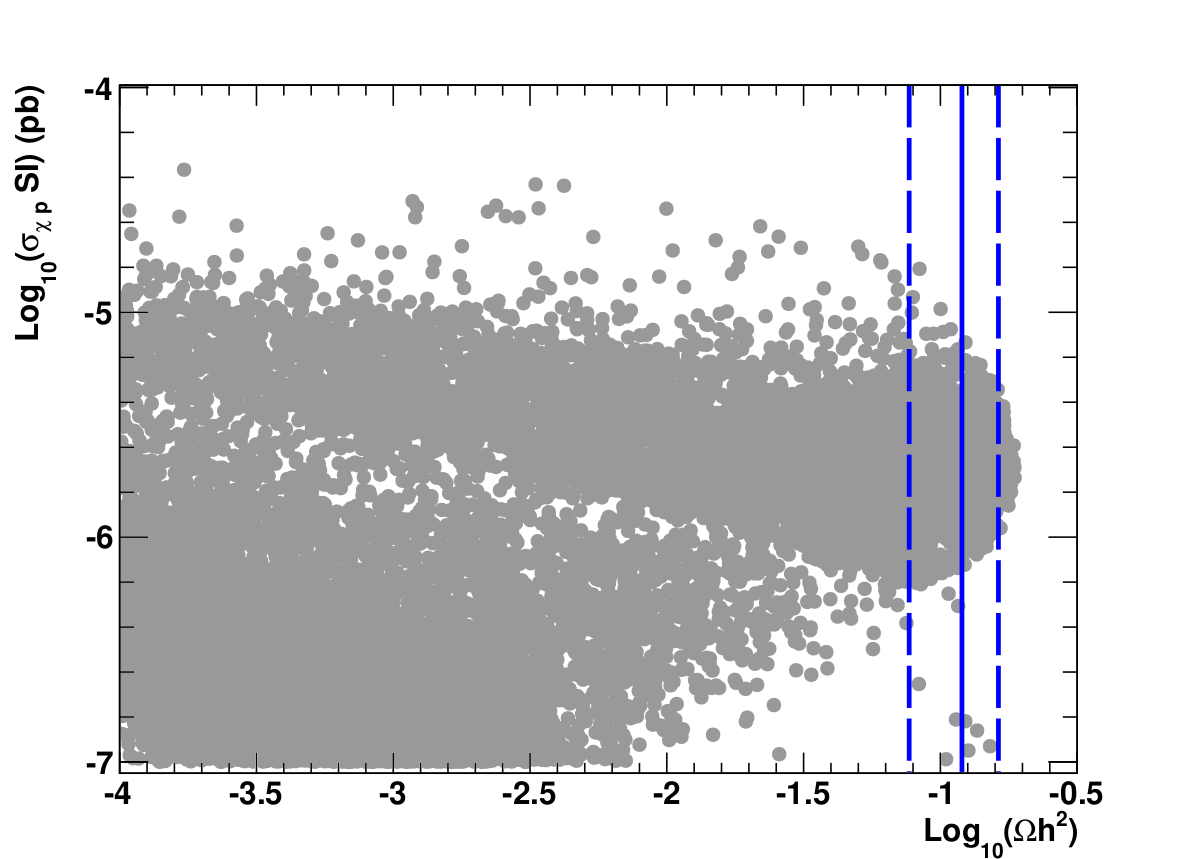}\\
  \includegraphics[width=8.cm]{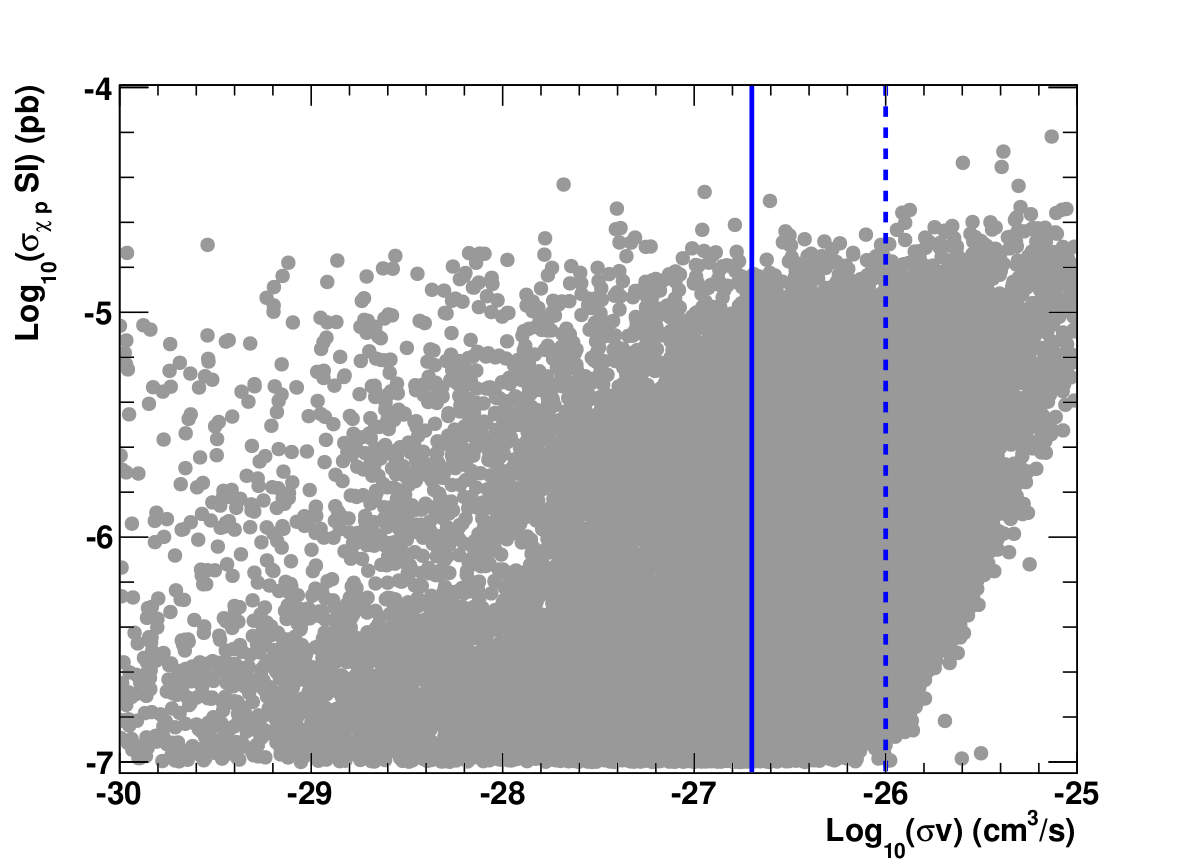}
 \end{center}
\caption{In the upper panel, DM direct detection scattering cross section as a function of the neutralino relic density. The vertical lines show the current experimental DM density value. In the lower panel, scattering cross section as a function of the DM indirect detection total annihilation cross section for selected pMSSM points. The vertical dashed and solid lines show the $\gamma$-ray strongest upper limit on the $\tilde{\chi} \tilde{\chi} \rightarrow b \bar b$ and the $\bar p$ strongest upper limit on $\tilde{\chi} \tilde{\chi} \rightarrow b \bar b g$ annihilation cross sections, respectively \cite{Arbey:2012na}.
\label{fig:lightdm}}
\end{figure}

In addition, the mass splitting of 5 GeV is at the right value, leading to a relic density compatible with the cosmological observations and to a scattering cross section with matter, mediated by the sbottom in a $t$-channel, large enough to be consistent with the data by the direct detection experiments seeing signals for a light DM particle. In addition no signal would be found in the cosmic ray spectra of indirect detection. This result is presented in Fig.~\ref{fig:lightdm}.

\begin{figure}[t!]
\begin{center}
\includegraphics[width=8.cm]{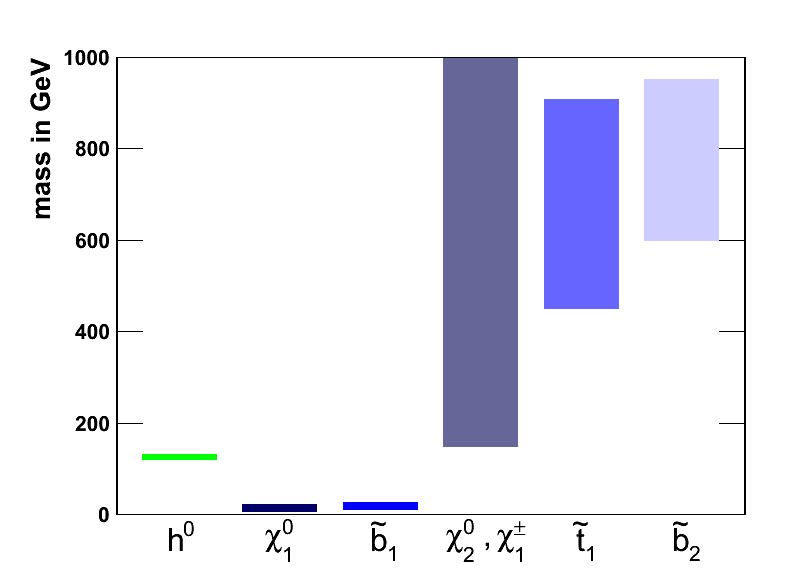}
\end{center}
\caption{Range of the masses of the relevant SUSY particles in the pMSSM scenario with a light neutralino and nearly degenerate sbottom \cite{Arbey:2012na}.\label{fig:spectrum}}
\end{figure}

A schematic representation of the typical spectrum for this kind of model point is given in Fig.~\ref{fig:spectrum}.

It was extensively checked that such a scenario is not excluded by all the available experimental data. The list of imposed constraints can be found in \cite{Arbey:2012na,Arbey:2013aba}. It is also remarkable that even if the production cross section of two sbottoms at the LHC can be very large, because the mass splitting between the sbottom and the neutralino is close to the bottom mass, the produced $b$-jets would be soft enough to escape detection. The monojet searches are also not sensitive enough as the production of two sbottoms and one jet would be strongly suppressed by kinematics, because the masses of the sbottom, bottom and neutralino are of the same order, falsifying the narrow-width approximation. In addition, simultaneous cuts on the jet-$p_T$ and missing $E_T$ would reject the events. The next run however would be able to probe this specific region thanks to the increased production cross section. Finally, the monophoton searches at PETRA \cite{Behrend:1986wj}, TRISTAN \cite{Abe:1995fq} and LEP \cite{Abdallah:2009ab} were not able to probe these scenarios because of the reduced production cross sections, consequence of the suppressed coupling of the sbottom to the $Z$ boson as well as the kinematic suppression due to the small mass splitting between the sbottom and the neutralino.

%%%%%%%%%%%%%%
\section{Effective field theories for Flavour, Higgs, and Dark Matter}
\label{sec:EFT}
\subsection{Introduction$^5$}
\label{sec:EFTintro}
\addtocounter{footnote}{1}

\footnotetext{Contributing authors: Yuval Grossman and Tobias Huber}

Effective Field Theories (EFTs) have many applications in contemporary
particle physics and quantum field theory. They are applicable
whenever there are two widely separated scales $S_H \gg S_L$ in a
physical problem, 
and serve to efficiently describe processes at scales  of order $S_L$, 
with or without knowledge of physics at $S_H$. The EFT is then valid at scales $S_L \ll S_H$.

The approaches to EFTs can be classified into two main categories. In the ``top-down'' approach the physics at $S_H$ (i.e.\ the ``full theory'') is known and the matching from the full to the effective theory is perturbative. This is for instance the case for the Fermi theory or the weak effective hamiltonian ${\cal H}_{\rm eff}$ of flavour physics. In the ``bottom-up'' approach the physics at the scale $S_H$ is either unknown (as in the case of the SM which can also be seen as an EFT of a more fundamental theory) or the matching is non-perturbative (as e.g.\ in Chiral perturbation theory).

The question arises why one should use an EFT in the ``top-down''
approach where the full theory is known. In this case it might be more
appropriate to formulate the physical problem in fewer or more
suitable degrees of freedom, which in turn leads to a reduced number
of scales. In addition, a systematic expansion in a small parameter of
${\cal O}(S_L/S_H)$ might simplify the problem consideraly. Moreover,
new approximate symmetries might appear which are hidden if the
problem is approached in the full theory. One example is HQET where at
leading power a spin and flavour symmetries show up. Finally, and most importantly, a systematic resummation of large logarithms $\log(S_H/S_L)$ which are generated in the full theory can be achieved conveniently in the EFT and allows for precision predictions in multi-scale problems.

In the ``top-down'' approach one divides the fields into high- and low-frequency modes, $\phi = \phi_H + \phi_L$, and integrates out the high-frequency modes $\phi_H$ via path-integral techniques, yielding a non-local effective action $S(\phi_L)$. Expanding the latter into products of local operators $O_i$ with matching coefficients (Wilson coefficients) $C_i$ is known as the ``Operator Product Expansion'' (OPE),
\begin{equation}
\displaystyle {\mathcal L}_{\rm eff} = \sum\limits_i \, C_i(\mu) \, O_i(\phi_L,\mu) \; ,
\end{equation}
where the separation between long distances (encoded in the matrix elements of the $O_i$) and short distances (contained in the Wilson coefficients $C_i$) is controled by the renormalisation scale $\mu$. For details see~\cite{Buras:1998raa}.

The operators $O_i$ are classified according to their canonical dimension $D=[O_i]$. Operators with 
$D<4$ and $D=4$ are called relevant and marginal, respectively. Contrary to what their names suggest, they don't tell us much about physics at some high scale. Operators with $D>4$, on the other hand, are named irrelevant, yet these are the operators which are the really interesting ones since they can teach us something about physics at the scale $S_H$. Neubert therefore comments on the above nomenclature as ``without a doubt one of the worst misnomers in the history of Physics''~\cite{Neubert:2005mu}.

In the following we are giving several examples of EFTs and their applications, in
particular, in relation to flavour, Higgs, and dark matter topics. In
some cases the scale separation is very large and the use of EFTs are
fully justified. In other cases, however, it is not
clear if the high and low scales have large separation between them. Nevertheless, in
all the examples below, the physics become much clearer due to the use
of EFTs.

\subsection{Effective field theories for Flavour and Higgs$^6$}

\addtocounter{footnote}{1}

\footnotetext{Contributing author: Tobias Huber}

\subsubsection{EFT in flavour physics}
\label{sec:EFTflavour}

We give four examples of EFTs in flavour physics.

a) The \emph{effective weak Hamiltonian} ${\cal H_{\rm eff}}$~\cite{Buchalla:1995vs}, which is the modern extension of the \emph{Fermi theory}. One makes use of the fact that in weak decays of mesons or baryons the masses and momentum transfers are much smaller than the masses of the weak gauge bosons and of the top quark, $m_i, \, \sqrt{q^2} \ll M_{W,Z}, \, m_t$, and uses the Fermi constant $G_F$ as expansion parameter. This yields higher dimension operators, that is with $d>4$, which are local interactions of four fermions or two fermions and a photon/gluon. Calculations consist of three steps: matching, running, and on-shell matrix elements, and have reached a highly sophisticated level. The former two steps are process independent, whereas the last one is process dependent.

b) \emph{Heavy-quark effective theory} (HQET)~\cite{Mannel:1991mc,Neubert:1993mb} is an expansion in inverse powers of the heavy quark mass $m_Q$ and is mainly applied for $Q=b,c$. The momentum $p_Q^\mu$ of the heavy quark gets separated according to $p_Q^\mu = m_Q \, v^\mu + k^\mu$ with $v^2=1$ and $k \sim {\cal O}(\Lambda_{\rm QCD})$. The leading order Lagrangian
\begin{equation}
\displaystyle {\mathcal L}_{\rm HQET} = \bar h_v \, i \, v\cdot D \, h_v
\end{equation}
possesses a heavy-quark flavour symmetry and a heavy-quark spin symmetry, which are both broken by subleading terms,
\begin{equation}
\displaystyle  - \bar h_v \frac{D^2_\perp}{2 m_Q} \, h_v - g \, \bar h_v \frac{\sigma_{\mu\nu}G^{\mu\nu}}{4 m_Q} \, h_v \, .
\end{equation}

c) \emph{Soft-collinear effective theory} (SCET)~\cite{Bauer:2000yr,Bauer:2001yt,Beneke:2002ph,Beneke:2002ni} is applicable for jet-like objects of large energy and small invariant mass. Momenta are decomposed according to $p^\mu = (n p) \, \frac{\bar n^\mu}{2} + (\bar n p) \, \frac{n^\mu}{2} + p_\perp^\mu$, with $\bar n^2 = n^2 = 0$ and $\bar n n = 2$. Depending on the scaling of the individual components one distinguishes hard, $n$-collinear, $\bar n$-collinear, and soft modes (in SCET$_I$). The Lagrangian density reads
\begin{align}
{\mathcal L}_{\rm SCET} &=  \bar{\xi} \frac{\bar{n}\!\!\!\slash}{2} \left[ i n \cdot D
+ i D\!\!\!\!\slash_{c \perp} \frac{1}{i \bar{n} \cdot D_c} i D\!\!\!\!\slash_{c \perp}\right] \xi \nonumber \\
&+\bar{\psi}_s i D\!\!\!\!\slash_s \psi_s -\frac{1}{4} \left( F_{\mu \nu}^{s, a} \right)^2 -\frac{1}{4} \left(F_{\mu \nu}^{c, a}\right)^2 \, ,
\end{align}
where $s,c$ stand for soft and collinear, respectively. An excellent review on SCET was recently released in~\cite{Becher:2014oda}. SCET has been applied to many problems in flavour physics, such as inclusive $b \to X_u\ell\nu$ decays in the presence of a cut on $m_X$~\cite{Greub:2009sv} (see also~\cite{Gambino:2007rp,Andersen:2005mj,Aglietti:2007ik}). The SCET language is suitable for deriving factorisation theorems which have the form
\begin{equation}
\Gamma \sim H \, \cdot \, J \otimes S \, , \label{eq:fac}
\end{equation}
i.e.\ the product of a hard function with the convolution of a jet- and a universal soft function. SCET is also applied to non-leptonic $B$-decays~\cite{Bell:2009fm,Beneke:2009ek}. More recently SCET has been vastly applied to collider physics problems. Just to mention a few, inclusive hadron-collider cross sections, Drell-Yan production, transverse momentum resummation, IR structure of gauge theory amplitudes, event shapes (thrust, jet-broadening, C-parameter), jet physics, electroweak Sudakov logarithms, Glauber gluons, and gravity. A more complete list of application, including the full list of references, can be found in~\cite{Becher:2014oda}. The interplay between flavour and collider physics might also help to get insight into yet unsolved problems in SCET. As is well-known, convolution integrals in QCD factorisation (QCDF) diverge at subleading power~\cite{Beneke:2003zv} (so-called endpoint divergencies). Recent analyses on the collinear anomaly~\cite{Becher:2010tm}, together with analytic regularisation in SCET~\cite{Becher:2011dz}, might help to better understand and eventually resolve the problem of endpoint divergences in QCDF.

d) \emph{Chiral perturbation theory} (ChPT) is written in terms of mesons and baryons, the QCD degrees of freedom which are present below the scale of spontaneous chiral symmetry breaking, i.e.\ at energies at or below $\sim 1$~GeV~\cite{Gasser:1983yg,Gasser:1984gg,Gasser:1987rb,Jenkins:1990jv}. The Lagrangian preserves the (approximate) chiral symmetry of QCD and is an expansion in meson momenta and masses. To lowest order, it reads
\begin{equation}
\displaystyle {\mathcal L}_{\rm ChPT} = \frac{F_\phi^2}{4} \, {\rm Tr}[(D_{\mu} U)(D^{\mu} U)^\dagger ]
+ \frac{F_\phi^2}{4} \,{\rm Tr}[\chi^\dagger U+\chi U^\dagger]
\end{equation}
The matrix $U=\exp(i\sqrt{2}\Phi/F_\phi)$ is a non-linear representation of the axial generators and contains the pseudoscalar Goldstone bosons $\Phi = \sqrt{2} T^a\phi^a$. Among the numerous applications are $\pi\pi$-scattering, $\eta$-decays, non-leptonic, semi-leptonic and radiative $K$ decays, pion-nucleon scattering and so on. Reviews can be found in~\cite{Ecker:1994gg,Kubis:2007iy,Scherer:2005ri}.

\subsubsection{EFT in Higgs physics}
\label{sec:EFTHiggs}

After the Higgs discovery~\cite{Aad:2012tfa,Chatrchyan:2012ufa} one has to measure its properties and continue the search for new physics (NP) beyond the SM (BSM). This is done in a model-independent way in EFTs, for example:

a) Higgs production in the $gg$ channel proceeds mainly via a top triangle loop, which can be integrated out to yield an effective $ggH$ vertex,
\begin{equation}
\displaystyle {\mathcal L}_{\rm ggH} = - \frac{C_1(\alpha_s)}{4v} \, H \, G^{\mu\nu}_a G_{a,\mu\nu} \, .
\end{equation}
The matching coefficient $C_1$ is known~\cite{Kramer:1996iq,Chetyrkin:1996ke}, as well as three-loop QCD corrections to the $ggH$ vertex~\cite{Baikov:2009bg,Gehrmann:2010ue}. The results were used recently in the prediction of the Higgs-production cross-section via gluon-fusion to NNNLO~\cite{Anastasiou:2015vya}.
The limit $m_t \to \infty$ works very well for inclusive observables, even for $m_H \! \ll\!\!\!\!\!\!\!\slash \; 2 m_t$. Corrections for finite top-mass have also been calculated~\cite{Harlander:2009my,Pak:2009dg}.

b) Recently the complete dimension-six Lagrangian for the SM was formulated in~\cite{Grzadkowski:2010es,Jenkins:2013zja,Jenkins:2013wua,Alonso:2013hga,Buchalla:2012qq,Buchalla:2013rka,Buchalla:2013eza,Elias-Miro:2013mua,Pomarol:2013zra}, (see also~\cite{Kilian:2003pc,Giudice:2007fh,AguilarSaavedra:2010zi}). They all build on earlier work from~\cite{Buchmuller:1985jz}. The operators can be classified into bosonic, single-fermionic-current, and four-fermion operators. Among them are CP-even and CP-odd, baryon-number conserving and violating ones. They allow to investigate processes in a model-independent way, using only the SM gauge group and unitarity. Although the total number of operators is quite large ($\sim 60$), only a few of them contribute to a given process, e.g.\ in Higgs physics or in the context of anomalous gauge boson couplings. We shall give a few examples below.

\subsubsection{Applications in flavour physics}
\label{sec:applflavour}

\begin{figure}
\includegraphics[width=0.95\columnwidth]{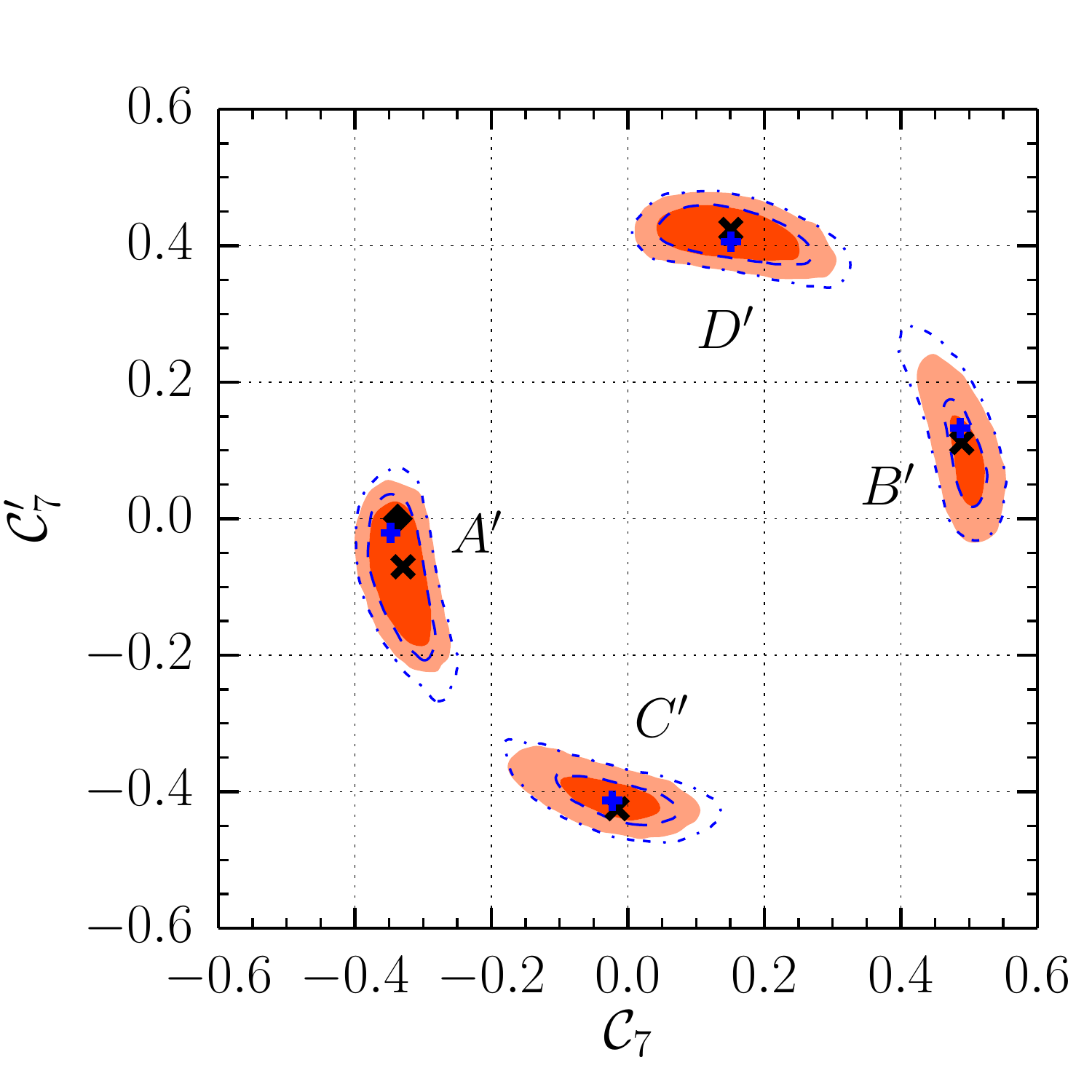}
\caption{\label{Huber_fig1} Bayesian fit to ${\cal C}_7$ and ${\cal C}_7^\prime$ in the (SM+SM${}^\prime$) scenario. The black diamond shows the projection of the SM point. The crosses mark the best-fit point. Figure taken from~\cite{Beaujean:2013soa}.}
\end{figure}

\begin{figure}
\includegraphics[width=0.95\columnwidth]{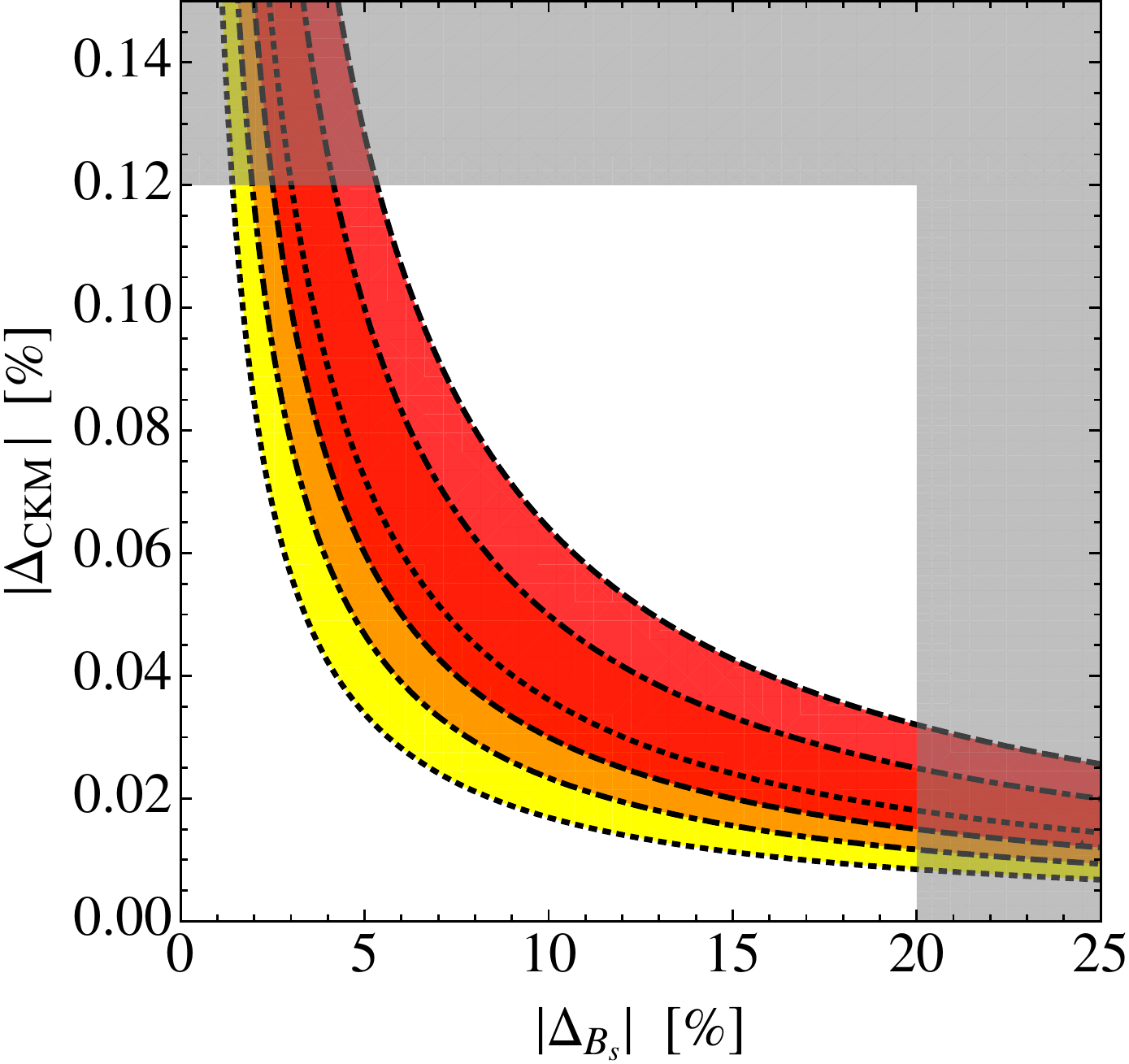}
\caption{\label{Huber_fig2} Parameter space in the $|\Delta_{B_s}| - |\Delta_{\rm CKM}|$
plane that is favoured by the $\bar B \to K^*\ell\ell$ anomaly. The shaded
region indicates the parameter range disfavoured at 95\% CL. The coloured bands correspond to different values of $M_{Z^\prime}$. Figure taken from~\cite{Gauld:2013qba}.}
\end{figure}

We can present here only a small selection of applications:

a) The Wilson coefficients, being real in the SM, become complex in general NP models. By introducing the ratios
\begin{equation}
\displaystyle R_i = C_i(\mu_0) / C^{\rm SM}_i(\mu_0) \, ,
\end{equation}
one can write the observables in terms of the $R_i$ and derive model-independent constraints on the Wilson coefficients at the matching scale $\mu_0$, as was done in~\cite{Huber:2005ig,Huber:2007vv}. For a recent update, see~\cite{Huber:2015sra} and section 5.

b) In exclusive $\bar B \to K^*\ell\ell$ decays, LHCb~\cite{Aaij:2013qta} measured in one bin of the observable $P_5^\prime$ a $3.7\sigma$ discrepancy between experiment and the SM prediction, a feature which got essentially confirmed using an increased data set of 3fb$^{-1}$~\cite{LHCb:2015dla}. A possible impact of NP was analysed both model-independently (see e.g.~\cite{Descotes-Genon:2013wba,Beaujean:2013soa,Hurth:2013ssa}) and in specific NP models (see below). Ref.~\cite{Beaujean:2013soa} uses data on many FCNC observables and combines them in a Bayesian analysis. The fits include theory uncertainties explicitly through nuisance parameters. The fit is done for the SM alone, and for the SM supplemented by chirality-flipped operators (SM+SM${}^\prime$), see Fig.~\ref{Huber_fig1}. One concludes that the SM provides adequate description of the available $|\Delta B|=|\Delta S| = 1$ data, when permitting subleading power corrections of $\sim 15\%$ at large hadronic recoil.

c) The $\bar B \to K^*\ell\ell$ anomaly has also been analysed in specific NP models (see e.g.~\cite{Altmannshofer:2013foa,Gauld:2013qba,Mahmoudi:2014mja}). Ref.~\cite{Gauld:2013qba} investigates a triple correlation between observables in $\bar B \to K^*\ell\ell$, $B_s-\bar B_s$-mixing and the CKM unitarity, in a minimal $Z^\prime$ model. Assuming that NP alters $C_9$, one can set constraints on
\begin{eqnarray}
\displaystyle \Delta_{B_s} &=& \frac{\Delta M_{B_s}}{\Delta M_{B_s}^{\rm SM}} - 1 \, , \nonumber \\
\displaystyle \Delta_{\rm CKM} &=& \textstyle \sum_{q=d,s,b} |V_{uq}|^2 - 1 \, ,
\end{eqnarray}
see Fig.~\ref{Huber_fig2}.

d) A review on flavour constraints from $b \to s$, $b \to d$, and $c \to u$ transitions can be found in~\cite{Isidori:2010kg}. Besides model-independent constraints, minimal-flavour violation (MFV), supersymmetry, and extra dimensions are investigated. For instance, in the MFV framework, bounds are set on the NP scale $\Lambda$ (assuming an effective coupling of $\pm 1/\Lambda^2$) for various $\Delta F=1$ and $\Delta F=2$ MFV operators. The lowest bound $\Lambda=1.5$~TeV is obtained for the operator $(\bar Q_L Y^u Y^{u\dagger} \gamma_\mu Q_L) (e D_\nu F^{\mu\nu})$, where observables from inclusive $\bar B \to X_s\ell\ell$ were used to set the bound. A similar analysis was done in~\cite{Porod:2014xia}, where flavour observables for BSM studies are implemented in {\tt SARAH}~\cite{Staub:2008uz} and {\tt SPheno}~\cite{Porod:2011nf}.

e) One can write down dimension-six operators which are singlets under the SM gauge group and which are baryon-number violating (BNV). This was first done in~\cite{Weinberg:1979sa} and later refined in~\cite{Grzadkowski:2010es,Alonso:2014zka}. The latter reference also takes into account two operators with right-handed (SM singlet) neutrino fields, and computes the one-loop renormalisation group equations (RGEs) for the dimension-six BNV operators. It turns out that the one-loop RGEs conserve baryon-number, so the dimension-six BNV operators only mix among themselves. The most stringent bounds on BNV operators come from the non-observation of proton decay.

\subsubsection{Applications in Higgs physics}
\label{sec:applHiggs}

\begin{figure}
\includegraphics[width=0.95\columnwidth]{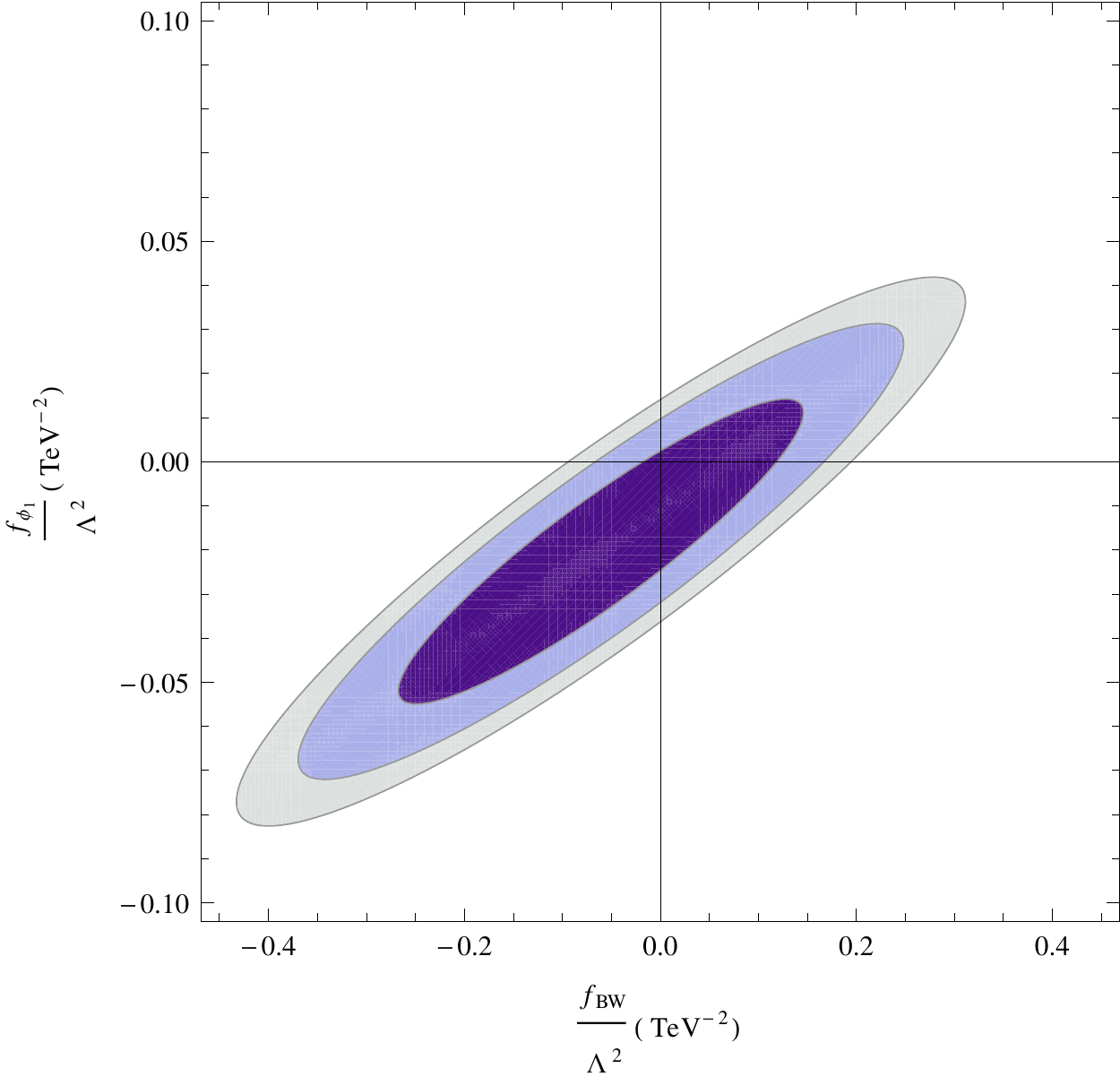}
\caption{\label{Huber_fig3} Limits from the oblique parameters on $f_{\Phi,1}$ and $f_{BW}$ for $\Lambda= 1$~TeV, for different confidence levels. Figure taken from~\cite{Chen:2013kfa}.}
\end{figure}

\begin{figure}
\includegraphics[width=0.95\columnwidth]{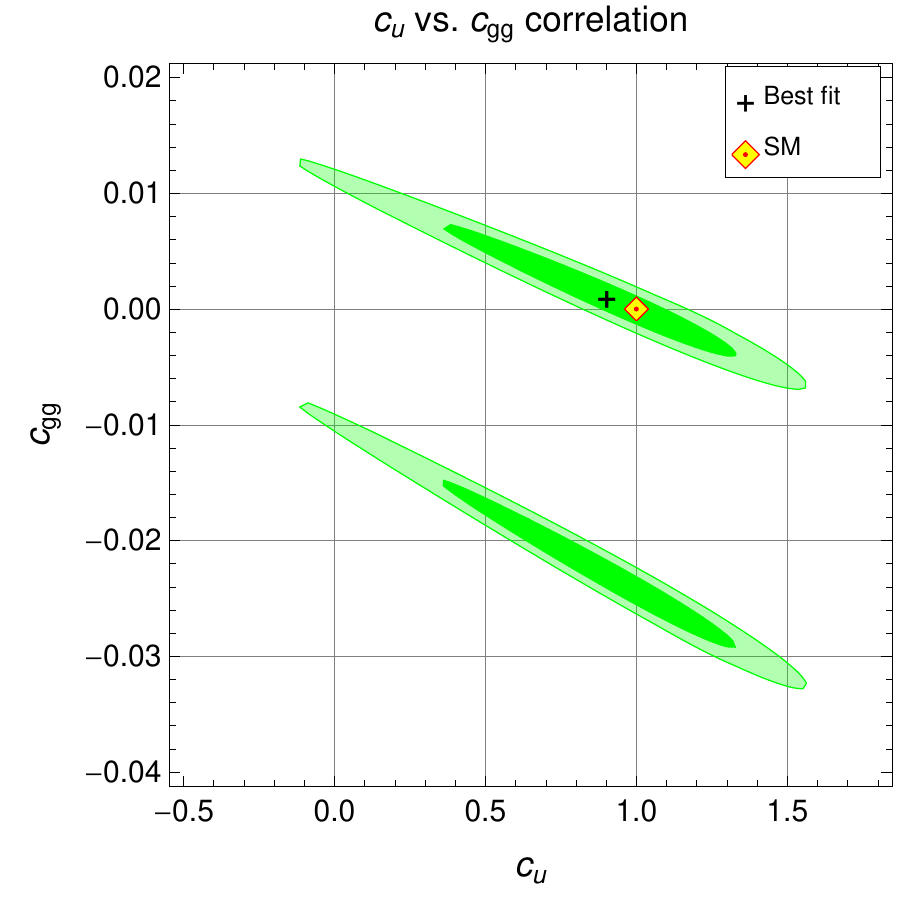}
\caption{\label{Huber_fig4} A fit in the $c_u - c_{gg}$ plane with the other couplings fixed at their SM values, for different confidence levels. Figure taken from~\cite{Belusca-Maito:2014dpa}.}
\end{figure}

Also in Higgs physics the applications of EFT are numerous and we have to stick to a few examples.

a) In ref.~\cite{Contino:2013kra} constraints on the operators from electroweak precision observables are investigated. For instance, for the operators
\begin{eqnarray}
\displaystyle O_W &=& i \bar c_W g/(2m_W^2) [H^\dagger \sigma^i \! \stackrel{\leftrightarrow}{D_\mu} H] [D_\nu W^{\mu\nu}]^i \, , \nonumber \\
\displaystyle O_B &=& i \bar c_B g^\prime/(2m_W^2) [H^\dagger \! \stackrel{\leftrightarrow}{D_\mu} H] [\partial_\nu B^{\mu\nu}] 
\end{eqnarray}
the 95\% CL constraint
\begin{equation}
\displaystyle -1.4 \times 10^{-3} < \bar c_W(m_Z) + \bar c_B(m_Z) < 1.9 \times 10^{-3}
\end{equation}
is provided (see also~\cite{Baak:2012kk}). For the fermionic operator
\begin{eqnarray}
\displaystyle O_{tG} &=& \bar c_{tG} \, g_s \, y_t/m_W^2 \, H^c \sigma^{\mu\nu} \lambda_a t_R \, G^a_{\mu\nu}\, , 
\end{eqnarray}
the 95\% CL constraints
\begin{eqnarray}
\displaystyle -6.12 \times 10^{-3} &<& {\rm Re}(\bar c_{tG}) < 1.94 \times 10^{-3} \, ,\nonumber \\
\displaystyle -1.39 \times 10^{-4} &<& {\rm Im}(\bar c_{tG})< 1.21 \times 10^{-4} 
\end{eqnarray}
could be derived from the $t\bar t$ cross section and the limits on the neutron EDM, respectively.

b) Ref.~\cite{Chen:2013kfa} studies effects on the oblique parameters $S$, $T$, and $U$, including one-loop corrections in the effective theory. Bounds on different operators, for instance
\begin{eqnarray}
\displaystyle O_{BW} &=& -g g^\prime/4 \, \Phi^\dagger B_{\mu\nu} \sigma^a W^{a,\mu\nu} \Phi\, , \nonumber \\
\displaystyle O_{\Phi,1} &=& (D_\mu \Phi)^\dagger (\Phi\Phi^\dagger) (D^\mu \Phi)
\end{eqnarray}
could be derived, see Fig.~\ref{Huber_fig3}. Limits on the coefficients in the EFT from loop contributions to the oblique parameters yield complementary information compared to direct Higgs production measurements.

c) In ref.~\cite{Belusca-Maito:2014dpa} a global fit to the effective operators  was performed using all available experimental data. For instance, constraints on
\begin{equation}
\displaystyle \frac{H^\dagger H}{v^2} \, \bar c_u \, y_u \, \bar q_L H^c u_R + \frac{\bar c_{gg} \, g_s^2}{m_W^2} H^\dagger H G^a_{\mu\nu} G^{a,\mu\nu}
\end{equation}
were obtained, see Fig.~\ref{Huber_fig4}. Similar results were shown for CP-odd parameters. A similar global fit was done in~\cite{Pomarol:2013zra} where correlations between different observables were investigated.

d) Constraints on anomalous gauge couplings and non-standard Higgs couplings were investigated in the EFT approach in~\cite{Degrande:2012wf} and~\cite{Buchalla:2013mpa}, respectively.

\subsection{Effectively Understanding Dark Matter$^7$}

\addtocounter{footnote}{1}

\footnotetext{Contributing author: William Shepherd}

We now move to show example of the use of EFTs in applications related to dark matter (DM).
%Effective Field Theory techniques can be fruitfully applied in many cases. In addition to %encoding known effects of heavy particles on physics at lower scales we can also use
Here we use them to stand in for the effects of as-yet unknown heavy particles. Many people have used this technique to study the possible interactions of DM with particles of the standard model. Unlike the more complicated cases often studied, it is straightforward in these theories to relate the different experimental probes of DM physics to one another and understand the interplay and complementarity of these disparate probes \cite{Arrenberg:2013rzp}.

\begin{figure}
\includegraphics[width=0.95\columnwidth]{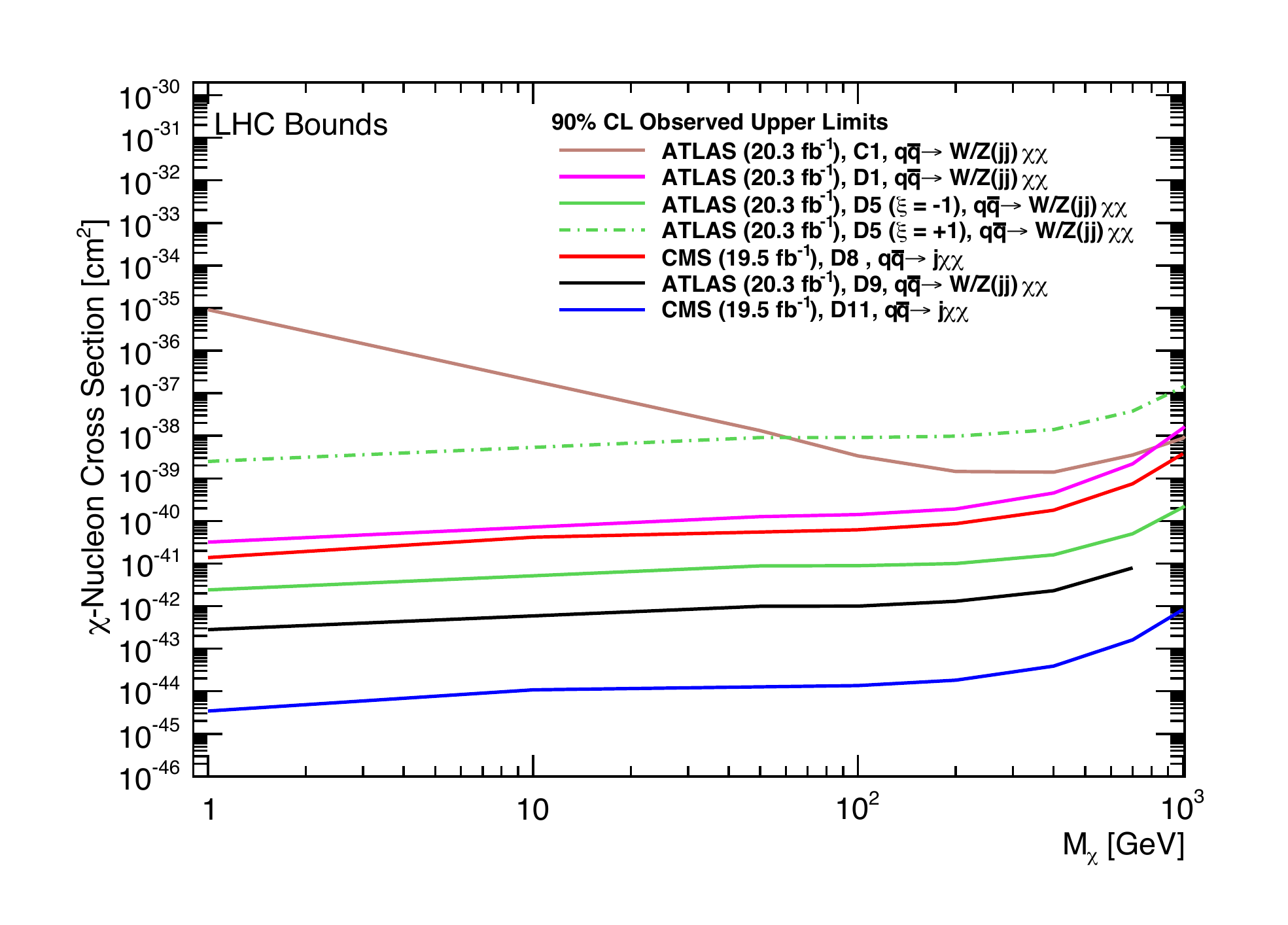}
\caption{\label{Shepherd_fig} The leading bounds from collider searches on dark matter operators from collider searches. The naming convention for the operators is due to \cite{Goodman:2010qn}, and the searches leading to the bounds are cited in the text. Figure taken from \cite{Askew:2014kqa}}
\end{figure}

The best known application of these techniques is to collider searches for dark matter\cite{Goodman:2010yf,Bai:2010hh,Goodman:2010ku, Fox:2011fx, 
Rajaraman:2011wf, Fox:2011pm, Friedland:2011za, Cheung:2012gi, 
Fox:2012ee, Bai:2012xg, Chae:2012bq, Fox:2012ru, Zhou:2013fla, Buckley:2013jwa, Essig:2013vha, Cotta:2013jna,
Crivellin:2014qxa, Busoni:2014sya, Mao:2014rga, Davidson:2014eia, Lopez:2014qja, 
Artoni:2013zba}, where one generally searches for some particle radiated from the initial state (generally a quark or gluon \cite{ATLAS:2012zim, CMS:rwa}, although photons \cite{Aad:2012fw, Chatrchyan:2012tea} and weak bosons \cite{CMS:2013iea, Aad:2013oja, Aad:2014vka} have also been searched for) and missing energy arising from the produced DM pair. Since their original proposal, these searches have been very quickly adopted by the LHC collaborations, leading to bounds such as those shown in Figure \ref{Shepherd_fig}. One can also perform searches for other operators, where radiation from the initial state is not required as the operator itself gives rise to a visible particle in addition to the DM pair \cite{Carpenter:2012rg, Cotta:2012nj, Chen:2013gya, Petrov:2013nia, Carpenter:2013xra}. EFT interactions of DM can also be used to understand the requirements for dark matter arising from our improving understanding of cosmology and from current searches for dark matter annihilation products \cite{Goodman:2010qn, Cornell:2013rza, Agrawal:2013hya, Nelson:2013pqa, Lopez:2014qja}.

Of course, the use of EFT techniques at the LHC are somewhat dubious, as one might hope that not only the DM itself but also the new particles which couple it to SM particles are kinematically accessible to the LHC, in which case it seems suspect to treat the interaction of DM and SM particles using contact interactions. Naturally, once the particle mediating the interaction is able to be produced at the LHC the true bounds can be either weaker or stronger than those derived from the contact operator approximation, and this problem has been studied by many \cite{Shoemaker:2011vi, Goodman:2011jq, Busoni:2013lha, 
An:2013xka, DiFranzo:2013vra, Buchmueller:2013dya, Chang:2013oia}. The CMS collaboration has already imrpoved one of its searches \cite{CMS:rwa} to include the possible effects of an accessible mediator, studying the bounds as a function of an assumed $Z^\prime$ mediator's mass. Interestingly, these techniques can also miss important points even in more low-energy probes, particularly when the DM interacts with different fields for the purposes of annihilation and direct detection, for instance \cite{Profumo:2013hqa}. Ultimately, the EFT techniques provide a dictionary for understanding the interplay between the various experiments searching for DM, and provide a simple benchmark to search for, as well as to search for deviations from once a signal is available.
%%%%%
\section{Extended Standard Model}
\label{sec:EFT}
\subsection{\boldmath $(g-2)_\mu$ versus BR$(\mu \to e \gamma)$ in the MSSM$^8$}
\label{test92}
\addtocounter{footnote}{1}

\footnotetext{Contributing authors: J\"orn  Kersten and  Liliana Velasco-Sevilla}

%\addtocounter{footnote}{1}

The measured value of the muon anomalous magnetic moment
$a_\mu = (g-2)_\mu/2$ deviates from the Standard Model prediction by more
than $3\sigma$
\cite{Bennett:2006,Davier:2010nc,Hagiwara:2011af,Kinoshita2012,Benayoun:2012wc,Gnendiger:2013pva},
which could be due to the contributions from light sleptons, charginos,
or neutralinos in the MSSM\@.
Besides, the Feynman diagrams for the SUSY contributions to $a_\mu$
and to the branching ratio $\text{BR}(\mu \to e \gamma)$ are
identical up to the flavour transition appearing only in the latter case.
For this reason, correlations between the two observables have been
studied \cite{Graesser:2001ec,Chacko:2001xd,Isidori:2007jw,Kersten:2014xaa}.

\begin{figure*}
\centering
\includegraphics{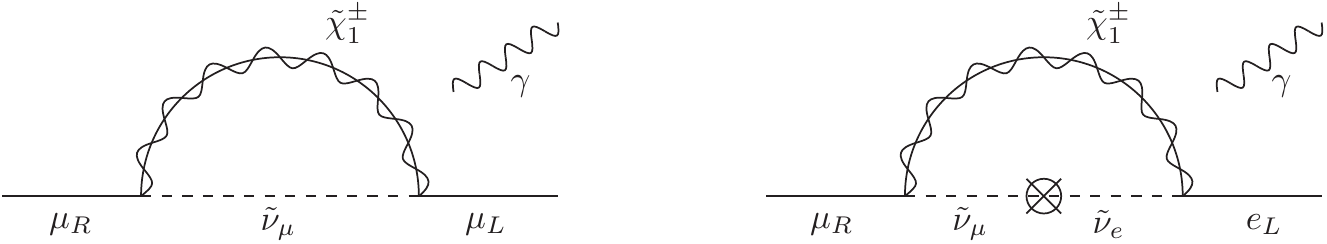}
\caption{Chargino--sneutrino contributions to $a_\mu$ and $\text{BR}(\mu \to e\gamma)$.
The cross denotes the flavour-violating parameter $m_{\tilde L_{12}}^2$.}
\label{fig:FeynmanCh}
\end{figure*}

As an example, Fig.~\ref{fig:FeynmanCh} shows the contributions from
sneutrinos and the lighter chargino.
As the couplings and loop functions involved in both diagrams are the
same, we expect for their ratio
\begin{equation} \label{eq:PerfectCorr}
\frac{a_{\mu e\gamma\text{L}}^{\tilde\chi^\pm_1}}{a_\mu^{\tilde\chi^\pm_1}} \propto
\frac{m^2_{\tilde L_{12}}}{m_{\tilde p}^2} \;,
\end{equation}
where $m^2_{\tilde L_{12}}$ is the flavour-violating (FV) entry in the soft
mass squared matrix of the left-handed sleptons that mixes selectrons
and smuons, and $m_{\tilde p}$ is the mass of a superparticle or a
combination of masses.  The branching ratio
$\text{BR}(\mu \to e\gamma) \propto |a_{\mu e\gamma\text{L}}|^2+|a_{\mu e\gamma\text{R}}|^2$
is obtained after summing over all diagrams and adding the amplitude
$a_{\mu e\gamma\text{R}}$ involving FV in the
right-handed sector.

Both the proportionality constant and the mass scale $m_{\tilde p}$
appearing in Eq.~\eqref{eq:PerfectCorr} are different for different
diagrams in general.  Consequently, we can expect a strong correlation
between $a_\mu$ and $\text{BR}(\mu \to e\gamma)$ only if a single
diagram dominates the SUSY contributions \cite{Graesser:2001ec,Chacko:2001xd}.
In the following we will study to which extent and in which parts of the
MSSM parameter space this is possible, summarizing results of
\cite{Kersten:2014xaa}.

\begin{figure}
\centering
\includegraphics[width=6cm]{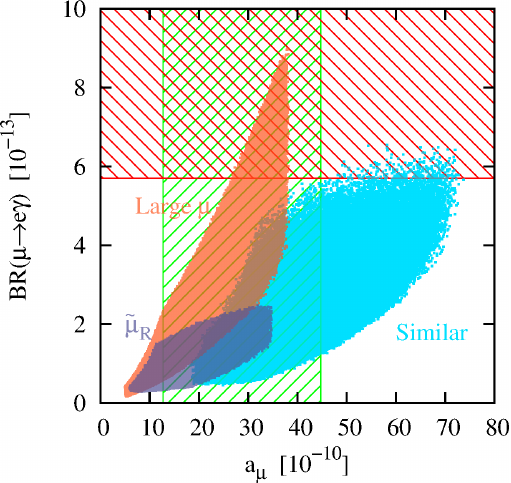}
\caption{Supersymmetric contribution to the anomalous magnetic moment of
the muon versus $\text{BR}(\mu \to e\gamma)$ for similar SUSY masses (light
blue), large $\mu$ (red), and heavy left-handed sleptons
($\tilde\chi^0$--$\tilde\mu_\text{R}$ dominance, violet).  In each case,
$\delta_\text{LL}=\delta_\text{RR}=2\cdot10^{-5}$ and $\tan\beta=50$.
The vertical hatched band corresponds to the experimentally favored
$2\sigma$ range for $a_\mu$, while the horizontal band marks the region
excluded by MEG\@.  Taken from \cite{Kersten:2014xaa}, which also
contains the details about the random scans.
}
\label{fig:Islands}
\end{figure}

As a first try, we randomly varied the relevant mass parameters $M_1$,
$M_2$, $\mu$, $m_{\tilde L_{11}}$, $m_{\tilde L_{22}}$, $m_{\tilde R_{11}}$, and $m_{\tilde R_{22}}$
between $300$\,GeV and $600$\,GeV, fixing
\begin{equation}
\delta_\text{LL} =
\frac{m^2_{\tilde L_{12}}}{m_{\tilde L_{11}} m_{\tilde L_{22}}}
\end{equation}
and the analogous parameter $\delta_\text{RR}$ to values compatible with
the experimental bound on $\text{BR}(\mu \to e\gamma)$ from MEG~\cite{Adam:2013mnn}.%
\footnote{For simplicity we assume all mass parameters to be real and positive.}
The result is the light blue region in Fig.~\ref{fig:Islands}.
We observe that for fixed $a_\mu$ the branching ratio varies by a
factor of about $10$.  Thus, in the considered scenario with similar
SUSY masses the correlation is sufficiently strong to correctly
determine the order of magnitude of $\text{BR}(\mu \to e\gamma)$ if
$a_\mu$ and the FV parameters are known.%
\footnote{Note that the variation of $\text{BR}(\mu \to e\gamma)$ becomes much
smaller if one imposes the constraints $M_2 \simeq 2 M_1$ and
$m_{\tilde L_{11}} = m_{\tilde L_{22}} = m_{\tilde R_{11}} = m_{\tilde R_{22}}$
\cite{Isidori:2007jw}.}

This allows to determine a limit $\delta_\text{LL} \lesssim 2 \cdot 10^{-5}$
below which $\text{BR}(\mu \to e\gamma)$ is guaranteed to satisfy the
MEG bound if our $7$ SUSY masses vary by up to $30\,\%$ around a
mass scale $M$ chosen such that we obtain the best-fit value of $a_\mu$ 
if all masses equal $M$.  Larger values up to
$\delta_\text{LL} \simeq 8 \cdot 10^{-5}$ are possible for some but not
all combinations of the masses and thus remain allowed.  The bounds on
$\delta_\text{RR}$ are much weaker.

The order-of-magnitude correlation found so far is not negligible but
nevertheless relatively weak.  This can be traced back to two reasons.
First, for similar SUSY masses cancellations between different diagrams
are always significant, reducing the value of $a_\mu$ by at least
$33\,\%$ compared to the contribution from the leading diagram.  In
other words, the leading diagram does not dominate sufficiently.
Second, $\delta_\text{LL}$ is not always the best choice for
parameterizing the FV\@.  Consequently, we can expect stronger
correlations in parameter space regions featuring hierarchies among the
SUSY masses which lead to the domination of a single diagram and
determine the optimal choice for $m_{\tilde p}$ in
Eq.~\eqref{eq:PerfectCorr}.  Let us consider three examples.

We can obtain \emph{chargino dominance} for $M_1 \sim \mu$ and
$\mu < M_2$ with a mass difference sufficiently large to make the
contribution of the heavier chargino small.  In this case the
correlation depends on the mass ratios
$x_{1,2} = ( m_{\tilde\chi^\pm_1}/m_{\tilde\nu_{1,2}} )^2$.
For $5$ out of the $9$ possible hierarchies, we can approximate the ratio 
$a_{\mu e\gamma\text{L}}/a_\mu$ as in Eq.~\eqref{eq:PerfectCorr} with an
accuracy of better than $50\,\%$, corresponding to an approximation for
the branching ratio that is accurate up to a factor of about $2$.
For example, for $x_1 \sim x_2 \sim 1$ we find
\begin{equation}
\left| \frac{a_{\mu e\gamma\text{L}}}{a_\mu} \right| \simeq
\frac{1}{4} \frac{m^2_{\tilde L_{12}}}{m_{\tilde\nu}^2} \simeq
\frac{1}{4} \delta_\text{LL} \;.
\end{equation}
Thus, $\delta_\text{LL}$ is indeed the most suitable FV parameter here.
If $x_1, x_2 \gg 1$, we find instead
\begin{equation} \label{eq:CorrLargex}
\left| \frac{a_{\mu e\gamma\text{L}}}{a_\mu} \right| \simeq
\frac{m^2_{\tilde L_{12}}}{m_{\tilde\chi_1^\pm}^2} 
%\not\simeq \delta_\text{LL}
\simeq
\delta_\text{LL} \frac{m_{\tilde\nu_1} m_{\tilde\nu_2}}{m_{\tilde\chi_1^\pm}^2}
\;.
\end{equation}
Now $\delta_\text{LL}$ is not suited well to parameterize the FV\@.
Rather, the correlation is governed by the chargino mass, the mass of
the heaviest particle in the diagram.  In order to quantify this, let us
compare two cases.  First, we fix $m_{\tilde\chi_1^\pm}$ (using the
measured value of $a_\mu$) as well as
$m^2_{\tilde L_{12}}/m_{\tilde\chi_1^\pm}^2$, and vary all remaining
parameters by up to a factor of $2$.  Then $a_{\mu e\gamma\text{L}}$
does not change by more than $50\,\%$.  Second, we fix
$m_{\tilde\chi_1^\pm}$ as well as $\delta_\text{LL}$, and vary the
remaining parameters.  In this case, $a_{\mu e\gamma\text{L}}$ changes
by a factor of $4$ according to Eq.~\eqref{eq:CorrLargex}, emphasizing
that $\delta_\text{LL}$ is not the decisive parameter determining this
amplitude.

In the \emph{large $\mu$} limit and for large $\tan\beta$, 
which was also studied in \cite{Endo:2013lva}, the dominant
SUSY contribution stems from the bino-like lightest neutralino and
charged sleptons.  For example, we can obtain the measured value of $a_\mu$
for $\tan\beta=50$, $\mu \simeq 4$\,TeV, $M_2 \simeq 1.8$\,TeV,
$M_1 \simeq 300$\,GeV, and slepton masses around $500$\,GeV\@.
We find that the approximation
\begin{equation} \label{eq:CorrLargeMu}
\left| \frac{a_{\mu e\gamma\text{L}}}{a_\mu} \right| \simeq
\frac{2}{3} \frac{m^2_{\tilde L_{12}}}{m_{\tilde e_\text{L}}^2} 
\end{equation}
is quite accurate, as is its analogue for the right-handed sector.
Again, $\delta_\text{LL}$ is not the parameter entering the analogue of
Eq.~\eqref{eq:PerfectCorr} (and neither is $\delta_\text{RR}$), and the
decisive superparticle mass $m_{\tilde e_\text{L}}$ is not even
necessarily the heaviest mass in the diagram.
In addition, $a_\mu$ is not very sensitive to the selectron masses, so 
it does not restrict the mass scale governing $\mu \to e\gamma$.  Hence,
the value of the correlation is somewhat limited in this case.  It could
still be used to place a lower limit on the selectron masses if
$m^2_{\tilde L_{12}}$ and $m^2_{\tilde R_{12}}$ were determined by an
additional source of information, such as a family symmetry.

\emph{Neutralino--$\tilde\mu_\text{R}$ dominance} occurs for very heavy
left-handed sleptons and $M_1, m_{\tilde\ell_\text{R}} < M_2, |\mu|$.
In this case we need a negative $\mu$ parameter to obtain the correct
positive sign of $a_\mu$ \cite{Stockinger:2006zn,Grothaus:2012js}.
We find a strong correlation that is well-approximated by
\begin{equation} \label{eq:CorrSmuR}
\left| \frac{a_{\mu e\gamma\text{R}}}{a_\mu} \right| \simeq
\frac{m^2_{\tilde R_{12}}}{m_{\tilde e_\text{R}}^2} \;.
\end{equation}

Fig.~\ref{fig:Islands} also shows points from random scans in the
large-$\mu$ and $\tilde\chi^0$--$\tilde\mu_\text{R}$ dominance regions.
As in the first scan with similar SUSY masses, we fixed
$\delta_\text{LL}$ and $\delta_\text{RR}$, leading to weaker
correlations than could be obtained by fixing the ``correct'' mass ratios
given in Eqs.~\eqref{eq:CorrLargeMu} and \eqref{eq:CorrSmuR}.

To summarize this section, we have found that for similar SUSY masses
determining both $a_\mu$ and $\text{BR}(\mu \to e\gamma)$, cancellations
are typical and the correlation between the two observables is
relatively weak.  However, there are interesting parameter space
islands with characteristic mass hierarchies and strong correlations.
In such regions the experimental limit on $\text{BR}(\mu \to e\gamma)$
can imply strong constraints on lepton FV parameters that cannot be
evaded by raising the overall SUSY mass scale, since the measured value
of $a_\mu$ fixes the mass scale of the contributing superparticles.

\subsection{ Effects of vectorlike leptons on Higgs decays and muon g-2$^9$}
\addtocounter{footnote}{-1}

\footnotetext{Contributing author: Radovan Derm\' \i\v sek}

Among simplest extensions of the SM are those with extra vectorlike fermions near the electroweak (EW) scale. Vectorlike fermions can acquire masses independently  of their Yukawa couplings to the Higgs boson and thus are not strongly constrained (compared to chiral fermions) by experiments. They can modify the evolution of  gauge couplings so that the couplings unify,   thus providing a framework that can be embedded into simple grand unified models (GUTs). 
Moreover, even small Yukawa couplings between  SM fermions and vectorlike fermions can affect a variety of processes, including the muon g-2 and  Higgs boson decays.

Extending the SM by three (or more) complete vector-like families  (VFs) with masses of order  1 TeV - 100 TeV allows for unification of gauge couplings~\cite{Dermisek:2012as, Dermisek:2012ke}. Predictions for gauge couplings at the EW scale are highly insensitive to fundamental parameters, and 
 ratios of  observed values  are to a large extent understood from the particle spectrum itself. The GUT scale can be sufficiently large to avoid the problem with fast proton decay, thus resurrecting simple non-supersymmetric GUT models.

The way this scenario works can be summarized in few steps. First,  extra 3VFs make all gauge couplings asymptotically divergent which opens a possibility for 
a unification with large (but still perturbative) unified gauge coupling. Consequently, in the RG evolution to lower energies gauge couplings run to the infrared fixed point.
Second,  the ratios of gauge couplings far from the GUT scale depend mostly on the particle content of the theory and they happen to be not far from the observed values. %at a chosen scale. 
Finally, the discrepancies between values of gauge couplings predicted from closeness to  the infrared fixed point and corresponding observed values  can be fully explained by  threshold effects  from  masses of particles originating from 3VFs. 
Note that  the first part is  similar to attempts to explain observed values of gauge couplings from infrared fixed point with 8 to 10 chiral families~\cite{Maiani:1977cg, Cabibbo:1982hy}  before the number of chiral families and  values of gauge couplings were tightly constrained.

 The evolution of  Higgs quartic  and  top Yukawa couplings is also significantly modified.  In the SM, the top Yukawa coupling already drives Higgs quartic coupling to negative values at a high scale. Additional sizable Yukawa couplings accelerate this behavior and thus the stability of the EW minimum sets a limit on the size of extra Yukawa couplings.
In the case of the SM extended by 3VFs the Higgs quartic coupling can remain positive all the way to the GUT scale even with additional Yukawa couplings. 
  The difference  comes from larger values of all gauge couplings compared to the SM above the scale of vectorlike fermions. Larger gauge couplings slow down the running of Higgs quartic coupling, and eventually turn the beta function of Higgs quartic coupling positive. This effect is further amplified by the fact that  the top Yukawa is driven fast to much smaller values compared to the SM (again due to larger gauge couplings)  and its contribution to the running of Higgs quartic coupling becomes small.

Extra vectorlike fermions near the EW scale can also contribute to many processes and observables involving SM particles and they have often been  considered to explain various anomalies. Examples include  attempts to explain the anomaly in   the forward-backward asymmetry of  the b-quark~\cite{Choudhury:2001hs, Dermisek:2011xu, Dermisek:2012qx} and the muon g-2 anomaly~\cite{Czarnecki:2001pv, Kannike:2011ng, Dermisek:2013gta}. In what follows I will focus on vector like leptons, the muon g-2 anomaly and possible modifications of other properties of the muon.

  If the muon mixes with vectorlike leptons originating from an $SU(2)$ doublet  L and a singlet E, 
 the deviation of the measured value of the muon anomalous magnetic moment from the standard model prediction can be completely explained. This mixing simultaneously contributes to the muon mass (we label this contribution by $ m_\mu^{LE}$), and the correlation between contributions to the muon mass and muon g-2 is controlled by the mass of the neutrino originating from the  doublet L,  that is given by the vectorlike mass parameter $M_L$~\cite{Dermisek:2013gta}. The possibility of explaining the muon g-2 anomaly by mixing of the muon with extra heavy leptons  was previously noticed  in  Ref.~\cite{Czarnecki:2001pv} and 
 the correlation between contributions from mixing to the muon mass and muon g-2 was also explored in Ref.~\cite{Kannike:2011ng}.

 \begin{figure}[h]
\includegraphics[width=3.5in]{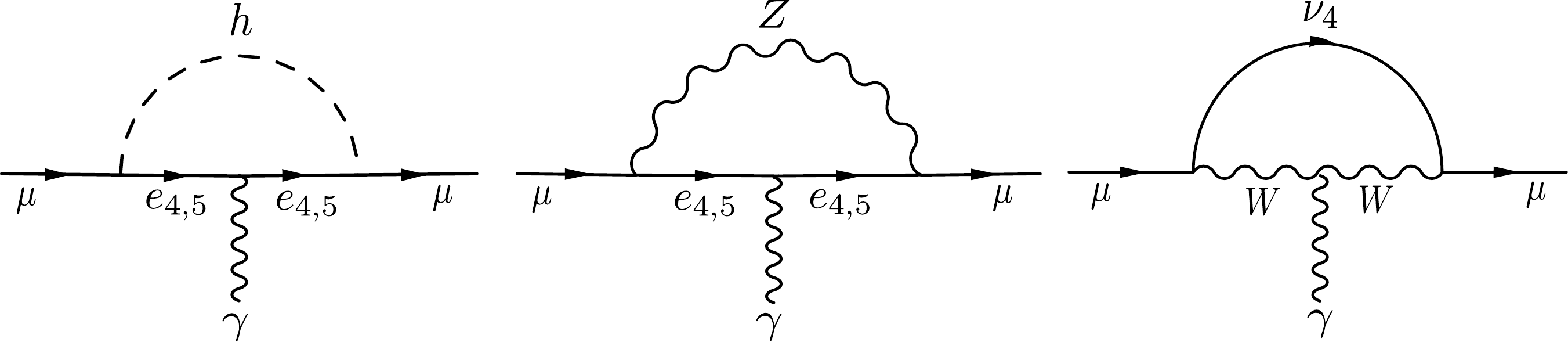} \caption{Feynman diagrams contributing to the muon magnetic moment that involve loops of new leptons and the Higgs, Z and W bosons.}
\label{fig:g-2}
\end{figure}

Feynman diagrams contributing to the muon magnetic moment that involve loops of new leptons are shown in Fig.~\ref{fig:g-2}. 
Depending on the mass of the heavy neutrino, there are two generic solutions: the asymptotic one,  $M_L \gg M_Z$, in which case the Higgs loop dominates and the measured value of the muon g-2 is obtained for $ m_\mu^{LE}/m_\mu \simeq -1$; and the second one with a light extra neutrino, $M_L \simeq M_Z$, in which case the $W$ loop dominates and the measured value of the muon g-2 is obtain for $ m_\mu^{LE}/m_\mu \simeq +1$. In the first case, about twice as large contribution from the direct Yukawa coupling of the muon is required to generate the correct muon mass, while in the second case, the muon mass can fully originate from the mixing with heavy leptons~\cite{Dermisek:2013gta}. 

As a result of the mixing, the Higgs coupling to the muon is not given by the physical muon mass. Therefore the branching ratio of $h \to \mu^+ \mu^-$ is modified  and can be significantly enhanced. Depending on additional Yukawa coupling, the branching ratio for $h \to \gamma \gamma$ can be also modified. 

The sizes of possible contributions to the muon g-2, muon mass  and other observables depend on the upper limit on Yukawa couplings (of vectorlike leptons or those that mix vectorlike leptons with the muon) that we allow in the model.
With the upper limit on Yukawa couplings being 0.5,  motivated by a simple UV embedding of this scenario  with three complete vectorlike families, the muon g-2 can be explained within one standard deviation either with  $M_L \lesssim 130$ GeV (the mass of the lightest extra charged lepton is $m_{e_4} \lesssim 150$ GeV), or with  $M_L \gtrsim 1$ TeV. The small $M_L$ case predicts the $h \to \mu^+ \mu^-$ in the range 5 -- 9 times the standard model prediction.
The branching ratio for $h \to \gamma \gamma$
can be enhanced by   $\sim$15\% or lowered by $\sim$25\% from its SM prediction. The asymptotic case predicts only very small modifications of  $h \to \mu^+ \mu^-$ and $h \to \gamma \gamma$ compared to the SM.  Results for larger Yukawa couplings being  allowed can be found in Ref.~\cite{Dermisek:2013gta}.

The small $M_L$ solution 
to the muon g-2 simultaneously explaining the muon mass completely from the mixing of the muon with vectorlike leptons is particularly interesting since it requires very light charged lepton.  In a large range of the parameter space this solution predicts the existence of $e_4$ below the Higgs mass and thus $h \to e_4^\pm \mu^\mp$ could be kinematically open and potentially significant. Subsequent decays of the heavy lepton, $e_4^\pm \to Z\mu^\pm$ and $e_4^\pm \to W^\pm \nu$,  lead to the same final states as $h \to ZZ^* \to Z \mu^+\mu^-$ and    $h \to WW^* \to W \mu\nu$, thus possibly affecting measurements of these processes~\cite{Dermisek:2014cia}. Since the partial width of $h \to Z\mu^+\mu^-$ is much smaller than $h \to W \mu\nu$ in the SM,  it is expected that the effect of the new lepton would show up in $h \to Z\mu^+\mu^-$ first  unless BR($e_4^\pm \to Z\mu^\pm$) is very small.

\begin{figure}[h]
\includegraphics[width=1.5in]{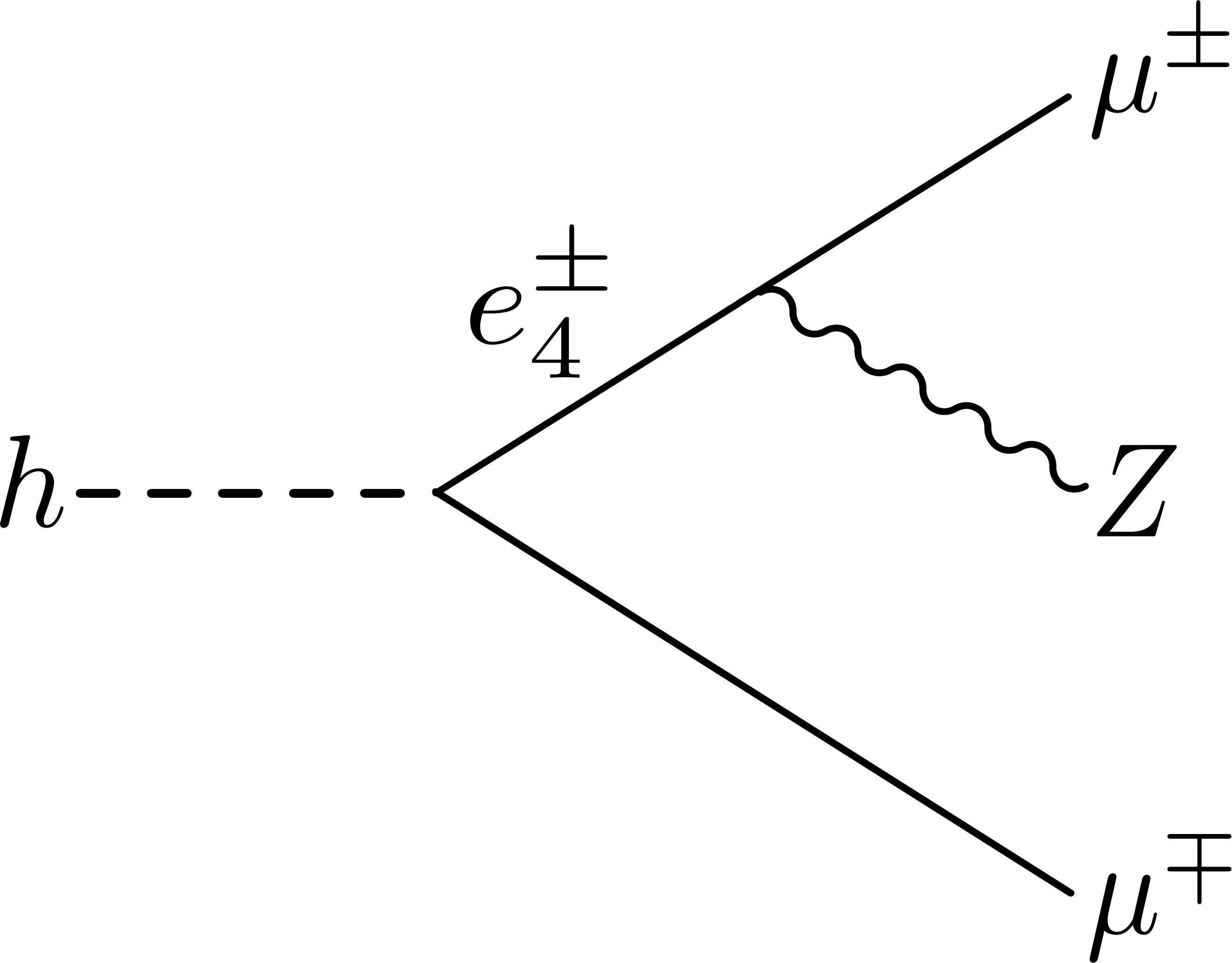} 
\caption{The Feynman diagram for $h \to e_4^\pm \mu^\mp \to Z \mu^+ \mu^-$ contributing to the same final state as $h\to ZZ^* \to Z \mu^+ \mu^-$.}
\label{fig:htoZmumu}
\end{figure}

The $e_4 - \mu - h $, $e_4 - \nu - W $ and $e_4 - \mu - Z $ couplings needed to explain the muon g-2 anomaly, see Fig.~\ref{fig:g-2},  are sufficient to modify the Higgs decays in $4\ell$, see Fig.~\ref{fig:htoZmumu}, and $2\ell 2\nu$ channels. Thus the contributions to the muon g-2 and $h\to 4\ell$ can be connected without any further assumptions. If only the muon mixes with vector like leptons, the new charged lepton can contribute to the $h\to 4\mu$ and $h\to 2e 2\mu$ processes. Without additional couplings it cannot contribute to $h\to 2\mu 2e$ (the first pair of leptons originating from the on-shell Z) or $h\to 4e$ decay modes.

Although the $4\ell$ final states originating from $h \to e_4^\pm \mu^\mp$ and  $h \to ZZ^*$ are identical, the kinematical distribution of final state leptons is not. The muon that accompanies the $e_4$ is somewhat soft, and if the mass of the $e_4$ is close to the Higgs mass, this muon does not pass the cuts used in the $h\to ZZ^*$ analysis. 
The acceptance drops significantly within about 6 GeV below the Higgs mass. For lighter  $e_4$ the $h \to e_4^\pm \mu^\mp$ can easily dominate over $h \to Z\mu^+\mu^-$ for values of couplings allowed by precision EW data and thus any enhancement in $h\to 4\mu$ and $h\to 2e 2\mu$ allowed by current limits is possible. 
 
There are also many scenarios which can explain the muon g-2 anomaly within 1 sigma and simultaneously significantly  enhance $h \to 4 \mu$ and $h\to 2e 2\mu$. Limiting Yukawa couplings to 0.5, the mass of $e_4$ has to be larger than about 113 GeV in order not to be ruled out by $h \to 4\mu$. Increasing the Yukawa couplings up to 1, the $m_{e_4}$can be close to the LEP limit for the small $M_L$ case. For the asymptotic case, $m_{e_4} $ is required to be larger than about 119 GeV.

If the $e_4$ is heavier than the Higgs boson then its contribution to $h \to 4 \mu$ and $h\to 2e 2\mu$ is very small. However the muon g-2 anomaly can still be fully explained. In this case the only solid connection of the muon g-2 anomaly and Higgs decays is through $h \to \mu^+\mu^-$. 

%New vectorlike leptons generically predict a variety of flavour violating processes. The existing limits set strong constraints on other possible couplings in the model besides those needed for the explanation of the muon g-2 anomaly.  
%In addition, extra charged leptons provide a variety of signatures at the LHC. They can be pair produced or can modify Higgs decays. Some searches are already excluding parts of the parameter space and others are getting close. Covering all possible decay modes of extra leptons should allow us to fully explore  the small $M_L$ case at the LHC with already available data.

\subsection{Very Minimal Composite Higgs Models$^{10}$}
\addtocounter{footnote}{1}

\footnotetext{Contributing author: Adri\'an Carmona}

%\subsubsection{Introduction}

Models of Composite Higgs \cite{Terazawa:1976xx, Terazawa:1979pj, Dimopoulos:1981xc, Kaplan:1983sm, Kaplan:1983fs, Contino:2003ve, Agashe:2004rs} provide one of the most compelling solutions to the hierarchy problem. In these models, the quadratic sensitivity of the Higgs boson mass to the ultra-violet is saturated by new physics at some scale $\Lambda\ll \Lambda_C\sim 4\pi f_{\pi}$, with $f_{\pi}\sim$~TeV, before the new strong interaction featuring the Higgs as a bound state starts to be resolved. Moreover, the small hierarchy existing between the scale of compositeness and the electroweak (EW) scale,  $\Lambda\gg M_{\rm EW}\sim v$,  can be alleviated within this framework if one assumes that the Higgs is the pseudo Nambu-Goldstone boson (pNGB) associated to the spontaneous breaking of some global symmetry $G$, analogously to what happens with pions in QCD. Thus, the Higgs boson can be effectively described at low energies by a non-linear $\sigma$-model parametrizing the breaking $G\to \mathcal{H}$, with $\mathcal{H}\subset G$. The Higgs boson %, which is then identified as the pNGB associated to the coset $G/\mathcal{H}$, 
 gets then a mass at the quantum level from weakly gauging just the EW subgroup of $G\supset G_{\rm EW}$, as well as from the interaction of the composite sector with the elementary fermions, transforming under the SM  group. %which are needed to generate the SM fermion masses within the framework of partial compositeness . 
 However, the relatively small value of the Higgs mass observed by the ATLAS and CMS experiments
 \cite{Aad:2012tfa, Chatrchyan:2012ufa}, 
%they have already been quoted before in the manuscript
$m_{H}\approx 125$ GeV, together with the large top mass value,  require the masses of some of the composite states mixing with the elementary top chiralities via linear mixings to be rather small, $m_{f}\sim f_{\pi}\ll \Lambda$. Otherwise, the coefficients of these linear mixings would become too big in order to accommodate the top mass, leading then to an excessively large breaking of the Goldstone symmetry and thus to a Higgs heavier than observed. For most minimal scenarios and natural values of $f_{\pi}\lesssim 1$~TeV, the presence of such ``anomalously'' light top partners leads to considerable tension with current LHC data \cite{Chatrchyan:2013uxa}. 

Until very recently, the only viable way of lifting the masses of these ultra-light top partners (without increasing the scale of compositeness) required the embedding of quarks in the symmetric representation of $SO(5)$ \cite{Panico:2012uw,Pappadopulo:2013vca}. However, these models  suffer  generically from an ``ad-hoc'' tuning, as the predicted Higgs mass sits in principle close to $f_{\pi}$, thus asking for a cancellation of unrelated parameters to bring it down. Even though, quantitatively, the tuning resulting in these setups is similar to the one existing in the most economical scenarios MCHM$_{5,10}$, demanding such large representations without any fundamental reason seems to go against the principle of minimality. In particular, these models populate the scale of fermionic resonances $m_{\psi}\equiv g_{\psi}f_{\pi}$ with a large number of colored particles. However, as it was shown recently \cite{Carmona:2014iwa}, the inclusion of a realistic lepton sector can in some cases change dramatically this picture,  lifting considerably  the masses of the lightest top-partner resonances without adding any new light degree of freedom to the spectrum. Moreover, it was also shown that extremely minimal realizations of the lepton sector could still avoid the need of light partners due to its sizable contributions to the Higgs potential, triggered by a type-III seesaw mechanism. In the following, we will review some of these scenarios, trying to shed some light on the deep relation existing in these models between the flavor pattern in the lepton sector and the predicted value of the Higgs mass. In order to fix some notation and illustrate some key ingredients of these models, we will first review briefly their generic five-dimensional (5D) descriptions.

\subsubsection{General (5D) Setup}
We consider a slice of AdS$_5$ with metric
\begin{eqnarray}
	\mathrm{d}s^2=a^2(z)\left(\eta_{\mu\nu}\mathrm{d}x^{\mu}x^{\nu}-\mathrm{d}z^2\right),
\end{eqnarray}
where $z \in [R, R^{\prime}]$ is the coordinate of the extra dimension, $R$ and $R^{\prime}$ are the positions of the ultra-violet (UV) and infra-red (IR) branes, respectively, and $a(z)\equiv R/z$. The bulk of the extra dimension is symmetric under the gauge group $SO(5)\times U(1)_X$,\footnote{Larger cosets can also be considered, which may even lead to the presence of Dark Matter candidates as explicitly studied in \cite{Gripaios:2009pe, Frigerio:2012uc, Chala:2012af}.} which is broken by boundary conditions to the EW group $SU(2)_L\times U(1)_Y$ on the UV brane and to $SO(4)\times U(1)_X$ on the IR one. More explicitly, this setup correspond to the following choice of boundary conditions 
\begin{eqnarray}
	L_{\mu}^{a}(+,+), \qquad R_{\mu}^{b}(-,+), \qquad B_{\mu}(+,+), \nonumber\\
	\qquad Z_{\mu}^{\prime}(-,+), \qquad C_{\mu}^{\hat{a}}(-,-),
\end{eqnarray}
where $a=1,2,3$,  $b=1,2$, $\hat{a}=1,2,3,4$ and $-/+$ denote Dirichlet/Neumann boundary conditions at the corresponding brane. The respective (4D) scalar components, i.e.,  $\mu \to 5$, have opposite boundary conditions, allowing for zero modes only in $C_{5}^{\hat{a}}$.
In the above equation,  $L_{\mu}^{1,2,3}$ and $R_{\mu}^{1,2,3}$  are the 4D vector components of the 5D gauge bosons associated to $SU(2)_L$ and $SU(2)_R$, respectively, both subgroups of $SO(5)$. We have also defined the linear combinations
\begin{eqnarray}
	B_{\mu}&\equiv &s_{\phi}R_{\mu}^3+c_{\phi} X_{\mu},\qquad Z^{\prime}_{\mu}\equiv c_{\phi} R_{\mu}^3-s_{\phi} X_{\mu},\nonumber\\
c_{\phi}&\equiv  &\frac{g_5}{\sqrt{g_5^2+g_X^2}},\qquad   \ \ \ \  s_{\phi}   \equiv  \frac{g_X}{\sqrt{g_5^2+g_X^2}},
\end{eqnarray}
with $g_5$ and $g_X$  being the dimensionfull 5D gauge couplings of $SO(5)$ and $U(1)_X$, respectively, and $X_{\mu}$ the gauge boson associated with  $U(1)_X$. Finally, $C_{\mu}^{\hat{a}}$ are the gauge bosons corresponding to the broken generators $\in SO(5)/SO(4)$, whose scalar counterparts provide zero-modes $C_{5,(0)}^{\hat{a}}(x,z)\equiv f_h^{\hat{a}}(z)h^{\hat{a}}(x)$ spanning a $SU(2)_L\times U(1)_Y$ doublet, with the proper quantum numbers to be identified with the Higgs boson doublet.

We fix $1/R\sim 10^{16}$~TeV and, for each value of  $1/R^{\prime}\sim \mathcal{O}(1)$~TeV 
addressing the hierarchy problem, we obtain $g_5$, $s_{\phi}$ and $\langle h^{\hat{a}}\rangle =v \delta_{\hat{a}4}$ in terms of $\{\alpha_{\rm QED}$, $M_W$, $M_Z$\}. This implies that, besides the value of $R\sim M_{\rm Pl}^{-1}$ fixed by naturalness, the only free parameter in the 5D gauge sector  is  $R^{\prime}$,\footnote{We assume for simplicity no brane localized kinetic terms.} or equivalently, 
\begin{eqnarray}
	f_{\pi}\equiv \frac{\sqrt{2}}{g_5}\left[\int_R^{R^{\prime}}\mathrm{d}z~a^{-1}(z)\right]^{-1/2}\approx \frac{2R^{1/2}}{g_5R^{\prime}}.
\end{eqnarray}
With very good approximation, we obtain
\begin{eqnarray}
	g_{\ast}\approx \frac{e}{\sin\theta_W}\sqrt{\log(R^{\prime}/R)},\qquad  s_{\phi}\approx \tan\theta_W\,,
\end{eqnarray}
and
\begin{eqnarray}
	M_W\approx \frac{e}{2}\sin\theta_W f_{\pi}	\sin(v/f_{\pi}),
\end{eqnarray}
where $g_{\ast}\equiv g_5 R^{-1/2}$ is the dimensionless 5D gauge coupling, $e=\sqrt{4\pi\alpha_{\rm QED}}$ is the electric charge and $\theta_W$ is the Weinberg angle. 

The fermion sector will depend on the specific $SO(5)$ representations in which the 5D fields transform,  $\mathbf{1}, \mathbf{4}, \mathbf{5}, \mathbf{10}$ or $\mathbf{14}$. %According to their $SO(4)\cong SU(2)_L\times SU(2)_R$ decomposition, they read 
Taking into account that $Y=T_R^3+Q_X$ it would be straightforward to work out all possible embeddings of the SM fermions. Since fermions with a sizable degree of compositeness  are the only ones playing a non-negligible role in the generation of the Higgs potential and the Higgs mass, henceforth we will neglect UV localized fermions. However,  to reproduce the different charged lepton masses we will still include left-handed (LH) leptons if their right-handed (RH) counterparts are composite, as it will be the case (see below). Neglecting thus the first two quark generations as well as the RH bottom, the smallest embedding of the quark sector features a fundamental representation of $SO(5)$ and a full singlet,
\begin{eqnarray}
\zeta_{1}&=&\left(\begin{array}{r}\tilde{\Lambda}_{1}[-,+]~  t_1[+,+]\\ \tilde{t}_{1}[-,+]~ b_1[+,+]\end{array}\right)\oplus t^{\prime}_{1}[-,+]\sim \mathbf{5}_{2/3},\nonumber\\
	\zeta_{2}&=&t^{\prime}_{2}[-,-]\sim \mathbf{1}_{2/3},
\end{eqnarray}
where $Q_X=2/3$ and we have explicitly shown the decomposition under $SO(4)\cong SU(2)_L\times SU(2)_R$, with the bidoublet being represented by a $2\times 2$ matrix on which the $SU(2)_L$ rotation acts vertically and the $SU(2)_R$ one horizontally. In particular, the left and right columns correspond to fields with $T_R^3=\pm 1/2$, whereas the upper and lower rows have $T_L^3=\pm 1/2$. The signs in square brackets denote the boundary conditions on the corresponding branes. A Dirichlet boundary condition for the LH chirality is denoted by $[-]$ while the opposite sign denotes the same boundary condition for the RH one. Hence, before electroweak symmetry breaking (EWSB),  zero-modes with quantum numbers $\mathbf{2_{1/6}}$ and $\mathbf{1_{2/3}}$ under $SU(2)_L\times U(1)_Y$ are present in $\zeta_1$ and $\zeta_2$, respectively. For the sake of concreteness, we will call this setup MCHM$_{5-1}$,  where the first and second subscripts refer to the specific embedding of the  LH and RH SM-fermions, respectively.\footnote{If they are both equal we will use just one index as customary.}

However, in the usual paradigm where the top is almost fully responsible of triggering the EWSB, such a compact realization of the quark sector does not provide a viable Higgs potential (see e.g. \cite{Carmona:2014iwa}). This is typically solved by promoting the full singlet hosting the RH top to a fundamental of $SO(5)$, i.e., by going to the MCHM$_5$,
\begin{eqnarray}
\zeta_{1}&=&\left(\begin{array}{r}\tilde{\Lambda}_{1}[-,+]~  t_1[+,+]\\ \tilde{t}_{1}[-,+]~ b_1[+,+]\end{array}\right)\oplus t^{\prime}_{1}[-,+]\sim \mathbf{5}_{2/3},\nonumber\\
\zeta_{2}&=&\left(\begin{array}{r} \tilde{\Lambda}_{2}[+,-]~  t_2[+,-]\\ \tilde{t}_{2}[+,-]~ b_2[+,-]\end{array}\right) \oplus t^{\prime}_{2}[-,-]\sim \mathbf{5}_{2/3}.\qquad
\end{eqnarray}
Nevertheless, as it was shown in \cite{Carmona:2014iwa} and we will see below, the consideration of minimal leptonic sectors featuring a type-III seesaw results in additional sizable contributions to the Higgs potential that can render viable the MCHM$_{5-1}$. Henceforth we will just consider these two scenarios in the quark sector.

In both cases, the relevant part of the action reads
\begin{eqnarray}
	\mathcal{S}&\supset&\sum_{k=1,2}\int\mathrm{d}^4x\int_R^{R^{\prime}}\mathrm{d}z~a^4\left\{\bar{\zeta}_k\left[i\cancel{D}+\left(D_5+2\frac{a^{\prime}}{a}\right)\gamma^5 \right.\right.\nonumber\\
&&\left.\left.\phantom{\frac{1}{2}}-aM_k\right]\zeta_k\right\}+\mathcal{S}_{\rm UV}+\mathcal{S}_{\rm IR},
\end{eqnarray}
with\footnote{See \cite{Carmona:2014iwa} for explicit expressions of the $SO(5)$ generators.}
\begin{eqnarray}
	D_{M}&=&\partial_M-ig_5T_L^a L_M^a-ig_5 T_R^b R_M^b-ig_Y Y B_M\nonumber\\
			&&-i\frac{g_Y}{c_{\phi}s_{\phi}}Z_{M}^{\prime}\left(T_R^3-s_{\phi}^2Y\right) -ig_5 T^{\hat{a}} C_M^{\hat{a}},
\end{eqnarray}
where $M=\mu,5$ and $g_Y\equiv g_5g_X/\sqrt{g_5^2+g_X^2}$. $\mathcal{S}_{\rm UV}$ and $\mathcal{S}_{\rm IR}$ include possible brane localized terms. As usual, we have parametrized the bulk masses $M_k=c_k/R$ in terms of dimensionless bulk mass parameters $c_k$ and the fundamental scale $R$. The fifth component of the covariant derivative in the above action generates the Yukawa interactions
\begin{eqnarray}
	\mathcal{S}&\supset& -\sum_{k=1,2}ig_5\int\mathrm{d}^4x\int_R^{R^{\prime}}\mathrm{d}z~a^4\bar{\zeta}_k \gamma^5 T^4 \zeta_k C_5^4=\\
									   &-&\frac{i}{\sqrt{2}}g_5^2 f_{\pi}\sum_{k=1,2}\int\mathrm{d}^4x\int_R^{R^{\prime}}\mathrm{d}z~a^3\bar{\zeta}_k \gamma^5 T^4 \zeta_k h+\ldots,\quad \nonumber
\end{eqnarray}
where the dots stand for terms involving the non-physical Kaluza-Klein (KK) excitations of the Higgs boson and we have used that the Higgs profile is given by
\begin{eqnarray}
	f_h^{4}(z)=a^{-1}\left[\int_R^{R^{\prime}}\mathrm{d}z^{\prime}~a^{-1}\right]^{-1/2}=\frac{1}{\sqrt{2}}g_5 f_{\pi}a^{-1}.
\end{eqnarray}
 
Looking at the specific form of the Yukawa interactions, one can readily see that a non-zero mass for the zero-modes after EWSB requires the addition of  some IR brane terms splitting the zero-modes between the different multiplets. Therefore we consider the following IR localized action
\begin{eqnarray}
	\mathcal{S}_{\rm IR}&=&-\int\mathrm{d}^4x ~\left\{a^4\left[M_{S}^q\overline{\zeta}^{(\mathbf{1},\mathbf{1})}_{1L} \zeta^{(\mathbf{1},\mathbf{1})}_{2R}\nonumber\right.\right.\\
	&&+\left.\left.M_B^q\overline{\zeta}^{(\mathbf{2},\mathbf{2})}_{1L} \zeta^{(\mathbf{2},\mathbf{2})}_{2R}\right]\right\}_{z=R^{\prime}}+\mathrm{h.c.},
\end{eqnarray}
where we have used the $SO(4)$ decomposition $\zeta=\zeta^{(\mathbf{2},\mathbf{2})}+\zeta^{(\mathbf{1},\mathbf{1})}$, and $\zeta_{2}^{(\mathbf{2},\mathbf{2})}\equiv 0\equiv M_B^q$ for the MCHM$_{5-1}$. 

When considering minimal scenarios in the lepton sector, it is instructive to take a close look to the symmetric representation of $SO(5)$, whose decomposition under $SO(4)$ reads $\mathbf{14}=(\mathbf{1},\mathbf{1})\oplus (\mathbf{2},\mathbf{2})\oplus (\mathbf{3},\mathbf{3}) $.  One can easily see that it is the only one which can host at the same time a $P_{LR}$ protected $SU(2)_L \times U(1)_Y$ singlet and a triplet $\sim \mathbf{3}_0$. This implies in particular that, using the $\mathbf{14}$, we can build very economical models in the lepton sector, where the neutrino masses are generated via a type-III seesaw.  In the following, we will consider the most minimal of these scenarios, the so called mMCHM$^{\rm III}$, realized with LH and RH leptons transforming as $\mathbf{5_{-1}}$ and $\mathbf{14_{-1}}$, respectively, under $SO(5)\times U(1)_X$.  The corresponding boundary conditions  read
\begin{eqnarray}
\xi_{1\tau}&=&\tau^{\prime}_{1}[-,+]\oplus \left(\begin{array}{r}\nu_{1}^{\tau}[+,+]~  ~\tilde{\tau}_1[-,+]\\ \tau_{1}[+,+]~ \tilde{Y}_1^{\tau}[-,+]\end{array}\right)\sim \mathbf{5}_{-1},\nonumber\\
 \xi_{2\tau}&=&\tau^{\prime}_{2}[-,-]\oplus\left(\begin{array}{r} \nu_{2}^{\tau}[+,-]~~  \tilde{\tau}_2[+,-]\\ \tau_{2}[+,-]~ \tilde{Y}_2^{\tau}[+,-]\end{array}\right) \\
						   &\oplus&\left(\begin{array}{r} \hat{\lambda}^{\tau}_2[-,-]~~\nu_{2}^{\tau\prime\prime}[+,-]~  ~~\tau_2^{\prime\prime\prime}[+,-]\\ 
	\hat{\nu}_2^{\tau}[-,-]~~~\tau_{2}^{\prime\prime}[+,-]~Y_2^{\tau\prime\prime\prime}[+,-]\\ \hat{\tau}_2[-,-]~Y_2^{\tau\prime\prime}[+,-]~\Theta_2^{\tau\prime\prime\prime}[+,-] \end{array}\right)\sim \mathbf{14}_{-1},\nonumber\qquad
\end{eqnarray}
where for simplicity we have just shown the multiplets for the third generation, being the ones for the first two generations completely analogous. These boundary conditions imply the presence of the following $SU(2)_L\times U(1)_Y$ zero-modes,  $l_{\ell L}^{(0)}\sim \mathbf{2_{-1/2}}\subset \xi_{1\ell}$ and $\ell_R^{(0)}\sim\mathbf{1_{-1}},\Sigma_{\ell R}^{(0)}\sim \mathbf{3_0}\subset \xi_{2\ell }$, with $\ell =e,\mu,\tau$.  

In this case, we can write down the following UV Majorana mass, 
\begin{eqnarray}
	\mathcal{S}_{\rm UV}&=&-\frac{1}{2}\sum_{\ell}\int\mathrm{d}^4x\int_{R}^{R^{\prime}}\mathrm{d}z\left\{a^4M_{\Sigma}^{\ell}\mathrm{Tr}\left(\bar{\Sigma}_{\ell R} \Sigma_{\ell R}^c\right)\right\}_{z=R}\nonumber\\
	&&+\mathrm{h.c.},
\end{eqnarray}
where 
\begin{eqnarray}
\Sigma_{\ell}=\begin{pmatrix}\hat{\nu}_2^\ell/\sqrt{2}&\hat{\lambda}_2^\ell\\ \ell_2& -\hat{\nu}_2^\ell/\sqrt{2}\end{pmatrix}, \qquad \ell=e,\mu,\tau,
\end{eqnarray}
are the 5D $SU(2)_L\times U(1)_Y$ triplets hosting the $\Sigma_{\ell R}^{(0)}$ zero-modes. On the other hand, the IR brane masses read
\begin{eqnarray}
	\mathcal{S}_{\rm IR}&=&\sum_{\ell}\int\mathrm{d}^4x \left\{a^4\left[\frac{\sqrt{5}}{2}M_{S}^{\ell}\left(\overline{\xi}^{(\mathbf{1},\mathbf{1})}_{1\ell L} \xi^{(\mathbf{1},\mathbf{1})}_{2\ell R}\right)_{5}\right.\right.\nonumber\\
																		   &&\left.\left.\phantom{\frac{1}{2}}+\sqrt{2}M_B^{\ell}\left(\overline{\xi}^{(\mathbf{2},\mathbf{2})}_{1\ell L} \xi^{(\mathbf{2},\mathbf{2})}_{2jR}\right)_{5}\right]\right\}_{z=R^{\prime}}+\mathrm{h.c.},\qquad 
\end{eqnarray}
where for convenience we have added  prefactors $-\sqrt{5}/2$ and $-\sqrt{2}$, see \cite{Carmona:2014iwa}, and, for the sake of simplicity, we have assumed all brane masses $M_{\Sigma}^{\ell},M_S^{\ell}$ and $M_B^{\ell}$ to be  diagonal.

The Majorana mass matrix for the corresponding zero-modes reads
\begin{eqnarray}
	\mathcal{M}_{\rm M}^{\ell \ell^{\prime}}\approx \frac{f_{-c_2^{\ell}}^2}{R^{\prime}}\left(\frac{R^{\prime}}{R}\right)^{-2c_2^{\ell^{\prime}}}M_{\Sigma}^{\ell}\delta_{\ell \ell^{\prime}}, \quad \ell ,\ell^{\prime}\in\{e,\mu,\tau\},\quad
\end{eqnarray}
where
\begin{eqnarray}
	f_c\equiv \left[\frac{1-2c}{1-\left(\frac{R}{R^{\prime}}\right)^{1-2c}}\right]^{\frac{1}{2}}
\end{eqnarray}
is the zero-mode wave function at the IR brane. This mass matrix is typically too large, $\Vert \mathcal{M}_{\rm M}\Vert \sim \mathcal{O}(M_{\rm Pl})$, unless the corresponding zero-mode profiles are pushed away from the UV brane. This leads to values of  $c_2^{\ell}\in (-1/2,0)$ and thus IR localized RH zero-modes. Therefore, just the quantum numbers of the lepton sector and the overall scale of the neutrino masses lead naturally to IR localized leptons for all three generations. This will allow us to compensate the relative color suppression of the lepton sector in the contribution to the Higgs potential,  making this setup particularly interesting for lifting the masses of the top partners.

\subsubsection{Two Concrete Examples} 

In the following we will study in more detail two particular examples of highly economical composite Higgs models, both featuring the smallest implementation of a type-III seesaw in the lepton sector. In particular, we will consider the mMCHM$_{5}^{\rm III}$ and the mMCHM$_{5-1}^{\rm III>}$, where as before subscripts refer to the specific quark representations and the superscript ``$>$'' implies that there will be no additional hierarchy between the brane masses in the lepton and the quark sector (see below). In both cases, we perform a numerical scan over the different brane masses for fixed values of $R=10^{-16}$~TeV$^{-1}$ and $f_{\pi}=0.8$~TeV,  which correspond roughly to $g_{\ast}\approx 4.0$ and $s_{\phi}\approx \tan\theta_W$.  We assume brane masses fulfilling
\begin{eqnarray}
	| M_S^q|,|M_B^q|,|M_T^q|\le Y_\ast^q,\qquad  M_S^{q,l},M_{B}^{q,l},M_{T}^{q,l}\in\mathbb{C},\quad
\end{eqnarray}
and
\begin{eqnarray}
	|M_{\Sigma}^\ell|,| M_S^\ell|,|M_B^\ell|\le Y_\ast^l, \nonumber\\
	M_{\Sigma}^\ell,M_{S}^\ell,M_{B}^\ell\in \mathbb{R},  \quad \mathrm{with}\quad\ \ell=e,\mu,\tau,
	\label{twomass}
\end{eqnarray}
where we have taken real brane masses in the lepton sector  for the sake of simplicity, since due to the presence of the Majorana masses $M_{\Sigma}^\ell$ the size of the system of equations that we have to solve (which is already $10\times10$ in this case) would double. The numbers $Y_\ast^{q},Y_\ast^l\in\mathbb{R}^{+}$ are fixed to some benchmark values specified below. The quark bulk masses $c_1^q$ and $c_2^q$ are fixed requiring
\begin{eqnarray}
	\left.	\frac{\partial V(h)}{\partial h}\right|_{h=v}=0, \qquad m_{t}=m_t^{\rm ref},
\end{eqnarray}
with
\begin{eqnarray}
m_t^{\rm ref}\in [145, 155] ~\mathrm{GeV}
\end{eqnarray}
being the top mass evaluated at the high scale $f_{\pi}\sim \mathcal{O}(1)$~TeV. On the other hand, for each lepton generation $\ell$, the bulk masses are fixed by imposing the corresponding charged lepton masses and the following neutrino spectrum   
\begin{eqnarray}
	m_{\nu}^\ell=m_{\nu}^{\ell;\mathrm{ref}},
	\end{eqnarray}
	with\footnote{For our purposes it is enough to impose a reasonable neutrino mass scale. One could easily generalize this for a complete flavor model, reproducing  the neutrino mass squared differences as well as the observed PMNS mixing matrix.} 
	\begin{eqnarray}
 m_{\nu}^{\ell;\mathrm{ref}}=\varepsilon_\ell 10^{-p_\ell}~\mathrm{eV}\quad \mathrm{and}\quad  \varepsilon_\ell\in[0,1],~p_\ell\in[0,3].
\end{eqnarray}

The chosen value $f_{\pi}=0.8$~TeV ensures a reasonable agreement with electroweak precision data, since 
\begin{eqnarray}
	T=0,\quad S\approx \frac{3}{2}\pi v^2R^{\prime 2}=6\pi \left(\frac{v}{f_{\pi}}\right)^2g_{\ast}^{-2}, \quad U=0,\
\end{eqnarray}
in these models at tree level \cite{Agashe:2004rs, Csaki:2008zd}, which leads in our case to $S\approx 0.112$ and $T=0=U$.
In general, even though $\left.S\right|_{U=0}=0.05\pm 0.09$ at $95\%$ confidence level \cite{Baak:2012kk}, the large correlation with the predicted value of the $T$ parameter,  $\rho_{\rm corr.}=0.91$, would in principle point to a larger value of $f_{\pi}$.  However, as we are also neglecting non-oblique effects as well as radiative fermion   corrections which can give an non-negligible positive contribution   to   the  $T$   parameter  \cite{Anastasiou:2009rv}, we will be satisfied by the numbers shown above as a first approximation. Besides this, we have also checked that $|\delta g_{Z \bar{\ell} \ell}/g_{Z \bar{\ell}\ell}|\leq 2\permil$, with $~\ell=e,\mu,\tau,$ for all points on the scan.

\begin{figure}[!h]
\begin{center} 
		   \includegraphics[width=0.39\textwidth]{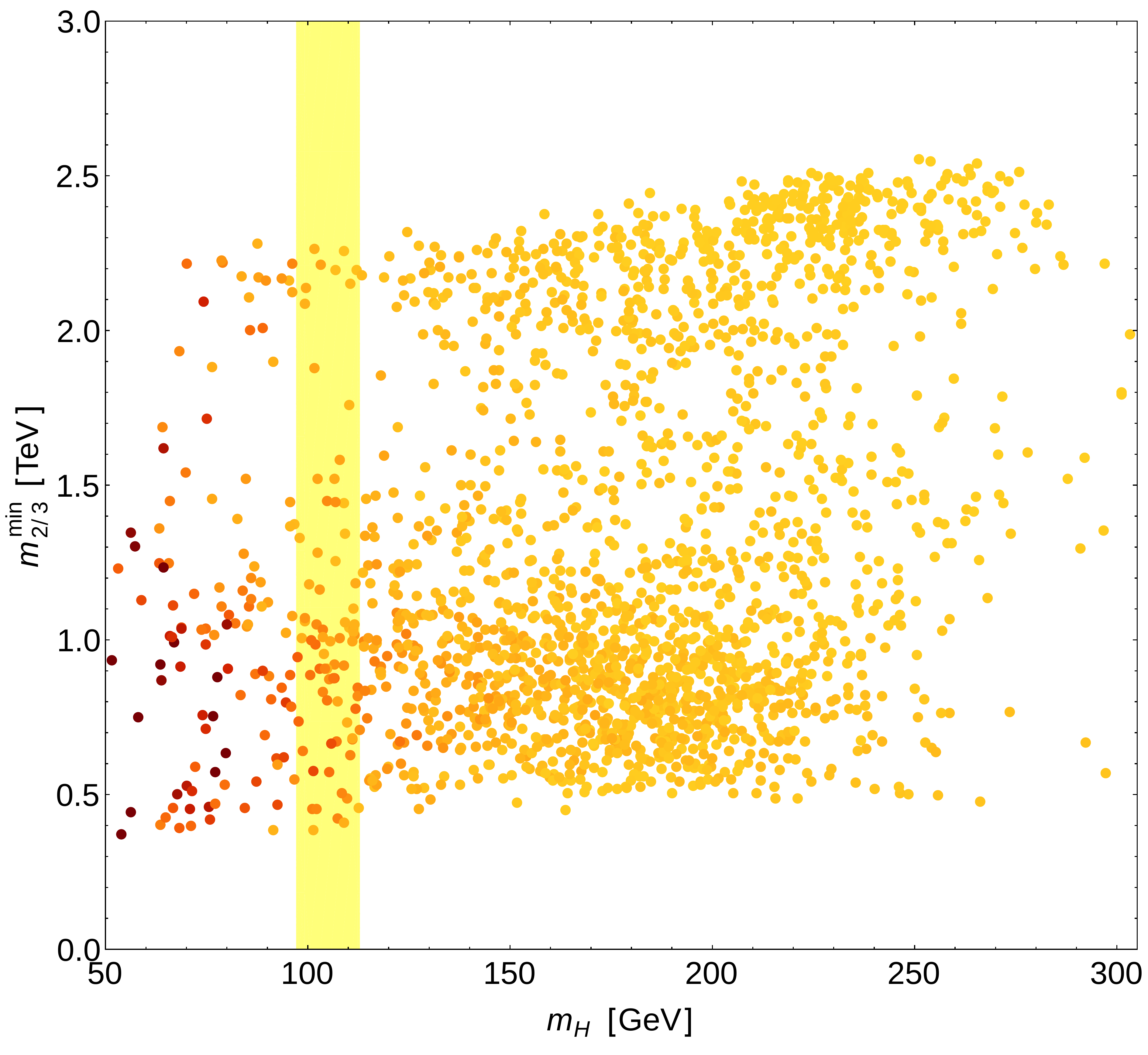}
		   \includegraphics[width=0.0785\textwidth]{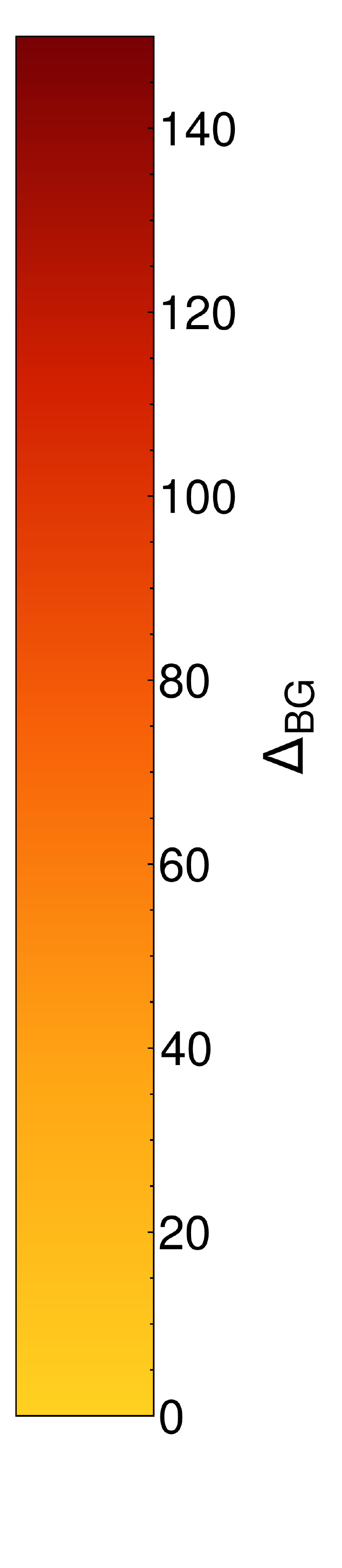}
		\caption{Mass of the first top partner as a function of the Higgs mass in the mMCHM$^{\rm III}_{5}$ for 
$Y_\ast^q=0.7$ and $Y_\ast^l=0.35$. Lighter points correspond to smaller values of $\Delta_{\rm BG}$ and therefore to less tuned points. }
\label{fig:14p5majmtpvsmh}
\end{center}
\end{figure}

\begin{figure}[!h]
\begin{center} 
			
			\includegraphics[width=0.43\textwidth]{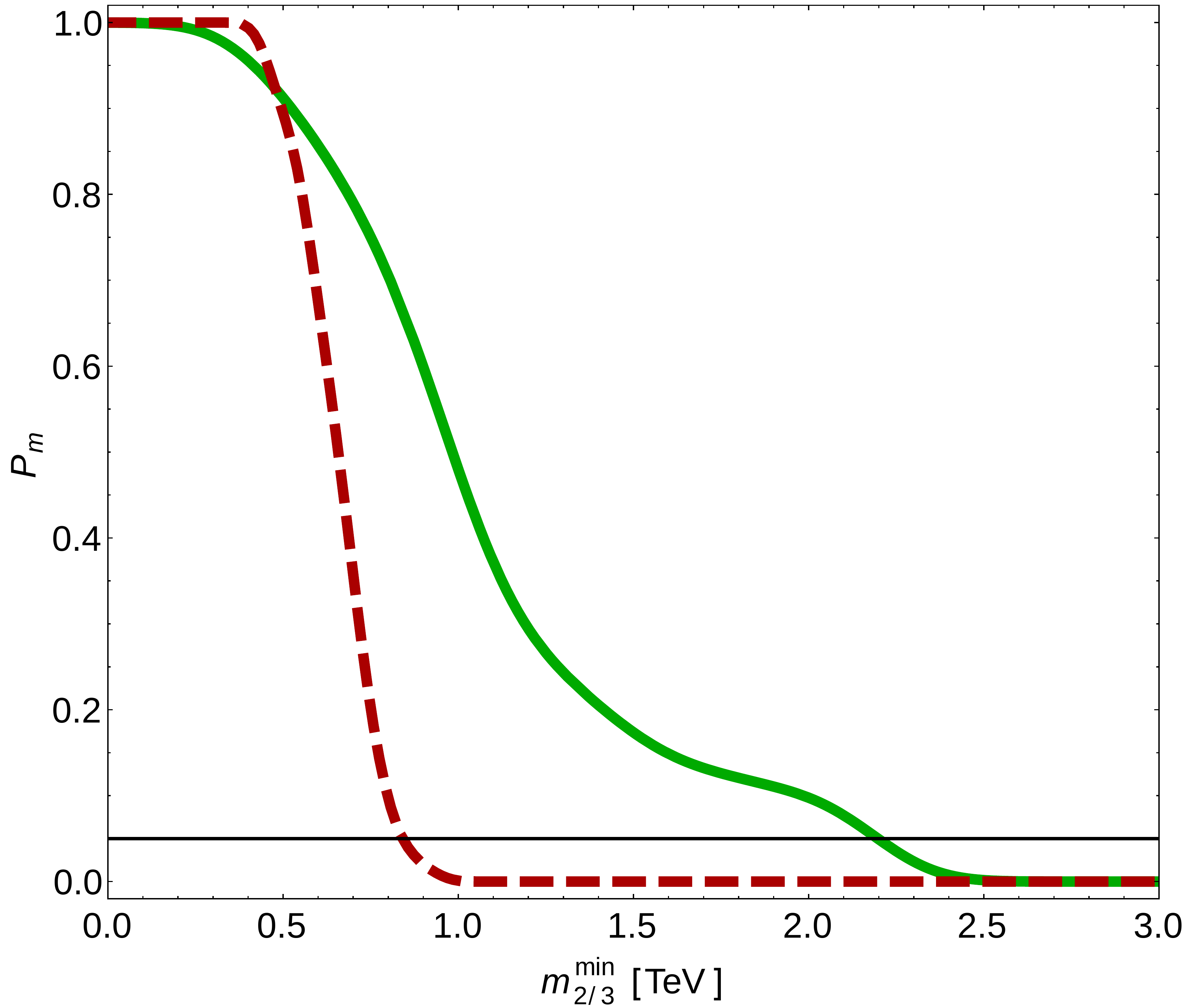}
			\caption{Survival function $\mathcal{P}_m$ of the first top partner mass 
				in the mMCHM$^{\rm III}_5$ with $Y_{\ast}^q=0.7$ and $Y_\ast^l=0.35$ (solid green) vs. the MCHM$_5$ (dashed red) with $Y_\ast^q=0.7$.}
\label{fig:14p5majsf}
\end{center}
\end{figure}

In Figure~\ref{fig:14p5majmtpvsmh} we show the mass of the lightest top partner, $m_{2/3}^{\rm min}$, as a function of the Higgs mass evaluated at the composite scale $f_{\pi}$  for the mMCHM$_{5}^{\rm III}$. The yellow band corresponds to the experimental value of the Higgs mass, $m_{H}(f_{\pi})=105\,\mathrm{GeV}\,(1\pm 7.5\%)$, with the allowed range accounting for the uncertainties in the running in a conservative way \cite{Carmona:2014iwa}. We also show the Barbieri-Giudice measurement of the tuning, $\Delta_{\rm BG}$,  by the color of every point in the $m_H-m_{2/3}^{\rm min}$ plane, where light yellow corresponds to a negligible tuning $\Delta_{\rm BG}\sim 0$, whereas dark red depicts a sizable tuning $\Delta_{\rm BG}\gtrsim 100$. This measure includes both the tuning to get a correct EWSB (the tuning entering the Higgs vev) as well as the potential ad-hoc tuning in the Higgs mass. Here, we assume some hierarchy between the brane masses in the quark and the lepton sector, or ``Yukawa suppression'', $Y_{\ast}^l=0.35$ versus $Y_{\ast}^q=0.7$, which may be motivated by the specific flavor pattern observed in the lepton sector \cite{delAguila:2010vg,Carmona:2014iwa}. We can see  from the plot that we can have masses for the lightest top partner well beyond the TeV, with a reasonably small tuning. In particular, a considerable amount of points with a viable Higgs mass feature $m_{2/3}^{\rm min}\sim (1-2.5)$~TeV together with a tuning $\Delta_{\rm BG}\lesssim (10-20)$. This is easy to understand since a negative lepton contribution to the Higgs mass can allow for a larger top breaking of the Goldstone symmetry and thus heavier partner masses. The fact that the RH charged leptons are embedded in $\mathbf{14}$s of $SO(4)$, with parametrically larger contributions to the Higgs mass, and that all three generations feature a similar degree of compositeness make in principle the lepton contribution important. However, the assumed Yukawa suppression, in agreement with the fact that $m_{\ell}\ll v$ in contrast to the top quark, avoids the otherwise expected increase of the tuning in the Higgs mass. Moreover, the masses of the lepton partners are in general at the KK scale $M_{\rm KK}\sim 3$~TeV, as the IR localization of the RH leptons is rather modest (and $m_{\ell}\ll m_t$). In order to be able to quantify more precisely the impact of leptons we display in Figure~\ref{fig:14p5majsf} the survival function $\mathcal{P}_m(x)$ of the first top partner mass, defined as the fraction of points  reproducing the correct Higgs mass for which $m_{2/3}^{\rm min}\ge x$, for the mMCHM$_{5}^{\rm III}$ (solid green) against the one of the MCHM$_5$ (dashed red). The curve has been obtained by smoothening the corresponding histograms and the black line depicts the $95\%$ quantile. One can see that while 95\% of the points of the MCHM$_5$ feature light partners below 800\,GeV, in the mMCHM$_5^{\rm III}$ the 95\% quantile is reached only for $m_{2/3}^{\rm min}=2.2$~TeV, with even $\gtrsim 10\%$ of the points having partner masses $m_{2/3}^{\rm min}\gtrsim 2$~TeV.

\begin{figure}[!t]
	\begin{center}
			\includegraphics[width=0.39\textwidth]{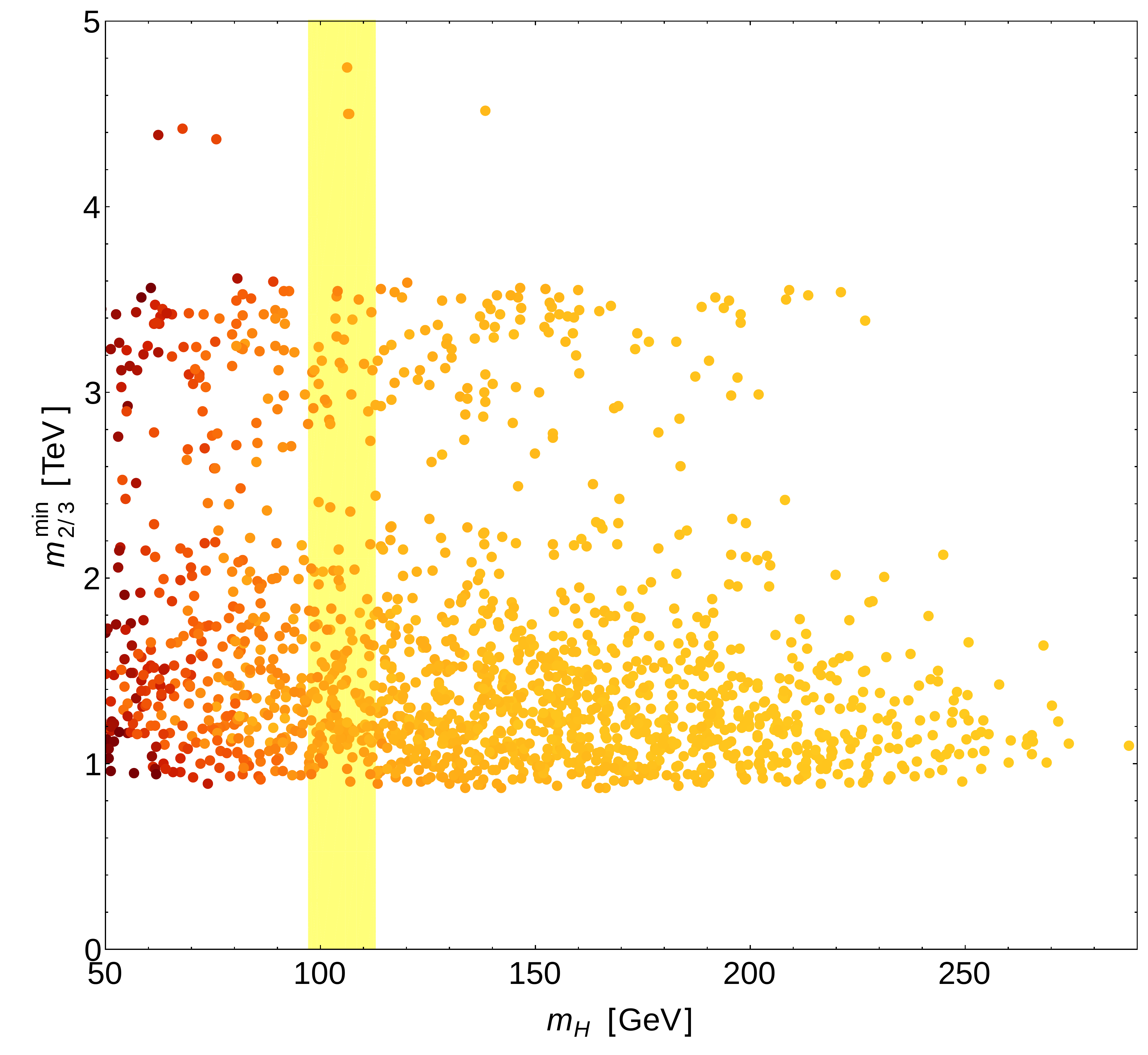}
			 \includegraphics[width=0.0785\textwidth]{bar_ft.pdf}
			
			 \caption{Mass of the first top partner as a function of the Higgs mass in the mMCHM$^{\rm III >}_{5-1}$ with $Y_\ast^l=Y_\ast^l=0.7$. Lighter points correspond to smaller values of $\Delta_{\rm BG}$ and therefore to less tuned points.
\label{fig:superminmtpvsmh}}
		 \end{center}
\end{figure}

\begin{figure}[t!]
\begin{center} 
			\includegraphics[width=0.43\textwidth]{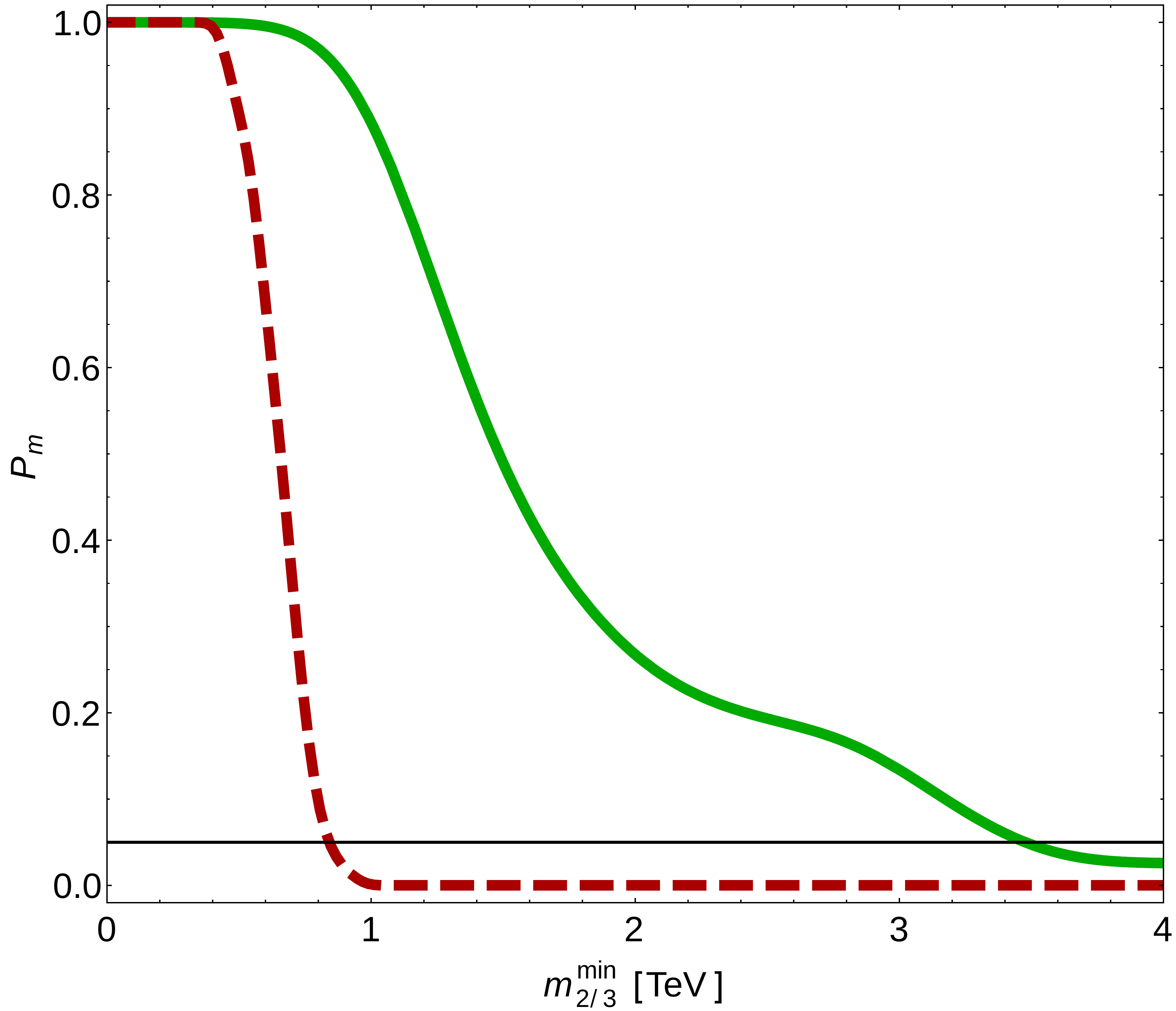}
			\caption{Survival function $\mathcal{P}_m$ of the first top partner mass in the mMCHM$^{\rm III>}_{5-1}$ with $Y_\ast^l=Y_{\ast}^q=0.7$ (solid green) vs. the MCHM$_5$ (dashed red) with $Y_\ast^q=0.7$.
\label{fig:superminsf}}
\end{center}
\end{figure}

The distribution of the top partner masses in the most minimal scenario mMCHM$_{5-1}^{\rm III>}$ is examined in Figure~\ref{fig:superminmtpvsmh}, where we show again the mass of the lightest top partner resonance $m_{2/3}^{\rm min}$ as a function of the Higgs mass  $m_H(f_{\pi})$. As denoted by the superscript ``$>$'', in this case we have lifted the previously assumed Yukawa suppression and taken equal maximum brane masses $Y_{\ast}^{q}=0.7=Y_{\ast}^l$, since the lepton contribution to the $\sin^2(h/f_{\pi})$ term of the potential is expected to cancel to a significant extend the sizable contribution of the top quark to allow for EWSB. In principle, this would also imply a considerable enhancement of the ad-hoc tuning in the Higgs mass, since the lepton contribution to the Higgs mass would increase accordingly. However, at the end such increase turns out to be rather moderate  and the correct Higgs mass can be reached with a modest tuning of $\Delta_{\rm BG} \sim (30-40) $. This is in particular a consequence of having a relatively fully composite 
$t_R$, not contributing to $V(h)$, which in turn allows for a less IR localized $t_L$ and thus a reduced 
top contribution to the Higgs potential. On the other hand, as it is clearly visible from the plot, the model does not even show ultra light partners below a TeV anywhere in its parameter space, being possible to lift these masses well above $3~$TeV. The corresponding survival function, depicted by the solid green lines in Figure~\ref{fig:superminsf}, do not drop under 
$5 \%$ even until $m^{\rm min}_{2/3}\gtrsim 3$\, TeV.

\begin{figure}[t!]
\begin{center} 
			\includegraphics[width=0.43\textwidth]{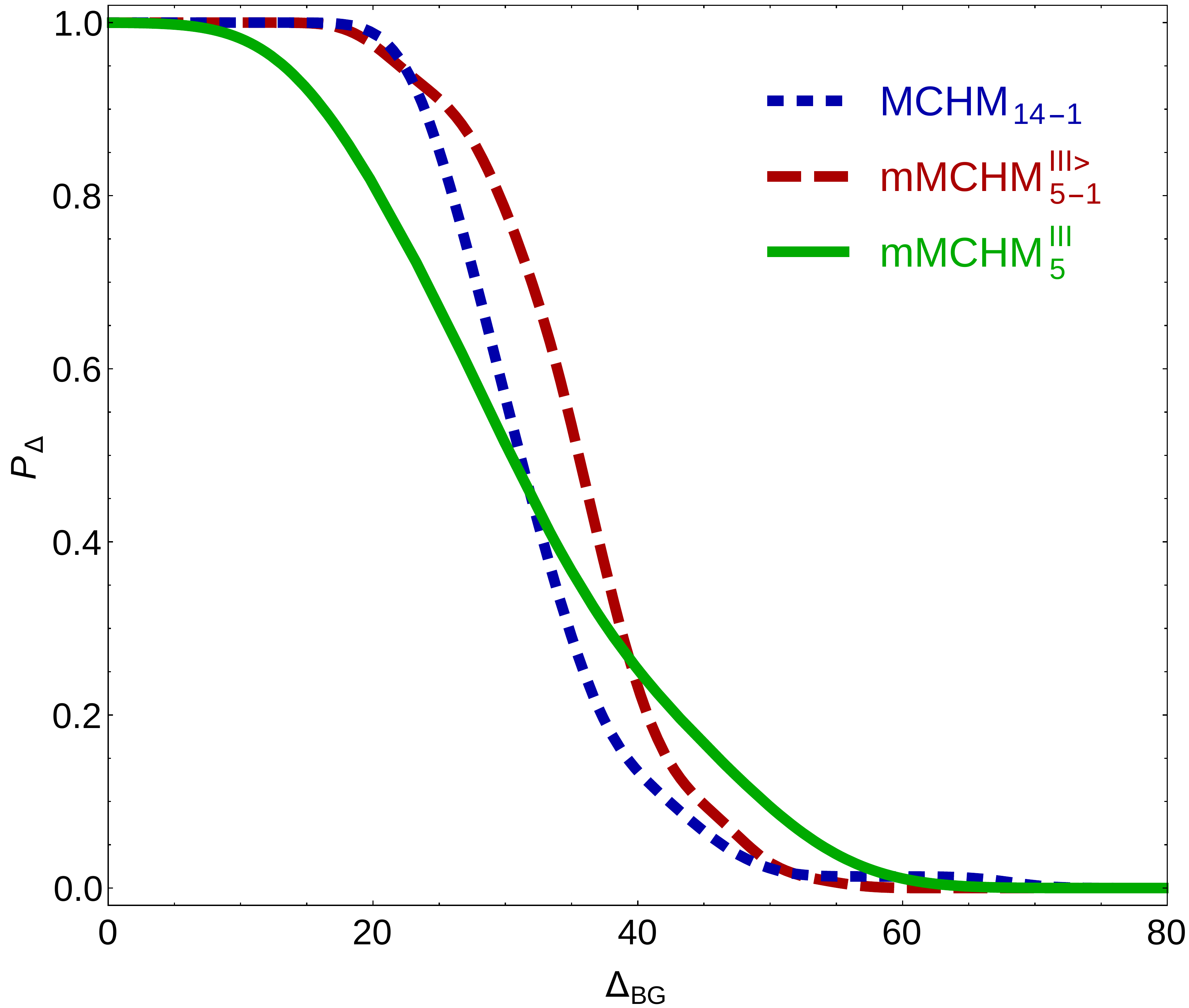}
		\caption{
		Survival function $\mathcal{P}_\Delta$ for the fine tuning $\Delta_{\rm BG}$ imposing $m^{\rm min}_{2/3}>1$~TeV 
in the MCHM$_{14-1}$ (dotted blue),  mMCHM$^{\rm III>}_{5-1}$ with $Y_\ast^l=0.7$ (dashed red), and mMCHM$^{\rm III}_5$ with $Y_\ast^l=0.35$  (solid green),   always employing   $Y_{\ast}^q=0.7$. }
\label{fig:FT}
\end{center}
\end{figure}

Finally, we confront in Figure~\ref{fig:FT} the fine tuning of the mMCHM$_{5}^{\rm III}$ and the mMCHM$^{\rm III >}_{5-1}$ with that of the MCHM$_{14-1}$, which is arguably the most competitive model in the quark sector avoiding the presence of light partners \cite{Panico:2012uw, Pappadopulo:2013vca, Carmona:2014iwa}. We display the survival function $P_\Delta(x)$, describing the fraction of points with a given fine tuning larger or equal than $x$, for all points in the viable Higgs-mass band and assuming $m_{2/3}^{\rm min}>1$~TeV. This plot confirms clearly that the  mMCHM$^{\rm III}_5$ opens for the first time the parameter space to allow for a minimal tuning of even less than $10\%$ while at the same time not predicting anomalously light partners. While already this model provides a motivation for the appearance of a symmetric representation of $SO(5)$ and does not 
introduce many new particles, a major virtue of the mMCHM$_{5-1}^{\rm III >}$ on the other hand is its highest degree of 
minimality and naturalness. This is true in the lepton sector, where it provides the most economical realization 
of the type-III seesaw, as well as in the quark one, where it embeds each SM fermion in the smallest $SO(5)$ multiplet one can imagine  (respecting custodial protection of the $Z$ couplings), leading to the least number of degrees of freedom in the full fermion sector for viable models \cite{Carmona:2014iwa}. Obviously, raising top partners through this model is much more minimal than in the MCHM$_{14-1}$, which adds many colored degrees of freedom at the TeV scale with a similar amount of tuning.

%\subsubsection{Conclusions}

To summarize, we have examined two particular examples of models that allow to lift the masses of the lightest top-partner resonances well above the region currently probed by the LHC in an orthogonal way to former studies, \emph{i.e.} without  significantly increasing the colored fermion sector while still predicting a naturally light Higgs. In particular, we pointed out the large minimality of the mMCHM$^{\rm III}$ models,  which, even though presenting a symmetric representation of $SO(5)$, allow for a smaller number of new particles than for instance the standard MCHM$_5$,
by unifying both LH and RH SM-leptons in a single multiplet, respectively. 
We also showed how, contrary to the quark sector, the lepton sector provides a compelling motivation for the emergence of a symmetric representation through the seesaw mechanism and how the  mMCHM$^{\rm III}_5$ allows to accommodate the absence of anomalously light partners with a minimal tuning. It is indeed the particular type-III seesaw that allows the unification of the RH  lepton fields, not possible in the quark sector due to the $SU(2)$ breaking masses and/or the quantum numbers. Furthermore, we have seen that the \emph{a priori} sizable lepton contribution in these models to the Higgs mass and the Higgs potential allow to render viable models where each top chirality is embedded in the smallest possible  $SO(5)$ representation (with custodially protected $Z$ couplings), enhancing thus the minimality of the setup also in the quark sector.  In particular, we have shown that the mMCHM$_{5-1}^{\rm III>}$ leads to an important rise in the masses of the lightest top partners with a modest amount of tuning, while featuring the least number of fermionic degrees of freedom of all viable models.

%%%%%
\section{Status of flavour physics$^{11}$}
\addtocounter{footnote}{-4}

\footnotetext{Contributing authors:  Enrico Lunghi and Tobias Hurth}

The status of flavour physics at the end of the $B$-factories/Tevatron era and after the first few years of data from LHCb is characterized by an overall confirmation of the CKM paradigm~\cite{Cabibbo:1963yz, Kobayashi:1973fv} and by a number of two-three sigma tensions. None of the latter is clean and/or significant enough to signify a clear breakdown of the Standard Model description of flavour. A selection of interesting deviations includes:
\begin{itemize}
\item A 1.5 sigma tension in unitarity triangle fits controlled by the rare decay $B\to \tau\nu$ and by the time dependent CP asymmetry in $B\to J/\psi K_s$ (that allows a clean extraction of the the angle $\beta$)~\cite{Charles:2015gya, CKMfitter, UTfit}.
\item Three sigma tensions in the the determinations of the CKM elements $V_{ub}$ and $V_{cb}$ from inclusive and exclusive semileptonic $B$ decays~\cite{Aoki:2013ldr, Amhis:2014hma, Aaij:2015bfa}. 
\item Tensions observed in the rare decays $B\to K^* \ell\ell \; (\ell = e,\mu)$ at low and high dilepton invariant mass~\cite{Aaij:2013qta, LHCb:2015dla}.
\item A puzzling deviation from lepton universality amongst the first two generations in the $B\to K \ell\ell\; (\ell = e,\mu)$ branching ratios~\cite{Aaij:2014ora}.
\item An anomalously large same sign di-muon charge asymmetry measured by D0~\cite{Hoeneisen:2014kka}.
\end{itemize}
In this short overview we review unitarity triangle fits and $B\to K^{(*)}\ell\ell$ rare decays. With regards to the latter we also discuss the impact of future inclusive $B\to X_s \ell\ell$ measurements at Belle~II.

\subsection{Unitarity Triangle Fits}
The standard global analysis of CP violation within the CKM framework is based on the unitarity relation
\begin{align}
V_{ub}^{*} V_{ud}^{} + V_{cb}^{*} V_{cd}^{} + V_{tb}^{*} V_{td}^{} = 0
\end{align}
that, when represented in the complex plan identify the so called unitarity triangle. Each of the CKM entries is a function of the four Wolfenstein parameters $\lambda$, $A$, $\rho$ and $\eta$ and the standard presentation consists in marginalizing over $\lambda$ and $A$ and present results in the $(\rho,\eta)$ plane (for a pedagogical review of these topics see, for instance, ref.~\cite{Buras:1998raa}). The constraints that we include in the fit are the $B_d$ and $B_s$ mass differences, $\varepsilon_K$, the direct determinations of $|V_{cb}|$ and $|V_{ub}|$ from inclusive and exclusive $b\to (c,u) \ell\nu$ decays, the time dependent CP asymmetry in $B\to J/\psi K_s$ ($\sin2\beta$), the determination of $\alpha$ from $B\to (\pi\pi, \rho\rho, \rho\pi)$ decays, the extraction of $\gamma$ from $B\to D^{(*)} K^{(*)}$ decays and the branching ratio for $B\to \tau\nu$. All of these quantities present exceptional challenges in their experimental determination, theoretical calculation or both. The most relevant inputs that we use in the fit are summarized in table~\ref{EL:inputs}. 

From the experimental point of view, the main difficulties lie in the measurement of $B\to  D^{(*)} K^{(*)}$ branching ratios and CP asymmetries (the extraction of $\gamma$ than follows from relatively clean isospin fits) and in the measurement of the $B\to \tau\nu$ branching ratio. While the former are expected to be measured with great precision in the next few years at LHCb, the latter require the clean environment of a $B$-factory to be reconstructed. The experimental precision that is expected from Belle~II with 50 ab${}^{-1}$ is about 3\% (compared to the 19\% of the current world average). 
\begin{table}
\begin{center}
\begin{tabular}{ll}
\hline
\hline
$\left| V_{cb} \right|_{\rm excl} =(39.36 \pm 0.75) \times 10^{-3}$ & $\hat B_K = 0.766 \pm 0.010$ \vphantom{$\Big($} \\
$\left| V_{ub} \right|_{\rm excl} = (3.42 \pm 0.21) \times 10^{-3}$  & $\kappa_\varepsilon = 0.944 \pm 0.015$ \vphantom{$\Big($}\\
$\left| V_{cb} \right|_{\rm incl} =(42.42\pm 0.86 ) \times 10^{-3}$ & $\xi  = 1.268 \pm  0.063 $ \vphantom{$\Big($}\\
$\left| V_{ub} \right|_{\rm incl} = (4.40 \pm 0.25) \times 10^{-3} $  & $\alpha = (87.7 \pm 3.4)^{\rm o}$\vphantom{$\Big($}\\
$\left| V_{cb} \right|_{\rm avg} =(40.7 \pm 1.5) \times 10^{-3}$ & $\gamma = (68.3 \pm 7.5)^{\rm o}$ \vphantom{$\Big($} \\
$\left| V_{ub} \right|_{\rm avg} = (3.82 \pm 0.48) \times 10^{-3}$ & \\
$f_{B_d} = (186.4 \pm 4.0) \; {\rm MeV}$\vphantom{$\Big($} \vphantom{$\Big($} & \\
$f_{B_s} \sqrt{\hat B_{B_s}} = (266 \pm 18) \; {\rm MeV} $ \vphantom{$\Big($} & \\
${\rm BR} (B\to \tau\nu) = (1.08 \pm 0.21) \times 10^{-4}$ \vphantom{$\Big($} & \\ 
\hline
\hline
\end{tabular}
\caption{Lattice-QCD and other inputs to the unitarity triangle analysis. The sources used a FLAG~\cite{Aoki:2013ldr}, PDG~\cite{Agashe:2014kda} and CKMfitter~\cite{CKMfitter} (for the determination of $\gamma$). \label{EL:inputs}}
\end{center}
\end{table}
\begin{figure}
\includegraphics[width=0.99 \linewidth]{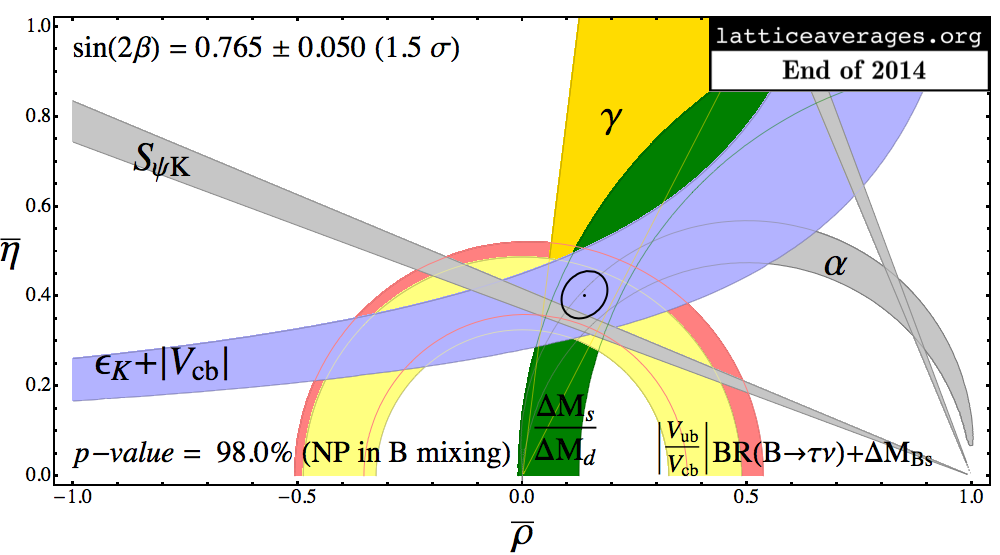}
\includegraphics[width=0.49 \linewidth]{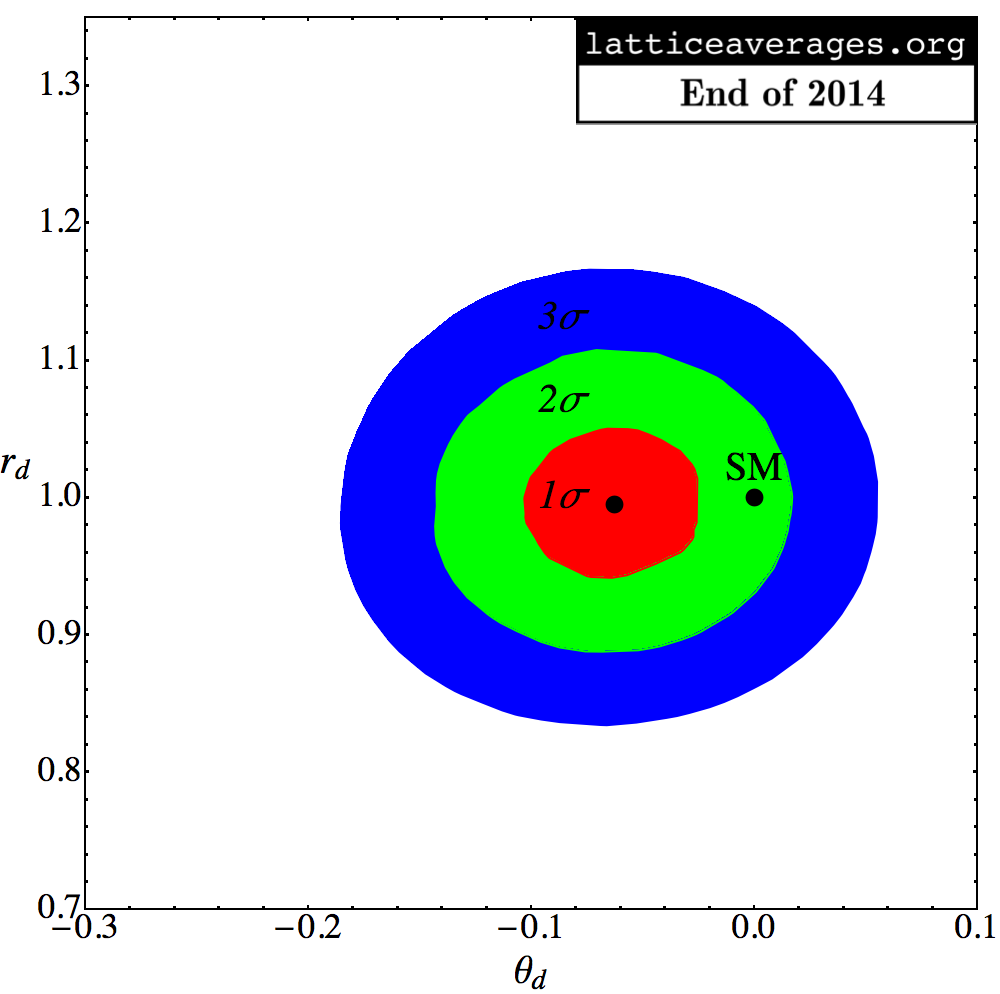}
\includegraphics[width=0.49 \linewidth]{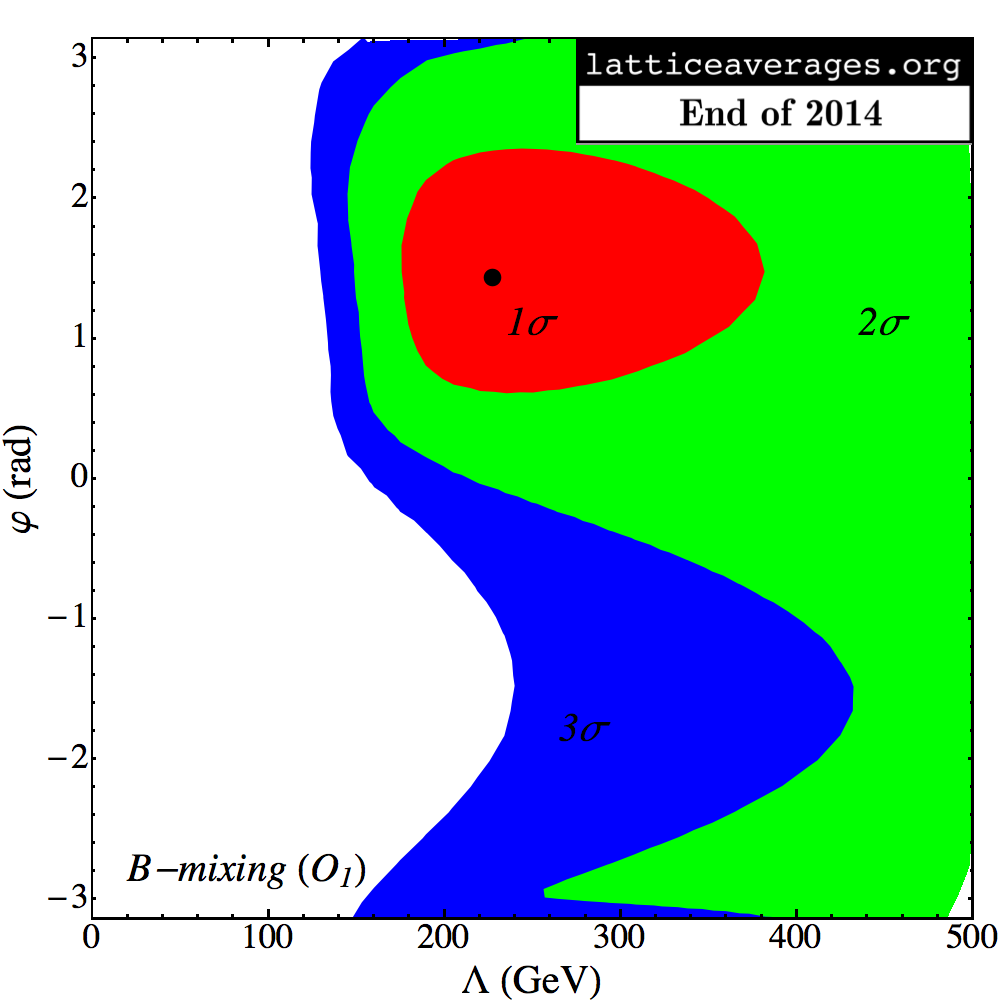}
\caption{Unitarity triangle fits.}
\label{EL:s2b}
\end{figure}

From the theoretical point of view, the strongest challenge remaining is the resolution of the conflicts between the determination of $|V_{cb}|$ and $|V_{ub}|$ from inclusive and exclusive $b\to (c,u) \ell\nu$ decays. In both cases, inclusive and exclusive modes yields CKM elements that differ at the about three sigma level (see table~\ref{EL:inputs}). Inclusive modes are in principle controlled by perturbative physics. While the status of $B\to X_c \ell\nu$ calculations is excellent (see for instance ref.~\cite{Gambino:2015ima}), the $B\to X_u\ell\nu$ transition is afflicted by a general breakdown of the OPE due to experimental cuts required to suppress the $B\to X_c \ell\nu$ background that result in large uncertainties related to the $B$ meson shape function~\cite{Lange:2005yw, Gambino:2007rp} (more recently BaBar and Belle presented results that take into account approximatively 90\% of the total available phase space~\cite{Beleno:2013jla}). Predictions for exclusive modes are controlled by the lattice QCD determination of the $B\to\pi$ and $B\to D^{(*)}$ form factors (the former, in particular, requires an extrapolation at low-$q^2$ that is usually performed using a $z$-parametrization and a simultaneous fit of lattice and experimental results). Very recently LHCb presented a determination of $V_{ub}$ from the baryonic process $\Lambda_b^0 \to p \mu \bar \nu_\mu$~\cite{Aaij:2015bfa} using a very recent lattice QCD calculation of the $\Lambda_b \to p$ form factor~\cite{Detmold:2015aaa}; this new result is in excellent agreement with the determination of $V_{ub}$ from $B\to \pi\ell\nu$ decays. The averages we adopt are presented in table~\ref{EL:inputs} where the uncertainties have been rescaled (using the PDG prescription) in order to take into account the three sigma tension amongst the inputs. Finally we should point out that recent impressive improvements of lattice QCD determinations of various matrix elements (e.g. the $K-\bar K$ matrix element $\hat B_K$ has an uncertainty of about 1\%) has made the remaining lattice inputs almost subdominant in the fit.

\begin{figure}
\includegraphics[width=0.95 \linewidth]{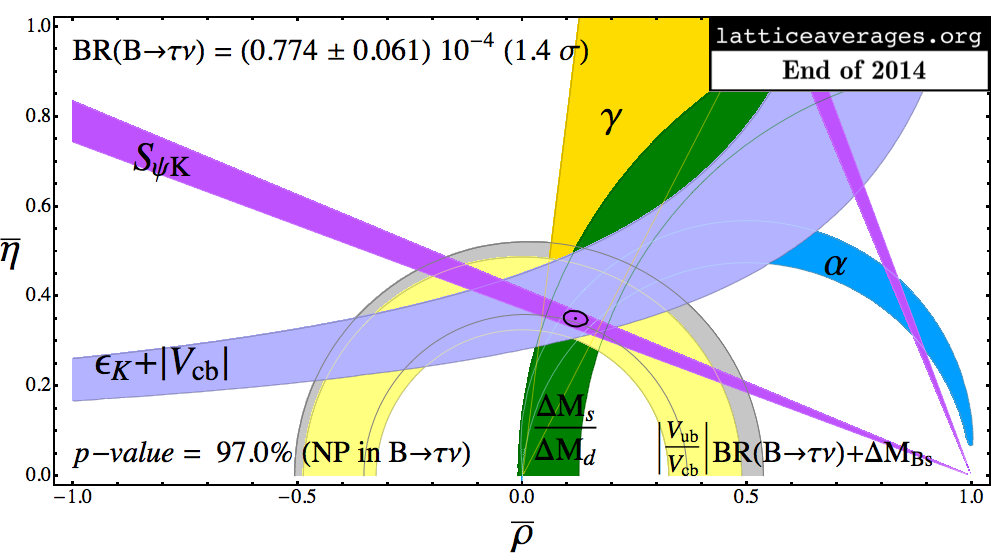}
\caption{Constraints on a generic parametrization of new physics contributions to $B_d$ mixing.}
\label{EL:btn}
\end{figure}

\begin{figure}
\includegraphics[width=0.95 \linewidth]{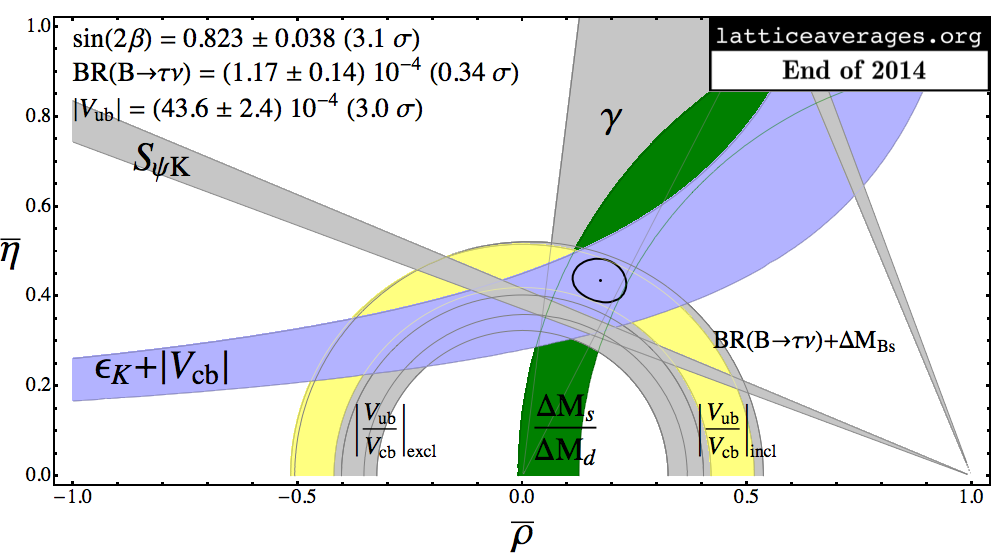}
\includegraphics[width=0.6 \linewidth]{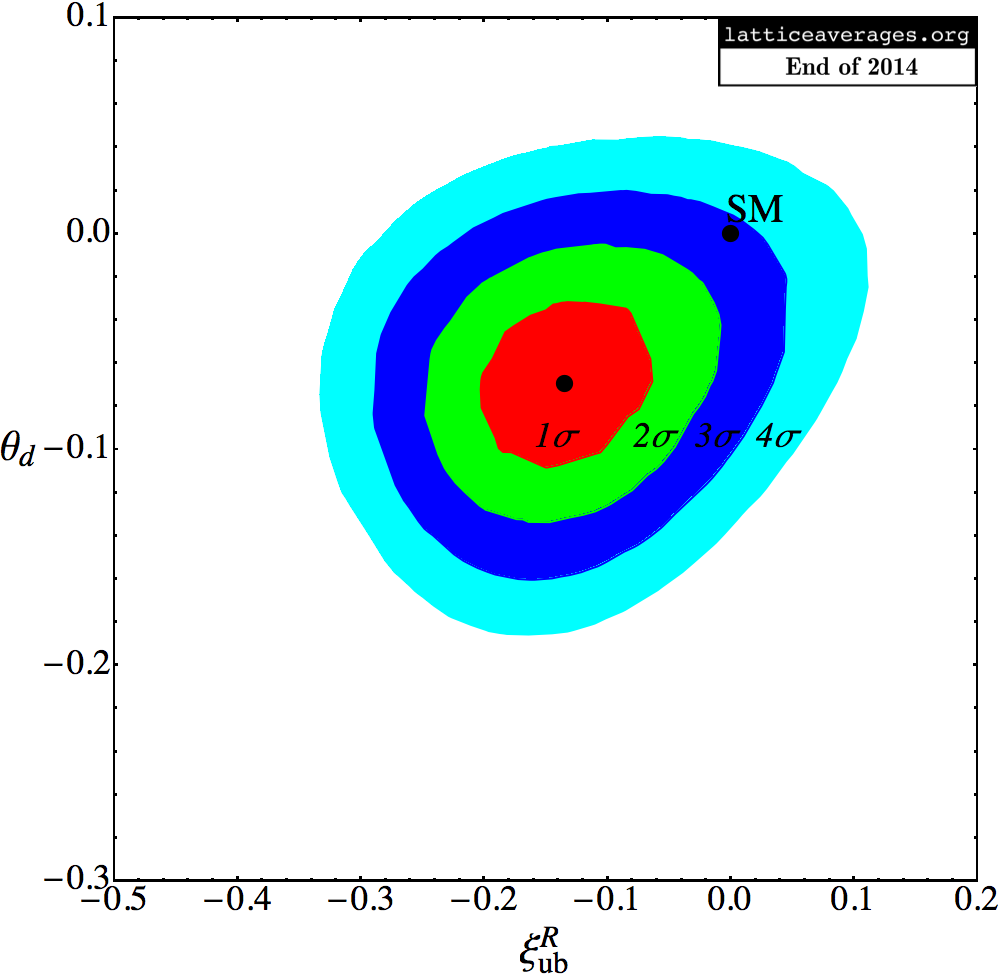}
\caption{Constraints on a generic parametrization of new physics contributions to $B_d$ mixing.}
\label{EL:RH}
\end{figure}

The results that we obtain are presented in Figs.~\ref{EL:s2b}-\ref{EL:RH} (the explicit formulae used can be found, for instance, in refs.~\cite{Lunghi:2009sm, Laiho:2009eu}). 

In Fig.~\ref{EL:s2b} we assume that high-scale new physics contributions are confined to the $B_d$ mixing sector thereby affecting the determinations of $\sin 2\beta$ and $\alpha$ (in principle also the ratio $\Delta M_{B_s}/\Delta M_{B_d}$ is affected by we find that the fit constraints this ratio to be very close to the SM prediction). We therefore remove these two constraints from the fit and extract a prediction for $\sin 2\beta$ (given in the plot) and find that deviates at 1.5 sigma level from its direct determination. The parametrization we adopt is $M_{12}^{d,{\rm NP}} / M_{12}^{d,{\rm SM}}  = r_d^2 \; e^{2 i \theta_d}$ where $M_{12}$ is the $B_d$-$\bar B_d$ matrix element (see for instance ref.~\cite{Buras:1998raa}) and in the Standard Model $r_d = 1$ and $\theta_d=0$. In presence of non-vanishing contributions to $B_d$ mixing the following observables are affected:
\begin{align}
S_{\psi K_S} &= \sin 2 (\beta + \theta_d)  \; , \\
\sin (2 \alpha_{\rm eff})  &= \sin 2 (\alpha - \theta_d)  \; , \\
 \frac{\Delta M_{B_s}}{\Delta M_{B_d}} &= r_d^{-2} \left(\frac{\Delta M_{B_s}}{\Delta M_{B_d}} \right)_{\rm SM} \; .
\end{align}
In the lower left plot of Fig.~\ref{EL:s2b} we show the results of this fit in the $[r_d,\theta_d]$ plane where we see that there is a slight tension that favors negative values of $\theta_d$ (the actual fit result is $\theta_d= -(3.6 \pm 2.3)^{\rm o}$). These results can be interpreted in terms of a new physics scale in an effective Hamiltonian framework:
\begin{align}
{\cal H}_{\rm eff} = \frac{G_F^2 m_W^4}{16 \pi^2} \left( V_{tb}^{} V_{td}^*\right)^2 C_1^{\rm SM} 
\left( \frac{1}{m_W^2} - \frac{e^{i\varphi}}{ \Lambda^2} \right) O_1 \; ,
\end{align}
In this parametrization $\Lambda$ is the scale of some new physics model whose interactions are identical to the Standard Model with the exception of an additional arbitrary $CP$ violating phase:  
\begin{align}
C_1 = C_1^{\rm SM} \left( 1 - e^{i\varphi} \frac{m_W^2}{\Lambda^2} \right) \;.
\end{align}
The corresponding results in the $[\Lambda,\varphi]$ plane are presented in the lower right plot of Fig.~\ref{EL:s2b} where we see that scales in the $[200,400]\; {\rm GeV}$ range are preferred. 

In Fig.~\ref{EL:btn} we entertain a complementary scenario in which new physics is allowed to affect only the $B\to\tau\nu$ branching ratio. Models in which this happens are fairly common because a simple charged Higgs tree level exchange can contribute sizably to the $B\to\tau\nu$ amplitude. Following the same strategy as above, we remove the $B\to\tau\nu$ constraint from the fit and compare the fit result to the direct measurement finding a 1.4 sigma tension.

Finally, in Fig.~\ref{EL:RH} we consider the more exotic possibility of taking the discrepancy between the inclusive and exclusive determinations of $|V_{ub}|$ at face value and introducing interactions whose impact on exclusive $B\to X_u \ell \nu$ decays is much larger than in inclusive ones.  The introduction of a right--handed effective $\bar u_R W\hskip-0.75em / b_R$ coupling offers the most elegant solution of the ``$V_{ub}$ puzzle'' (see for instance refs.~\cite{Chen:2008se, Crivellin:2009sd, Buras:2010pz, Blanke:2011ry, Laiho:2012ss}). In this scenario we have:
\begin{align}
V_{ub} \; \bar u_L W\hskip-0.75em / b_L \Longrightarrow V_{ub} \; \left( \bar u_L W\hskip-0.75em / b_L + \xi_{ub}^R \;  \bar u_R W\hskip-0.75em / b_R \right) \; .
\end{align}
The effective parameter $\xi_{ub}^R$ affects all $b\to u \ell\nu \; (\ell = e,\mu,\tau)$ transitions:
\begin{align}
\left| V_{ub} \right|_{\rm incl} &\Longrightarrow \sqrt{1+ \left| \xi_{ub}^R\right|^2} \; \left| V_{ub} \right| \; ,  \\
\left| V_{ub} \right|_{\rm excl} &\Longrightarrow \left| 1+  \xi_{ub}^R \right| \; \left| V_{ub} \right| \; ,\\
{\rm BR} (B\to \tau\nu) &\Longrightarrow  \left| 1-  \xi_{ub}^R \right|^2 \; {\rm BR} (B\to \tau\nu) \; .
\end{align}
The result of the fit to the unitarity triangle in which we allow $\xi_{ub}^R$, $\theta_d$ and $r_d$ to vary simultaneously yields
\begin{align}
\xi_{ub}^R &= -0.134 \pm 0.048  \; ,\\
\theta_d &= - (4.0 \pm 1.5)^{\rm o}  \; ,\\
r_d &= 1.000 \pm 0.057  \; ,
\end{align}
indicating deviations in $\xi_{ub}^R$ and $\theta_d$ at the three sigma level. In the upper plot in Fig.~\ref{EL:RH} we show the fit we obtain after excluding all quantities that are sensitive to $\xi_{ub}^R$ and $\theta_d$; the resulting predictions for $\sin 2\beta$ and ${\rm BR}(B\to\tau\nu)$ differ from their direct determination at the three sigma level. In the lower plot in Fig.~\ref{EL:RH} we show the corresponding two--dimensional allowed regions in the $[\xi_{ub}^R,\theta_d]$ plane. 

In conclusion, the overall status of unitarity triangle fits is in fairly good agreement with the SM. The few tensions we observe could be interpreted as a hint for new physics but could also be resolved in the near future by improvements on (1) the experimental determination of ${\rm BR}(B\to\tau\nu)$ at Belle~II, (2) lattice-QCD determinations of the semileptonic form factors, (3) the experimental uncertainty on $\gamma$ from LHCb.

\subsection{Exclusive $B\to (K,K^*)\ell\ell$}
Different theoretical concepts are used in the treatment of exclusive rare semileptonic decays  within the two alternative kinematic regimes: large recoil (i.e., to low dilepton invariant mass squared, $q^2$) and small recoil (i.e., high $q^2$). However, calculations in the $q^2$ region close to the narrow $c \bar c$ resonances are difficult.

Exclusive heavy-to-light $B \to K^* \mu^+\mu^-$ decays in  the low-$q^2$ region are described by  the method of QCD-improved Factorisation (QCDF) and its field-theoretical formulation of Soft-Collinear Effective Theory (SCET). The decay amplitude factorises to leading order in $\Lambda/m_b$ and to all orders in $\alpha_s$ into process-independent non-perturbative quantities in the combined limit of a heavy $b$-quark and of an energetic $K^*$ meson:  e.g., $B\to K^*$ form factors and light-cone distribution amplitudes (LCDAs) of the heavy (light) mesons and perturbatively calculable quantities, which are known to $O(\alpha_s^1)$~\cite{Beneke:2001at,Beneke:2004dp}:
\begin{equation}
  \label{eq:factorisation}
  \mathcal T 
   = C\, \xi + \phi_B \otimes T \otimes \phi_{M} 
     + O(\Lambda_{\rm QCD}/m_b)\,. 
\end{equation}
which is accurate to leading order in $\Lambda{\rm QCD}/m_b$ and to all orders in $\alpha_s$.

In addition, if $M$ is a vector $V$ (pseudoscalar $P$), the seven (three) {\it a priori} independent $ B \to V$ ($B \to P$) form factors reduce to two (one) universal {\it soft} form factors $\xi_{\bot,\|}$ ($\xi_P$) in QCDF/SCET~\cite{Charles:1998dr}. The factorisation formula Eq.~(\ref{eq:factorisation}) is well applicable in the dilepton mass range, $1 < q^2 < 6 \; {\rm GeV}^2$.

If we take all these  simplifications into account the various spin amplitudes at leading order in $\Lambda_{\rm QCD}/m_b$ and $\alpha_s$ get linear in the soft form factors 
$\xi_{\bot,\|}$ and also in the short-distance Wilson coefficients. These simplifications allow us to design a set of optimized observables, in which any soft form factor dependence (and its corresponding uncertainty) cancels out for all low dilepton mass squared $q^2$  at leading order in $\alpha_s$  and $\Lambda_{\rm QCD}/m_b$, as was explicitly shown in refs.~\cite{Egede:2008uy, Egede:2010zc}. In refs.~\cite{Matias:2012xw,Descotes-Genon:2013vna}, an optimized set of independent observables  was constructed. Here, nearly all  observables  are free from hadronic uncertainties which are related to the form factors.

However, in these angular observables, the soft form factors  are {\it not} the only source of hadronic uncertainties. Within the QCDF/SCET approach, a general and quantitative method to estimate the important $\Lambda_{\rm QCD}/m_b$ corrections to the heavy quark limit is missing.

The high $q^2$ (low hadronic recoil) region corresponds to dilepton invariant masses above the two narrow $c \bar c$ resonances ($q^2 > 14$ GeV$^2$).  Broad $c\bar c $-resonances can be treated using a local operator product expansion~\cite{Grinstein:2004vb,Beylich:2011aq}. One finds small sub-leading corrections which are suppressed by either $(\Lambda/m_b)^2$~\cite{Beylich:2011aq} or $\alpha_s  \Lambda/m_b$~\cite{Grinstein:2004vb}. This depends on whether full QCD or subsequent matching on heavy quark effective theory in combination with form factor symmetries~\cite{Isgur:1989ed} is adopted. Numerically, the sub-leading corrections to the amplitude have been estimated to be below $2$\,\%~\cite{Beylich:2011aq}. Those due to form factor relations are numerically suppressed by $C_7 / C_9 \sim O(0.1)$. In addition, duality violating effects have been estimated within a model of resonances. They were found to be at the level of $2\,\%$ of the rate, if sufficiently large bins in $q^2$ are selected~\cite{Beylich:2011aq}.  Moreover, the heavy-to-light form factors can be calculated using lattice calculations (see,  for instance, ref.~\cite{Horgan:2015vla}).
As a consequence, this region is theoretically well controlled. 
\begin{figure}
\includegraphics[width=0.8 \linewidth]{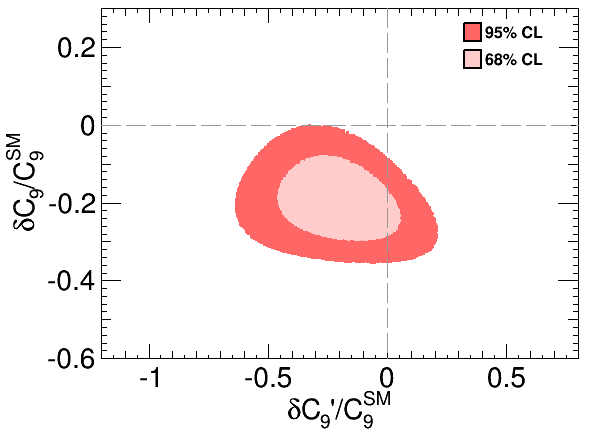}
\includegraphics[width=0.8 \linewidth]{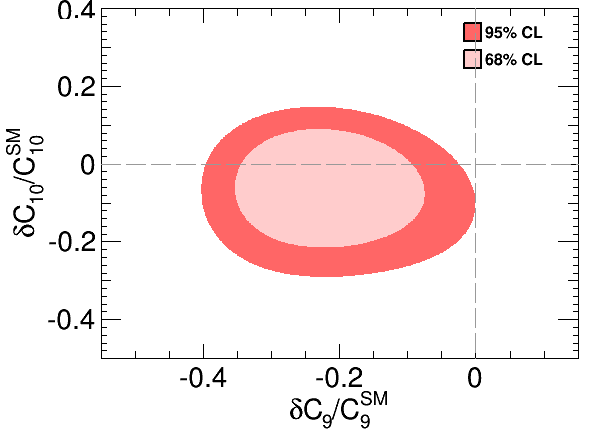}
\caption{Global fit results obtained allowing new physics contributions exclusively to $[C_9,C_9^\prime]$ (upper plot) and $[C_9,C_{10}]$ (lower plot).}
\label{TH:globalfits}
\end{figure}
\bigskip

Within  the first experimental findings on new angular observables in the exclusive decay $B \to K^*\mu^+\mu^-$ based on the 1~fb$^{-1}$ dataset, LHCb founds  a 3.7$\sigma$ local discrepancy  in one of the $q^2$ bins for one of the angular observables~\cite{Aaij:2013qta}, namely in the bin $q^2 \in               [ 4.3,8.63 ]$ GeV$^2$ of the observable $P_5^{'}$. The latter belongs to the set of optimized observables in which form factor dependence cancels out to first order. 
LHCb compared its experimental results with the theoretical predictions in ref.~\cite{Descotes-Genon:2013vna}.  

This observation has been confirmed  by the recent LHCb analysis using the $3 {\rm fb}^{-1}$ data sets.  The investigation of the observable $P_5^{'}$ shows again in the new bins $q^2 \in [4.0, 6.0]$ and  $[6.0, 8.0]$ deviations of 2.9$\sigma$. Therefore the significance has not increased, but the new measurement is compatible with the $1 {\rm fb}^{-1}$ measurement.

The deviation in the observable $P_5^{\prime}$ can be  consistently described by a smaller $C_9$ Wilson coefficient, together with a less significant contribution of a non-zero $C_9^{\prime}$ (see for example ref.~\cite{Descotes-Genon:2013zva}).  This is a challenge for the model-building. 
\begin{figure}
\includegraphics[width=0.8 \linewidth]{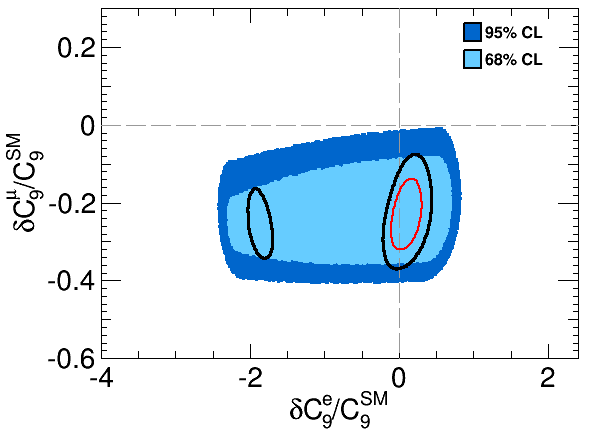}
\includegraphics[width=0.8 \linewidth]{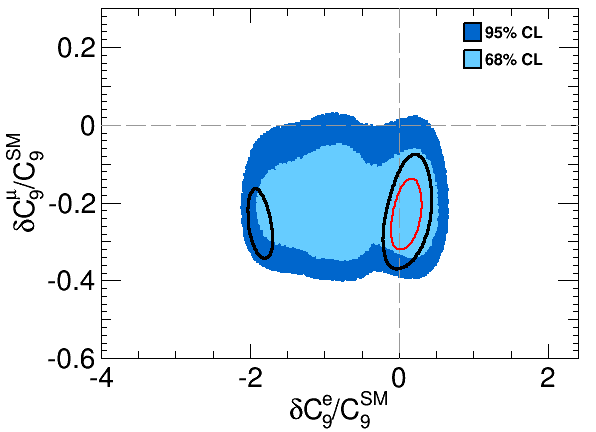}
\caption{Global fit results obtained allowing for different new physics contributions to operators involving electron and muon currents. In the upper (lower) plot it is shown the allowed region in the $[C_9^e,C_9^\mu]$ plane obtained marginalizing against variations in $C_{10}^{e,\mu}$ (${C^\prime}_{9}^{e,\mu}$).}
\label{TH:globalfitsEMU}
\end{figure}
A large number of references~\cite{Descotes-Genon:2013wba, Altmannshofer:2013foa, Hambrock:2013zya, Gauld:2013qba, Buras:2013qja,Gauld:2013qja, Datta:2013kja,Beaujean:2013soa, Buras:2013dea,Hurth:2013ssa, Mahmoudi:2014mja, Altmannshofer:2014cfa} discuss consistent SM and new physics interpretations of the measured deviation in the $B \to K^* \ell^+\ell^-$ mode. The results of a global fit to all LHCb $B\to K^{(*)} \ell\ell$ results is presented in Fig.~\ref{TH:globalfits}~\cite{Hurth:2014vma, Hurth:2014zja} using the SuperIso program~\cite{Mahmoudi:2007vz,Mahmoudi:2008tp}. Scenarios with new physics confined to the pair of Wilson Coefficient $[C_9,C_9^\prime]$ (upper plot) and $[C_9,C_{10}]$ (lower plot) are considered. In a model-independent analysis, the anomaly can be consistently described by smaller $C_9$ and $C'_9$ Wilson coefficients and, to a lesser extent, by a positive contribution to $C_{10}$ (we remind the reader that $C_{10}^{\rm SM} < 0$). 

Large contributions to the $C_9$ are difficult to accomodate in typical new physics models. Indeed, the $Z$-boson coupling to the leptonic vector current is suppressed by a factor $(1-4 s_W^2) \sim 0.04$ which implies that flavour changing effective $Z-b-s$ couplings (that are potentially enhanced in many BSM models) contribute only minimally to the operator $O_9 = \bar s_L \gamma_\mu b_L^{}\bar \ell \gamma^\mu \ell$. Therefore, main stream models, for instance the minimally flavour violating MSSM, warped extra dimension scenarios, or models with partial compositeness, cannot accommodate the deviation at the $1\sigma$ level. On the other hand, $\mbox{Z}^{'}$ models have been shown to be viable~\cite{Gauld:2013qba}.

In the MSSM, we cannot generate any sizable contribution to the coefficient $C'_9$, but also any significant contribution to $C_9$ is correlated to contributions to other Wilson coefficients affecting the other observables. Nevertheless, combining all the observables in a fit one can check the global agreement of the model with the available data~\cite{Mahmoudi:2014mja}: The overall (global) agreement is relatively good, with regions in SUSY parameter space where the absolute $\chi^2$ is sufficiently small and an agreement at the 1$\sigma$ level is obtained.

In our opinion it is too soon to conclude whether this anomaly is a sign of beyond the SM physics, is due to our lack of understanding of hadronic power corrections and/or non-perturbative charm effects, or is just a statistical fluctuation.

\bigskip
More recently, another small discrepancy was found. The ratio $R_K = {\rm BR}(B^+ \to K^+ \mu^+ \mu^-) / {\rm BR}(B^+ \to K^+ e^+ e^-)$ in the low-$q^2$ region has been measured by LHCb showing a $2.6\sigma$ deviation from the SM prediction~\cite{Aaij:2014ora}. In contrast to the anomaly in the rare decay $B \rightarrow K^{*} \mu^+\mu^-$ which is affected by unknown power corrections, the ratio $R_K$ is theoretically rather clean. This might be a sign for lepton non-universality.

A few recent studies~\cite{Alonso:2014csa, Hiller:2014yaa, Ghosh:2014awa, Biswas:2014gga, Hurth:2014vma, Glashow:2014iga}  address this discrepancy . Fig.~\ref{TH:globalfitsEMU} shows that a global fit to all observables considering separately new physics contributions to the electron and muon semileptonic Wilson coefficients $C_{9,10}^e$ and $C_{9,10}^\mu$ (and the corresponding chirality flipped coefficients) favors the non-universal solutions. However, in a two-operator fit lepton-universality, $\delta C_9^\mu = \delta C_9^e$, is disfavored by $2 \sigma$, while within the four-operator fit the agreement is improved~\cite{Hurth:2014vma}.

\subsection{Inclusive $B\to X_s\ell\ell$}
Finally let us discuss some recent progress on the calculation of inclusive $B\to X_s \ell\ell \; (\ell=e,\mu)$ decays. These inclusive modes are controlled by perturbative QCD and are under much stronger theoretical control than the corresponding exclusive $K$ and $K^*$ channels. In the rest of this section we follow the discussion presented in refs.~\cite{Huber:2005ig, Huber:2007vv, Huber:2015sra}.
\begin{figure}
\includegraphics[width=0.8 \linewidth]{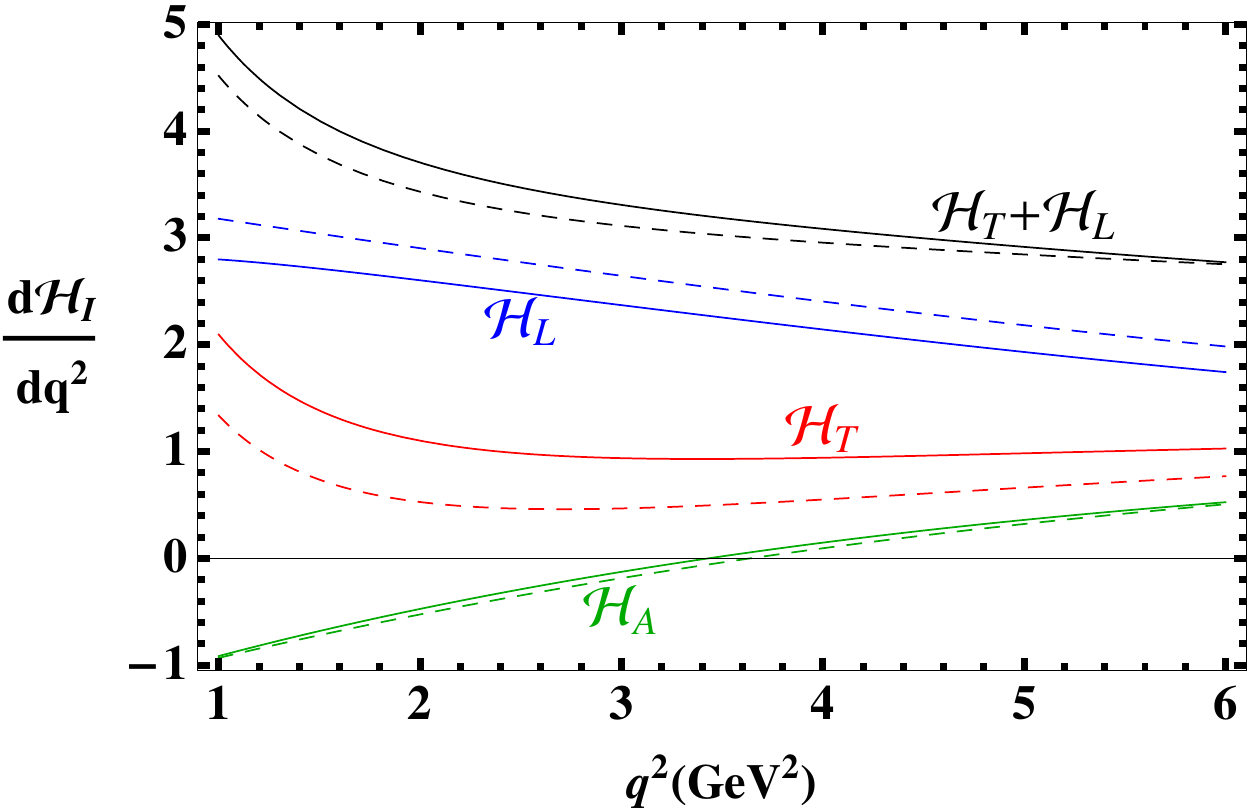}
\caption{Inclusive $b\to s \ell\ell$ spectrum and QED corrections.}
\label{EL:HI}
\end{figure}

In a fully inclusive transition the only observable kinematical variables are the invariant mass and scattering angle of the dilepton pair (the latter is defined with respect to the incoming $B$ direction in the dilepton center-of-mass frame). At leading order in electroweak interactions and to all orders in QCD the double differential rate can be written as
\begin{align}
\frac{d^2\Gamma}{d q^2\,  d z} \; = \;   \frac{3}{8} \Bigl[ &(1 + z^2) H_T(q^2) + 2(1 - z^2) H_L(q^2) \\ \nonumber 
 &  +  2 z H_A(q^2) \Bigr]\; ,
\end{align}
where $H_T$ and $H_L$ are related to the transverse and longitudinally polarized hadronic tensor. The differential width and the normalized forward--backward asymmetry are then:
\begin{align}
\frac{d \Gamma}{d q^2} &= H_T  (q^2)  + H_L (q^2) \; ,\\
\frac{d \overline A_{\rm FB}}{d q^2} &= \frac{3}{4} \frac{H_A (q^2)}{H_T (q^2)  + H_L (q^2)} \;.
\end{align} 
\begin{figure}
\includegraphics[width=0.7 \linewidth]{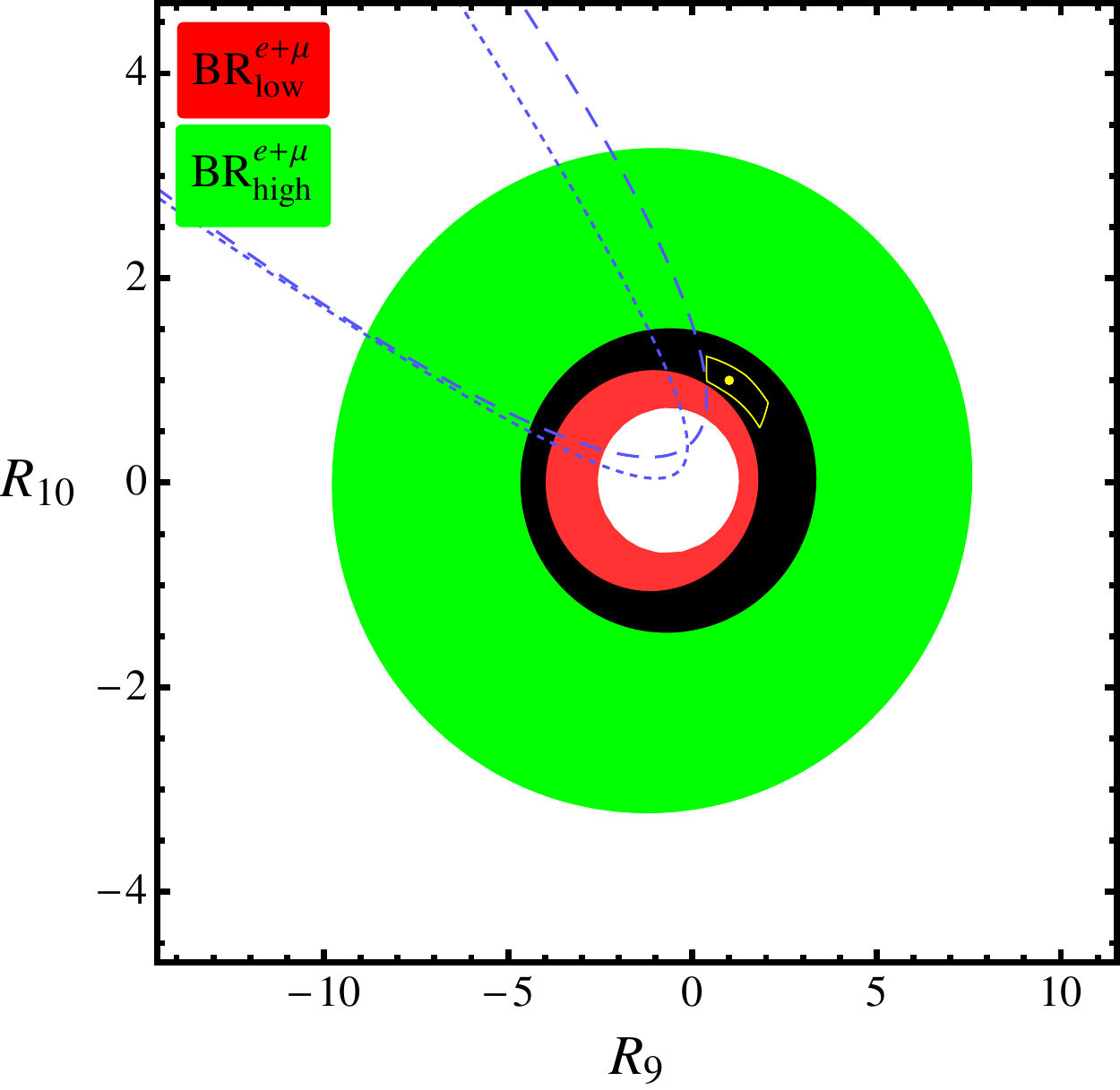}
\caption{Inclusive $b\to s \ell\ell$: constraints on Wilson coefficients.}
\label{EL:Ri}
\end{figure}
As for exclusive modes, it is necessary to cut the region around the two main charmonium resonances thus isolating two distinct regions at low-$q^2$ (further divided in the two bins $[1,3.5] \; {\rm GeV^2}$ and $[3.5,6] \; {\rm GeV^2}$) and high-$q^2$ ($q^2 > 14.4 \; {\rm GeV}^2$). The region below $1\; {\rm GeV}^2$ is mostly controlled by the almost real photon pole and is not very sensitive to the Wilson coefficients $C_9$ and $C_{10}$. The effect of resonances is included using the Kr\"uger-Sehgal method~\cite{Kruger:1996cv, Kruger:1996dt}. Note that the effect of charmonium resonances on the various  low-$q^2$ distributions is minimal (only the relatively well understood tail of the $J/\psi$ is relevant) while at high-$q^2$ they are much more prominent (see also the discussion in ref.~\cite{Lyon:2014hpa}). More importantly, the breakdown of the OPE at high-$q^2$ results in very large $1/m_b^2$ and $1/m_b^3$ power corrections that, because of our poor knowledge of the involved hadronic matrix elements, is a dominant source of uncertainty. Because of this the uncertainty on the total high-$q^2$ branching ratio (about 30\%) is much larger than the corresponding one at low-$q^2$ (about 6\%). As suggested in ref.~\cite{Ligeti:2007sn}, this problem can be ameliorated by normalizing the $B\to X_s \ell\ell$ width to the $B\to X_u \ell\nu$ width integrated over the same $q^2$ range. The total uncertainty on this observable (referred as ${\cal R}(s_0)$ in ref.~\cite{Huber:2015sra}) is then reduced to about 11\% (9\% of which is due to the error on $V_{ub}$ and is therefore expected to be sizably reduced in the future). 
\begin{figure}
\includegraphics[width=0.8 \linewidth]{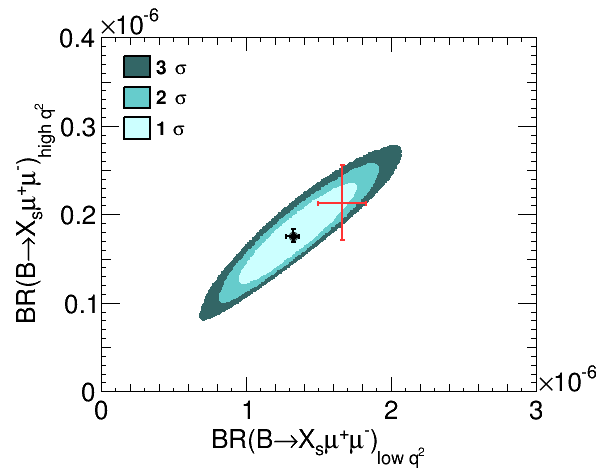}
\includegraphics[width=0.8 \linewidth]{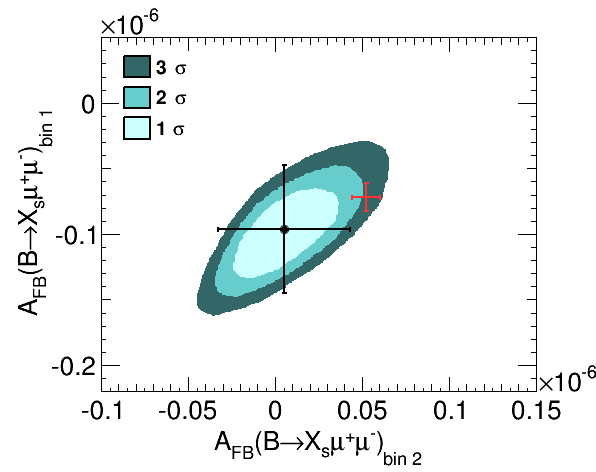}
\caption{Inclusive vs exclusive $b\to s \ell\ell$.}
\label{TH:inclexcl}
\end{figure}

An important point, that has been first noted in ref.~\cite{Huber:2005ig} is that impact of collinear photon emission (enhanced by the relatively large logarithm of the ratio $m_\ell^2/m_b^2$) from the final state leptons onto the differential double decay width. These effects vanish when integrated over the whole available $q^2$ phase space but not when one restricts the integration to various $q^2$ bins. As it was found in refs.~\cite{Huber:2005ig, Huber:2007vv, Huber:2015sra} these QED corrections, while affecting the total integrated branching ratio at low-$q^2$ at the 5\% level with respect to the NNLO QCD prediction, are extremely large when one considered the quantities ${\cal H}_L$ and ${\cal H}_T$ for which QED effects (for the electron channel) are -11\% and +73\%, respectively. In Fig.~\ref{EL:HI} we show the differential distributions for ${\cal H}_{T,L,A}$ (and for the branching ratio ${\cal H}_T+{\cal H}_L$) with (solid lines) and without (dashed lines) the inclusion of log-enhanced QED corrections. It is clear that the reason for the large effect on ${\cal H}_T$ is due to a suppression of the this observable with respect to ${\cal H}_L$ (this suppression is present already at tree-level and is not affected by QCD corrections) coupled with an accidental enhancement of QED effects (the relatively small positive shift on the branching ratio is obtained by a small negative contribution to ${\cal H}_L$ coupled with a larger positive shift on ${\cal H}_L$). This does not indicate a breakdown of the perturbative series because the large relative size of QED corrections is almost entirely due to the suppression of the tree-level plus QCD contribution, and not due to a large absolute value of the QED corrections. 
  
Experimental measurements of these inclusive observables (either by summing over exclusive final states or by fully/partially reconstructing the recoiling $B$ meson) can only be performed in the clean $B$-factories environments. Both BaBar~\cite{Aubert:2004it, Lees:2013nxa} and Belle~\cite{Iwasaki:2005sy} presented measurements of the rates at both low- and high-$q^2$ (Belle presented also a first measurement of the forward-backward asymmetry~\cite{Sato:2014pjr}). 

In Fig.~\ref{EL:Ri} we present the bounds on the ratios $R_{9,10} = C_{9,10} (\mu_0)/C_{9,10}^{\rm SM} (\mu_0)$ under the assumption of no new physics contributions to the magnetic and chromo-magnetic dipole operators (similar analyses were done, e.g., in~\cite{Ali:2002jg, Lee:2006gs}). The contours are the 95\% C.L.\ regions allowed by the BaBar and Belle experimental results; two sigma theoretical uncertainties are added linearly. We show the impact of the branching ratio measurement in the low-$q^2$ (red regions) and high-$q^2$ (green regions) and their overlap (black regions). The SM corresponds to the point $[R_9, R_{10}] = [1,1]$. The small yellow contour correspond to the Belle~II estimated reach with 50 ab${}^{-1}$ of integrated luminosity, assuming that the observed central values coincide with our predictions. The region outside the dashed and dotted parabola shaped regions are allowed by the Belle measurement of the normalized forward--backward asymmetry in the two low-$q^2$ bins. The resulting picture is in overall agreement with the SM expectations at the 95\% C.L.. We refer to ref.~\cite{Huber:2015sra} for a more extensive phenomenological discussion.

Finally, assuming that the anomalies in exclusive modes are indeed due to new physics in the semileptonic operators, one can extract allowed ranges for the inclusive branching ratios (low- and high-$q^2$) and forward--backward asymmetries (in the two low-$q^2$ bins) and check whether the Belle~II expected sensitivity to the inclusive modes will suffice to observe deviations from the SM predictions. The result of this study~\cite{Hurth:2014zja} is presented in Fig.~\ref{TH:inclexcl}. The shaded areas are the regions compatible with a new physics interpretation of the $B\to (K,K^*)\ell\ell$ anomalies at various confidence levels, the black point is the best fit result and the error bands correspond to the expected Belle~II total uncertainty. The red point and error bars indicate the SM predictions (under assumption of no new physics contributions). It is clear that the future measurement of the inclusive branching ratio and forward--backward asymmetry is able to detected the potential new physics contributions hinted at by the global fit.

%%%%%
\section*{Acknowledgments}

This document is an off-spring to a two-week long program on {\em Theory facing Experiment on Electroweak Symmetry Breaking, Flavour and Dark Matter\/} jointly organized by the Mainz Institute for Theoretical Physics (mitp.uni-mainz.de) and the Department of Physics of  the University of Naples Federico II. This event was held at the International Center for the Scientific Culture Villa Orlandi in the 
%village
small town  of Anacapri on the beautiful island of Capri. The program consisted of informal, black-board lectures in the (late) morning and extended discussion sessions in the afternoon. Besides physics, the participants enjoyed the gorgeous scenery of the island, its beaches and cultural beauties, amazing
% Italian
 food, and very good company. 
The very positive response
% which we have 
received on this first program has motivated the organization of a successor meeting.
%, which will take place at the same location from June 13-24, 2016.
%
\\[3 mm]
All the authors would like to thank both organizing  Institutions, the Chairpersons of the Institute, Matthias Neubert and Giulia Ricciardi, and the other members of the Steering Committee, Laura Covi, Csaba Csaki, Tobias Hurth and  Antonio Masiero.
Thanks are also due to the Physics  Department   of  the University of Naples Federico II, to the Rector Massimo Marrelli and to the Instituto Nazionale di Fisica Nucleare, for their support.
\\[3 mm]
G.R.   acknowledges partial support   by
Italian MIUR under project 2010YJ2NYW and INFN
under specific initiative QNP.
A.A. acknowledges partial support from the European Union FP7 ITN 
INVISIBLES (Marie Curie Actions, PITN-GA-2011-289442).
T.~Huber 
acknowledges support from Deutsche Forschungsgemeinschaft within research unit FOR 1873 (QFET).

\bibliography{biblioCapri11}

\end{document}